\documentclass[prd,preprint,tightenlines,floatfix,
preprintnumbers,nofootinbib,eqsecnum,superscriptaddress]{revtex4}

\usepackage[T1]{fontenc}		

\usepackage{amsmath,amsfonts,amssymb,amstext,mathrsfs}
\usepackage{mathpazo}

\usepackage[dvips]{graphicx}
\usepackage{epsf,float}
\usepackage{revsymb}

\usepackage{dcolumn}
\usepackage{braket}
\usepackage{color,xcolor}
\usepackage{graphicx}
\usepackage{subfigure}
\usepackage{multirow}
\usepackage{tabularx}
\usepackage{pstricks}
\usepackage[section]{placeins}
\usepackage{booktabs}
\usepackage{array}

\usepackage{hyperref}

\newcommand{\Pom}{\mathbb{P}}
\newcommand{\Ode}{\mathbb{O}}
\newcommand{\Reg}{\mathbb{R}}

\newcommand{\bdPt}{\mbox{\boldmath $dP_{t}$}}
\newcommand{\bqta}{\mbox{\boldmath $q_{t,1}$}}
\newcommand{\bqtb}{\mbox{\boldmath $q_{t,2}$}}
\newcommand{\bpta}{\mbox{\boldmath $p_{t,1}$}}
\newcommand{\bptb}{\mbox{\boldmath $p_{t,2}$}}

\begin{document}

\title{\boldmath
Central exclusive diffractive production of $K^{+} K^{-} K^{+} K^{-}$\\
via the intermediate $\phi \phi$ state in proton-proton collisions}

\vspace{0.6cm}

\author{Piotr Lebiedowicz}
 \email{Piotr.Lebiedowicz@ifj.edu.pl}
\affiliation{Institute of Nuclear Physics Polish Academy of Sciences, Radzikowskiego 152, PL-31342 Krak\'ow, Poland}

\author{Otto Nachtmann}
 \email{O.Nachtmann@thphys.uni-heidelberg.de}
\affiliation{Institut f\"ur Theoretische Physik, Universit\"at Heidelberg,
Philosophenweg 16, D-69120 Heidelberg, Germany}

\author{Antoni Szczurek
\footnote{Also at \textit{Faculty of Mathematics and Natural Sciences, University of Rzesz\'ow, Pigonia 1, PL-35310 Rzesz\'ow, Poland}.}}
\email{Antoni.Szczurek@ifj.edu.pl}
\affiliation{Institute of Nuclear Physics Polish Academy of Sciences, Radzikowskiego 152, PL-31342 Krak\'ow, Poland}

\begin{abstract}
We present a study of the exclusive $pp \to pp K^{+} K^{-} K^{+} K^{-}$ reaction at high energies.
We consider diffractive mechanisms with the intermediate 
$\phi \phi$ state with its decay into the $K^{+} K^{-} K^{+} K^{-}$ system.
We include the $\phi(1020)$ $\hat{t}/\hat{u}$-channel exchanges 
and the $f_{2}(2340)$ $s$-channel exchange mechanism.
This $f_{2}$ state is a candidate for a tensor glueball.
We discuss the possibility to use the $pp \to pp \phi \phi$ process 
in identifying the odderon exchange. 
An upper limit for the $\Pom \Ode \phi$ coupling is extracted from the WA102 experimental data.
The amplitudes for the processes are formulated
within the tensor-pomeron and vector-odderon approach.
We adjust parameters of our model to the WA102 data
and present several predictions for the ALICE, ATLAS, CMS and LHCb experiments.
Integrated cross sections of order of a few nb are obtained
including the experimental cuts relevant for the LHC experiments.
The distributions in the four-kaon invariant mass, 
rapidity distance between the two $\phi$ mesons,
special ``glueball filter variable'', 
proton-proton relative azimuthal angle are presented.
The distribution in rapidity difference of both $\phi$-mesons
could shed light on the $f_{2}(2340) \to \phi \phi$ coupling, not known at present.
We discuss the possible role of
the $f_{0}(2100)$, $\eta(2225)$, and $X(2500)$ resonances observed 
in the $\phi \phi$ channel in radiative decays of $J/\psi$.
Using typical kinematic cuts for LHC experiments
we find from our model that the odderon-exchange contribution
should be distinguishable from other contributions
for large rapidity distance between the $\phi$ mesons
and in the region of large four-kaon invariant masses.
At least, it should be possible to derive an upper limit 
on the odderon contribution in this reaction.
\end{abstract}


\maketitle

\section{Introduction}
\label{sec:intro}

Diffractive studies are one of the important parts of the physics program
for the RHIC and LHC experiments. 
A particularly interesting class is the central-exclusive-production (CEP) processes, where
all centrally produced particles are detected;
see Sec.~5 of \cite{N.Cartiglia:2015gve}. 
In recent years, there has been a renewed interest in exclusive production 
of $\pi^+ \pi^-$ pairs at high energies
related to successful experiments by 
the CDF \cite{Aaltonen:2015uva} and the CMS \cite{Khachatryan:2017xsi} collaborations.
These measurements are important in the context of resonance production, 
in particular, in searches for glueballs.
The experimental data on central exclusive $\pi^{+}\pi^{-}$ production
measured at Fermilab and CERN 
all show visible structures in the $\pi^{+}\pi^{-}$ invariant mass.
As we discussed in Ref.~\cite{Lebiedowicz:2016ioh}
the pattern of these structures has a mainly resonant origin and is very sensitive to
the cuts used in a particular experiment
(usually these cuts are different for different experiments).
In the CDF and CMS experiments, the large rapidity gaps around the centrally
produced dimeson system 
are checked, but the forward- and backward-going (anti)protons are not detected.
Preliminary results of similar CEP studies have been presented by the ALICE \cite{Schicker:2012nn}
and LHCb \cite{McNulty:2016sor} collaborations at the LHC.
Although such results will have a diffractive nature, 
further efforts are needed to ensure their exclusivity.
Ongoing and planned experiments at the RHIC (see, e.g., \cite{Sikora:2018cyk})
and future experiments at the LHC will be able to detect all particles
produced in central exclusive processes, 
including the forward- and backward-going protons.
Feasibility studies for the $p p \to p p \pi^+ \pi^-$ process with 
tagging of the scattered protons as carried out 
for the \mbox{ATLAS} and ALFA detectors are shown in \cite{Staszewski:2011bg}.
Similar possibilities exist using the CMS and TOTEM detectors; 
see, e.g., \cite{Albrow:2014lrm}.

It was known for a long time that the frequently 
used vector-pomeron model has problems from the point of view of field theory.
Taken literally it gives opposite signs for $pp$ and $\bar{p}p$ total 
cross sections.
A way to solve these problems was discussed in \cite{Nachtmann:1991ua},
where the pomeron was described as a coherent superposition of exchanges
with spin 2 + 4 + 6 + ... 
The same idea is realised in the tensor-pomeron model formulated in \cite{Ewerz:2013kda}.
In this model, pomeron exchange can effectively be treated as the
exchange of a rank-2 symmetric tensor.
In \cite{Ewerz:2016onn} it was shown that the tensor-pomeron model 
is consistent with the experimental data 
on the helicity structure of proton-proton elastic scattering at
$\sqrt{s} = 200$~GeV and small $|t|$ from the STAR experiment~\cite{Adamczyk:2012kn}.
In~Ref.~\cite{Lebiedowicz:2013ika} the tensor-pomeron model was applied to 
the diffractive production of several scalar and pseudoscalar mesons 
in the reaction $p p \to p p M$.
In \cite{Bolz:2014mya} an extensive study of the photoproduction reaction
$\gamma p \to \pi^{+} \pi^{-} p$ in the framework
of the tensor-pomeron model was presented.
The resonant ($\rho^0 \to \pi^{+}\pi^{-}$) and nonresonant (Drell-S\"oding)
photon-pomeron/reggeon $\pi^{+} \pi^{-}$ production in $pp$ collisions
was studied in \cite{Lebiedowicz:2014bea}.
The~central exclusive diffractive production of the $\pi^{+} \pi^{-}$ continuum 
together with the dominant scalar $f_{0}(500)$, $f_{0}(980)$, 
and tensor $f_{2}(1270)$ resonances was studied by us in \cite{Lebiedowicz:2016ioh}.
The $\rho^{0}$ meson production associated with 
a very forward/backward $\pi N$ system
in the $pp \to pp \rho^{0} \pi^{0}$ and $pp \to pn \rho^{0} \pi^{+}$ processes
was discussed in \cite{Lebiedowicz:2016ryp}.
Also the central exclusive $\pi^+ \pi^-\pi^+ \pi^-$ production 
via the intermediate $\sigma\sigma$ and $\rho^0\rho^0$ states in $pp$ collisions
was considered in \cite{Lebiedowicz:2016zka}.
In \cite{Lebiedowicz:2018sdt} the $pp \to pp p\bar{p}$ reaction was studied.
Recently, in \cite{Lebiedowicz:2018eui}, the exclusive diffractive production of the $K^{+} K^{-}$
in the continuum and via the dominant scalar $f_{0}(980)$, $f_{0}(1500)$, $f_{0}(1710)$,
and tensor $f_{2}(1270)$,  $f'_{2}(1525)$ resonances, 
as well as the $K^{+} K^{-}$ photoproduction contributions, was discussed in detail.
In \cite{Lebiedowicz:2019por} a possibility to extract 
the pomeron-pomeron-$f_{2}(1270)$ [$\Pom \Pom f_{2}(1270)$] couplings
from the analysis of angular distributions in the $\pi^{+}\pi^{-}$ rest system was studied.

The identification of glueballs in the $pp \to pp \pi^{+} \pi^{-}$ reaction,
being analysed by the STAR, ALICE, ATLAS, CMS, and LHCb collaborations,
can be rather difficult, as the di-pion spectrum is dominated by the $q \bar{q}$ states
and mixing of the pure glueball states with nearby $q \bar{q}$ mesons is possible.
The partial wave analyses of future experimental data could be used in this context.
Studies of different decay channels in central exclusive production would be very valuable.
One of the promising reactions is $p p \to p p \phi \phi$
with both $\phi \equiv \phi(1020)$ mesons decaying into the $K^{+} K^{-}$ channel.
The advantage of this process for experimental studies is the following.
The $\phi(1020)$ is a narrow resonance and it can be easily identified in the $K^{+} K^{-}$ spectra.
On the other hand, non-$\phi \phi$ backgrounds in these spectra should have a broad distribution.
However, identification of possible glueball-like states in this channel
requires calculation/estimation both of resonant and continuum processes.
It is known from the WA102 analysis of various channels that 
the so-called ``glueball-filter variable'' ($\rm{dP_{t}}$) \cite{Close:1997pj},
defined by the difference of the transverse momentum vectors of the outgoing protons,
can be used to select out known $q \bar{q}$ states from non-$q \bar{q}$ candidates.
It was observed by the WA102 Collaboration 
(see, e.g., \cite{Barberis:1996iq,Barberis:1997ve,Barberis:1998ax,Barberis:1999cq,Barberis:2000em},
\cite{Kirk:2000ws, Kirk:2014nwa}) 
that all the undisputed $q \bar{q}$ states 
are suppressed at small $\rm{dP_{t}}$ in contrast to glueball candidates.
It is therefore interesting to make a similar study of the $\rm{dP_{t}}$ dependence
for the $\phi \phi$ system decaying into $K^{+}K^{-}K^{+}K^{-}$ 
in central $pp$ collisions at the LHC.

Structures in the $\phi \phi$ invariant-mass spectrum were observed by several experiments.
Broad $J^{PC} = 2^{++}$ structures around 2.3~GeV were reported 
in the inclusive \mbox{$\pi^{-} Be \to \phi \phi + X$} reaction \cite{Booth:1985kv,Booth:1985kr},
in the exclusive $\pi^{-} p \to \phi \phi n$ \cite{Etkin:1985se,Etkin:1987rj} and
$K^{-} p \to \phi \phi \Lambda$ \cite{Aston:1989gx,Aston:1990wf} reactions,
in central production \cite{Armstrong:1986ky,Armstrong:1989hz,Barberis:1998bq},
and in $p \bar{p}$ annihilations \cite{Evangelista:1998zg}.
In the radiative decay $J/\psi \to \gamma \phi \phi$
an enhancement near ${\rm M}_{\phi \phi} = 2.25$~GeV 
with preferred $J^{PC} = 0^{-+}$ was observed 
\cite{Bisello:1986pt,Bai:1990hk,Ablikim:2008ac,Ablikim:2016hlu}.
The last partial wave analysis \cite{Ablikim:2016hlu} 
shows that the $\eta(2225)$ state is significant,
but a large contribution from the direct decay of $J/\psi \to \gamma \phi \phi$,
modeled by a $0^{-+}$ phase space distribution of the $\phi \phi$ system,
was also found there.
Also the scalar state $f_{0}(2100)$ and two additional pseudoscalar states, 
$\eta(2100)$ and the $X(2500)$, were observed.
Three tensor states, $f_{2}(2010)$, $f_{2}(2300)$, and $f_{2}(2340)$, 
observed previously in \cite{Etkin:1985se,Etkin:1987rj},
were also observed in $J/\psi \to \gamma \phi \phi$.
It was concluded there that the tensor spectrum is dominated by the $f_{2}(2340)$.
The nature of these resonances is not understood at present
and a tensor glueball has still not been clearly identified.
According to lattice-QCD simulations, the lightest tensor glueball 
has a mass between 2.2 and 2.4~GeV; 
see, e.g., \cite{Morningstar:1997ff,Morningstar:1999rf,Hart:2001fp,
Loan:2005ff,Gregory:2012hu,Chen:2005mg,Sun:2017ipk}.
The $f_{2}(2300)$ and $f_{2}(2340)$ states are good candidates to be tensor glueballs.
For an experimental work indicating a possible tensor glueball, see \cite{Longacre:2004jn}.
Also lattice-QCD predictions for the production rate of the pure gauge tensor glueball
in radiative $J/\psi$ decays \cite{Yang:2013xba}
are consistent with the large production rate of the $f_{2}(2340)$
in the $\eta \eta$ \cite{Ablikim:2013hq}, $\phi \phi$ \cite{Ablikim:2016hlu} 
and $K_{S}K_{S}$ \cite{Ablikim:2018izx} channels.

We have presented here some discussion of the role of resonances with masses around 2~GeV
in connection with their possible glueball interpretations.
With this we want to underline the importance of the study of resonances
in this mass range. Our present paper aims to facilitate such studies,
for instance, by investigating in detail the interplay of continuum
and resonance production of $\phi \phi$ states.
But we emphasize that in the following we make no assumptions
if the resonances considered are glueballs or not.

In the present paper we wish to concentrate on the CEP of four charged kaons
via the intermediate $\phi \phi$ state.
Here we shall give explicit expressions for the $pp \to pp \phi \phi$ amplitudes 
involving the pomeron-pomeron fusion to $\phi \phi$ ($\Pom \Pom \to \phi \phi$)
through the continuum processes, due to the $\hat{t}$- and $\hat{u}$-channel 
reggeized $\phi$-meson, photon, and odderon exchanges,
as well as through the $s$-channel resonance reaction ($\Pom \Pom \to f_{2}(2340) \to \phi \phi$).
The pseudoscalar mesons having $I^{G} = 0^{+}$ and
$J^{PC}=0^{-+}$ can also be produced in pomeron-pomeron fusion
and may contribute to our reaction if they decay to $\phi \phi$.
Possible candidates are, e.g., $\eta(2225)$ and $X(2500)$,
which were observed in radiative decays of $J/\psi$~\cite{Ablikim:2016hlu}.
The same holds for scalar states with $I^{G} = 0^{+}$ and $J^{PC}=0^{++}$,
for example, the scalar $f_{0}(2100)$ meson.
We will comment on the possible influence of these contributions 
for the CEP of $\phi \phi$ pairs.
Some model parameters will be determined from the comparison 
to the WA102 experimental data \cite{Barberis:1998bq,Barberis:2000em}.
In order to give realistic predictions we shall include absorption effects
calculated at the amplitude level and related to the $pp$ nonperturbative interactions.

\section{Exclusive diffractive production of four kaons}
\label{sec:section_two_meson}

In the present paper we consider the $2 \to 6$ process, CEP of four $K$ mesons,
with the intermediate $\phi(1020) \phi(1020)$ resonance pair,
\begin{eqnarray}
pp \to pp \, \phi \phi \to pp K^{+} K^{-}K^{+} K^{-}\,.
\label{2to6_reaction_phiphi}
\end{eqnarray} 

\begin{figure}
(a)\includegraphics[width=7.cm]{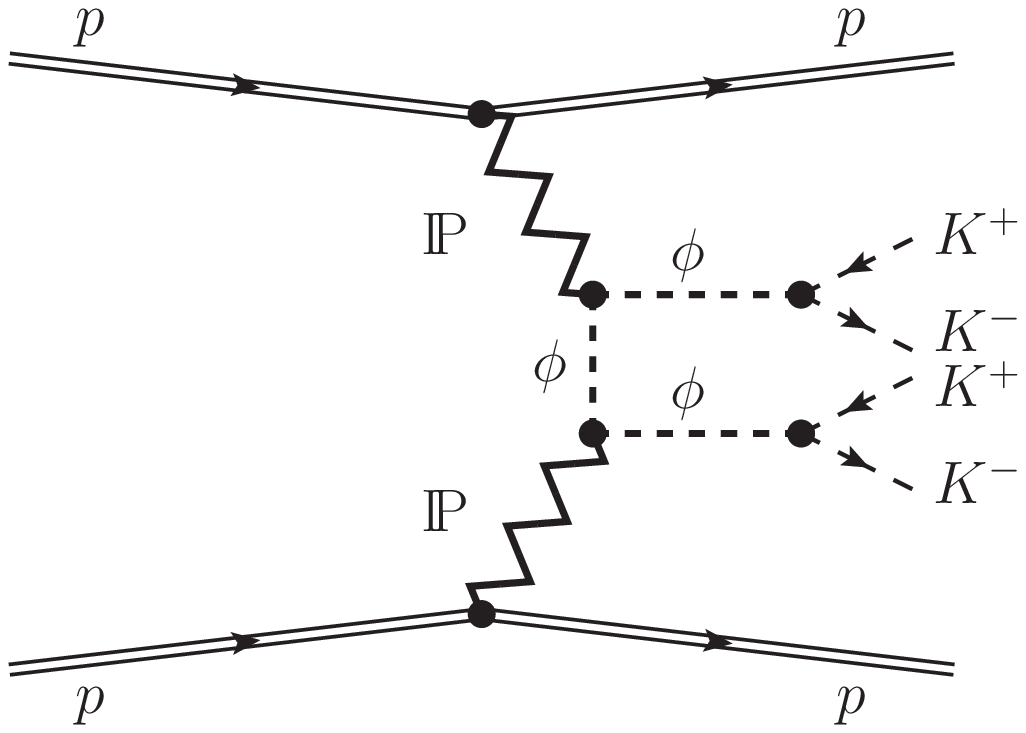}    
(b)\includegraphics[width=7.cm]{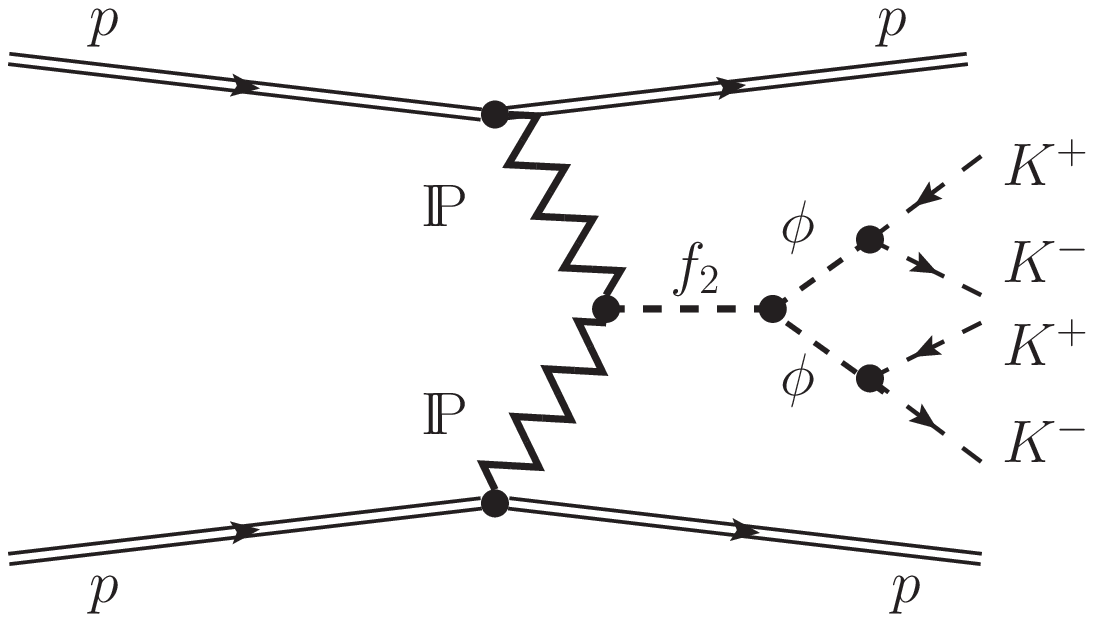}  
  \caption{\label{fig:4K_diagrams}
The ``Born-level'' diagrams for double pomeron central exclusive $\phi \phi$ production 
and their subsequent decays into $K^+ K^- K^+ K^-$ in proton-proton collisions:
(a)~continuum $\phi\phi$ production;
(b)~$\phi\phi$ production via an $f_{2}$ resonance.
Other resonances, e.g. of $f_{0}$- and $\eta$-type, can also contribute here.
}
\end{figure}

In Fig.~\ref{fig:4K_diagrams} we show diagrams for this process which are expected to be
the most important ones at high energies 
since they involve pomeron exchanges.
Figure~\ref{fig:4K_diagrams}~(a) shows the continuum process.
In Fig.~\ref{fig:4K_diagrams}~(b) we have the process with intermediate production of
an $f_{2}$ resonance,
\begin{eqnarray}
pp \to pp \, (\Pom \Pom \to f_{2} \to \phi \phi) \to pp \, K^{+} K^{-} K^{+} K^{-}\,.
\label{2to6_reaction_f2phiphi}
\end{eqnarray} 
In the place of the $f_{2}$ we can also have an $f_{0}$- and an $\eta$-type resonance.
That is, we treat effectively the $2 \to 6$ processes 
(\ref{2to6_reaction_phiphi}) and (\ref{2to6_reaction_f2phiphi})
as arising from the $2 \to 4$ process,
the central diffractive production of
two vector $\phi(1020)$ mesons in proton-proton collisions.

In Fig.~\ref{fig:4K_diagrams}~(a) we have the exchange
of a $\phi$ or $\phi_{\Reg}$ reggeon, depending on the kinematics,
as we shall discuss in detail below.
In place of the $\phi$ or $\phi_{\Reg}$ we can, in principle,
also have an $\omega$ or $\omega_{\Reg}$.
But these contributions are expected to be very small since the $\phi$ is
nearly a pure $s \bar{s}$ state, the $\omega$ nearly a pure $u \bar{u} + d \bar{d}$ state.
In the following we shall, therefore, neglect such contributions.

The production of $\phi \phi$ can also occur through diagrams of the type of
Fig.~\ref{fig:4K_diagrams} but with reggeons in the place of the pomerons.
For example, in Fig.~\ref{fig:4K_diagrams}~(a) we can replace the pomerons
by $\phi_{\Reg}$ reggeons and the intermediate $\phi$ by a pomeron.
In Fig.~\ref{fig:4K_diagrams}~(b) we can replace one or two pomerons
by one or two $f_{2 \Reg}$ reggeons. For high energies and
central $\phi \phi$ production such reggeon contributions
are expected to be small and we shall not consider them
in our present paper. We shall treat in detail the diagrams with pomeron exchange
(Fig.~\ref{fig:4K_diagrams}) and diagrams involving odderon
and also photon exchange; see Figs.~\ref{fig:odderon} and \ref{fig:gamma} below.

A resonance produced in pomeron-pomeron fusion must have $I^{G} = 0^{+}$
and charge conjugation $C = +1$, but it may have various
spin and parity quantum numbers.
See e.g. the discussion in Appendix~A of \cite{Lebiedowicz:2013ika}.

In Table~\ref{table:table} we have listed intermediate resonances
that can contribute to the $pp \to pp \phi \phi$ reaction (\ref{2to6_reaction_f2phiphi})
and to other processes with two vector mesons in the final state.
It must be noted that the scalar state $f_{0}(2100)$ 
and three pseudoscalar states, $\eta(2100)$, $\eta(2225)$, and $X(2500)$,
which were observed in the process $J/\psi \to \gamma \phi \phi$ \cite{Ablikim:2016hlu},
are only listed in PDG~\cite{Tanabashi:2018oca}
and are not included in the summary tables.
Clearly these states need confirmation.
\begin{table}[!h]
\caption{A list of resonances, up to a mass of 2500~MeV, that decay into a vector meson pair.
The~meson masses $m$ and their total widths $\Gamma$ 
are taken from PDG~\cite{Tanabashi:2018oca}. 
For $\eta(2100)$ and $X(2500)$, the information is taken from \cite{Ablikim:2016hlu}.
In the first column, the $\bullet$~symbol indicates rather established particles.
In the fifth column, the (?)~symbol denotes the states that need further experimental confirmation.}
\begin{tabular}{|r|c|l|l|c|c|c|c|}
\hline
Meson		&$I^{G}J^{PC}$& $m$ (MeV) 	& $\Gamma$ (MeV) & 
$\phi \phi$	& $K^{*0} \bar{K}^{*0}$ & $\rho^{0} \rho^{0}$ & $\omega \omega$ \\ 
\hline
$\bullet \, f_{1}(1285)$&$0^{+}1^{++}$&	$1281.9 \pm 0.5$ & $22.7 \pm 1.1$ 	 &  & & Seen & \\
$\bullet \, f_{0}(1370)$&$0^{+}0^{++}$&	$1200 - 1500$ & $200 - 500$ 	 &  &  & Dominant & Not seen \\
$\bullet \, f_{0}(1500)$&$0^{+}0^{++}$&	$1504 \pm 6$ & $109 \pm 7$ 	 &  &  & Seen &         \\ 
$f_{2}(1565)$&$0^{+}2^{++}$&	$1562 \pm 13$ & $134 \pm 8$ 	 &  &  & Seen     & Seen\\ 
$f_{2}(1640)$&$0^{+}2^{++}$&	$1639 \pm 6$ & $99^{+60}_{-40}$ 	 &  &  & & Seen \\ 
$\bullet \, f_{0}(1710)$&$0^{+}0^{++}$&	$1723^{+6}_{-5}$ & $139 \pm 8$ 	 &  &  & & Seen \\ 
$\eta(1760)$&$0^{+}0^{-+}$&	$1751 \pm 15$ & $240 \pm 30$ 	 &  &  & Seen & Seen \\ 
$f_{2}(1910)$&$0^{+}2^{++}$&	$1903 \pm 9$  & $196 \pm 31$         &      &      & Seen & Seen\\
$\bullet \, f_{2}(1950)$&$0^{+}2^{++}$&	$1944 \pm 12$ & $472 \pm 18$         &      & Seen &   
&     \\
$\bullet \, f_{2}(2010)$&$0^{+}2^{++}$&	$2011^{+60}_{-80}$ & $202 \pm 60$    & Seen &      &      &     \\
$f_{0}(2020)$&$0^{+}0^{++}$&	$1992 \pm 16$ & $442 \pm 60$         &      &      & Seen & Seen\\
$f_{0}(2100)$&$0^{+}0^{++}$&	$2101 \pm 7$ & $224^{+23}_{-21}$ &  Seen (?)    &      &  & \\
$\eta(2100)$&$0^{+}0^{-+}$&	$2050^{+30+75}_{-24-26}$ \cite{Ablikim:2016hlu}& $250^{+36+181}_{-30-164}$ \cite{Ablikim:2016hlu} & Seen (?) &      &      &     \\
$\bullet \, f_{4}(2050)$&$0^{+}4^{++}$&	$2018 \pm 11$ & $237 \pm 18$         &      &      &      & Seen \\
$f_{J}(2220)$&$0^{+}(2^{++} \, \rm{or} \, 4^{++})$&	$2231.1 \pm 3.5$ & $23^{+8}_{-7}$    & Not seen &  &      &     \\
$\eta(2225)$&$0^{+}0^{-+}$&	$2221^{+13}_{-10}$ & $185^{+40}_{-20}$         & Seen (?) &      &      &     \\
$\bullet \, f_{2}(2300)$&$0^{+}2^{++}$&	$2297 \pm 28$ & $149 \pm 40$         & Seen &      &      &     \\
$f_{4}(2300)$&$0^{+}4^{++}$&	$2320 \pm 60$ & $250 \pm 80$         &      &      & Seen & Seen\\
$\bullet \, f_{2}(2340)$&$0^{+}2^{++}$&	$2345^{+50}_{-40}$ & $322^{+70}_{-60}$ & Seen &      &      &     \\
$X(2500)$&$0^{+}0^{-+}$&	$2470^{+15+101}_{-19-23}$ \cite{Ablikim:2016hlu} & $230^{+64+56}_{-35-33}$ \cite{Ablikim:2016hlu} & Seen (?) &      &      &     \\
\hline
\end{tabular}
\label{table:table}
\end{table}

To calculate the total cross section for the $2 \to 4$ reactions
one has to calculate \mbox{the 8-dimensional} phase-space integral
\footnote{In the integration over four-body phase space
the transverse momenta of the produced particles ($p_{1t}$, $p_{2t}$, $p_{3t}$, $p_{4t}$),
the azimuthal angles of the outgoing protons ($\phi_{1}$, $\phi_{2}$)
and the rapidities of the produced mesons ($\rm{y}_{3}$, $\rm{y}_{4}$)
were chosen as integration variables over the phase space.} numerically \cite{Lebiedowicz:2009pj}.
Some modifications of the $2 \to 4$ reaction are needed
to simulate the $2 \to 6$ reaction with 
$K^{+} K^{-} K^{+} K^{-}$ in the final state.
For example, since the $\phi(1020)$ is an unstable particle 
one has to include a smearing of the $\phi$ masses due to their resonance distribution.
Then, the general cross-section formula can be written approximately as
\begin{eqnarray}
{\sigma}_{2 \to 6} &=&
[{\cal B}(\phi(1020) \to K^{+} K^{-})]^{2} \nonumber\\
&&\times \int_{2 m_{K}}^{{\rm max}\{m_{X_{3}}\}} \int_{2 m_{K}}^{{\rm max}\{m_{X_{4}}\}}
{\sigma}_{2 \to 4}(...,m_{X_{3}},m_{X_{4}})\,
f_{\phi}(m_{X_{3}})\, f_{\phi}(m_{X_{4}}) \,dm_{X_{3}}\, dm_{X_{4}}
\qquad
\label{4K_amplitude}
\end{eqnarray}
with the branching fraction ${\cal B}(\phi(1020) \to K^{+} K^{-}) = 0.492$ \cite{Tanabashi:2018oca}.
We use for the calculation of the decay process
$\phi(1020) \to K^{+} K^{-}$ the spectral function 
\begin{equation}
f_{\phi}(m_{X_{i}}) = C_{\phi}\,
\left( 1-\dfrac{4 m_{K}^{2}}{m_{X_{i}}^{2}} \right)^{3/2}
\frac{\frac{2}{\pi}{m_{\phi}^{2}} \Gamma_{\phi}}{(m_{X_{i}}^{2}-m_{\phi}^{2})^{2} + m_{\phi}^{2} \Gamma_{\phi}^{2}}\,,
\label{spectral_function}
\end{equation}
where $i  = 3, 4$, $\Gamma_{\phi}$ is the total width of the $\phi(1020)$ resonance,
$m_{\phi}$ its mass, and
$C_{\phi}= 64.1$ is found from the condition
\begin{equation}
\int_{2 m_{K}}^{\infty} f_{\phi}(m_{X_{i}}) dm_{X_{i}} = 1\,.
\label{spectral_function_aux}
\end{equation}
%
The quantity $\left( 1-4 m_{K}^{2}/m_{X_{i}}^{2} \right)^{3/2}$
smoothly decreases the spectral function when approaching the $K^{+}K^{-}$ threshold,
$m_{X_{i}} \to 2 m_{K}$, and takes into account the angular momentum $l=1$ of the $K^{+}K^{-}$ state.

To include experimental cuts on charged kaons
we perform the decays of $\phi$ mesons isotropically
\footnote{This is true for unpolarised $\phi$'s.
In principle our model also makes predictions for the polarisation of the $\phi$'s
and the anisotropies of the resulting $K^{+}K^{-}$ decay distributions.
Once a good event generator for our reaction is available, all
of these effects should be included.}
in the $\phi$ rest frames and then use relativistic
transformations to the overall center-of-mass frame.

In principle, there are other processes contributing to 
the $K^{+} K^{-} K^{+} K^{-}$ final state,
for example, direct $K^{+} K^{-} K^{+} K^{-}$ continuum production
and processes with $f_{0,2}$ resonances:
\begin{eqnarray}
&& pp \to pp \, K^{+} K^{-} K^{+} K^{-} \,,
\label{2to6_reaction_KKKK}\\
&& pp \to pp \, f_{0,2}\, K^{+} K^{-} \to pp \, K^{+} K^{-} K^{+} K^{-} \,,
\label{2to6_reaction_f0KK}\\
&& pp \to pp \, f_{0,2}\, f_{0,2} \to pp \, K^{+} K^{-} K^{+} K^{-} \,,
\label{2to6_reaction_f0f0}\\
&& pp \to pp \, (f_{2} \to f_{0}\, f_{0}) \to pp \, K^{+} K^{-} K^{+} K^{-}\,.
\label{2to6_reaction_f2f0f0}
\end{eqnarray} 
Here $f_{0,2}$ stands for one of the scalar or tensor mesons decaying to $K^{+} K^{-}$.
It should be noted that a complete theoretical model of the $pp \to pp K^{+}K^{-}K^{+}K^{-}$ process
should include interference effects of the processes 
(\ref{2to6_reaction_phiphi}), (\ref{2to6_reaction_f2phiphi}), and
(\ref{2to6_reaction_KKKK})--(\ref{2to6_reaction_f2f0f0}).
However, such a detailed study of the $pp \to pp K^{+}K^{-}K^{+}K^{-}$ reaction 
will only be necessary once high-energy experimental data 
for the purely exclusive measurements will be available.
We leave this interesting problem for future studies.
The GenEx Monte Carlo generator \mbox{\cite{Kycia:2014hea,Kycia:2017ota}}
could be used in this context.
We refer the reader to Ref.~\cite{Kycia:2017iij} where a first calculation
of four-pion continuum production in the $pp \to pp \pi^{+}\pi^{-}\pi^{+}\pi^{-}$ reaction
with the help of the GenEx code was performed.

\section{The reaction $pp \to pp \phi \phi$}
\label{sec:section_phiphi}
Here we discuss the exclusive production of 
$\phi \phi \equiv \phi(1020) \phi(1020)$ in proton-proton collisions,
\begin{eqnarray}
p(p_{a},\lambda_{a}) + p(p_{b},\lambda_{b}) \to
p(p_{1},\lambda_{1}) + \phi(p_{3},\lambda_{3}) + \phi(p_{4},\lambda_{4}) + p(p_{2},\lambda_{2}) \,,
\label{2to4_reaction_phiphi}
\end{eqnarray}
where 
$p_{a,b}$, $p_{1,2}$ and $\lambda_{a,b}, \lambda_{1,2} = \pm \frac{1}{2}$ 
denote the four-momenta and helicities of the protons and
$p_{3,4}$ and $\lambda_{3,4} = 0, \pm 1$ 
denote the four-momenta and helicities of the $\phi$ mesons, respectively.

The amplitude for the reaction (\ref{2to4_reaction_phiphi}) can be written as
\begin{equation}
\begin{split}
{\cal M}_{\lambda_{a}\lambda_{b}\to\lambda_{1}\lambda_{2}\phi \phi} = 
\left(\epsilon^{(\phi)}_{\rho_{3}}(\lambda_{3})\right)^*
\left(\epsilon^{(\phi)}_{\rho_{4}}(\lambda_{4})\right)^*
{\cal M}^{\rho_{3} \rho_{4}}_{\lambda_{a}\lambda_{b}\to\lambda_{1}\lambda_{2} \phi \phi}\,,
\end{split}
\label{amplitude_phiphi}
\end{equation}
where $\epsilon^{(\phi)}_{\mu}(\lambda)$ are the polarisation vectors of the $\phi$ meson.

We consider here unpolarised protons
in the initial state and no observation of polarisations in the final state.
Therefore, we have to insert in (\ref{4K_amplitude})
the cross section $\sigma_{2 \to 4}$ summed over the $\phi$ meson polarisations.
The spin sum for a $\phi$ meson of momentum $k$ and squared mass $k^{2}=m_{X}^{2}$ is
\begin{equation}
\begin{split}
\sum_{\lambda = 0, \pm 1}
\epsilon^{(\phi)\,\mu}(\lambda)
\left(\epsilon^{(\phi)\,\nu}(\lambda)\right)^* =
-g^{\mu \nu} + \dfrac{k^{\mu}k^{\nu}}{m_{X}^{2}}\,.
\end{split}
\label{spinsum}
\end{equation}
But in our model the $k^{\mu}k^{\nu}$ terms do not contribute to the cross section
since we have the relations
\begin{equation}
\begin{split}
p_{3\, \rho_{3}} {\cal M}^{\rho_{3} \rho_{4}}_{\lambda_{a}\lambda_{b}\to\lambda_{1}\lambda_{2}\phi\phi} =0\,, \qquad
p_{4\, \rho_{4}} {\cal M}^{\rho_{3} \rho_{4}}_{\lambda_{a}\lambda_{b}\to\lambda_{1}\lambda_{2}\phi\phi} =0
\,,
\end{split}
\label{spinsum_aux}
\end{equation}
which will be shown below in Secs.~\ref{sec:pompom_phiphi} and \ref{sec:f2_phiphi}.

Taking also into account the statistical factor $\frac{1}{2}$ due to the identity
of the two $\phi$ mesons we get for the amplitudes squared 
[to be inserted in $\sigma_{2 \to 4}$ in (\ref{4K_amplitude})]
\begin{equation}
\begin{split}
\frac{1}{2} \frac{1}{4} \sum_{\rm{spins}}
\Big|{\cal M}_{\lambda_{a}\lambda_{b}\to\lambda_{1}\lambda_{2}\phi \phi}
\Big|^{2}
=\frac{1}{8} \sum_{\lambda_{a},\lambda_{b},\lambda_{1},\lambda_{2}}
\left({\cal M}^{\sigma_{3} \sigma_{4}}_{\lambda_{a}\lambda_{b}\to\lambda_{1}\lambda_{2}\phi\phi}\right)^{*}
{\cal M}^{\rho_{3} \rho_{4}}_{\lambda_{a}\lambda_{b}\to\lambda_{1}\lambda_{2}\phi\phi}\,
g_{\sigma_{3}\rho_{3}} \,g_{\sigma_{4}\rho_{4}}
\,.
\end{split}
\label{amplitude_squared_phiphi}
\end{equation}

To give the full physical amplitude for the $pp \to pp \phi \phi$ reaction
we include absorptive corrections to the Born amplitudes discussed below.
For the details of how to include the $pp$-rescattering corrections 
in the eikonal approximation for the four-body reaction see Sec.~3.3 of \cite{Lebiedowicz:2014bea}.

\subsection{\boldmath{$\phi$}-meson exchange mechanism}
\label{sec:pompom_phiphi}

The diagram for the $\phi \phi$ production with an intermediate $\phi$-meson exchange
is shown in Fig.~\ref{fig:4K_diagrams}~(a).
The Born-level amplitude can be written as the sum
\begin{eqnarray}
{\cal M}^{(\phi{\rm-exchange})\,\rho_{3} \rho_{4}}_{\lambda_{a}\lambda_{b}\to\lambda_{1}\lambda_{2} \phi \phi} =
{\cal M}^{({\hat{t}})\,\rho_{3} \rho_{4}}_{\lambda_{a} \lambda_{b} \to \lambda_{1} \lambda_{2} \phi\phi} +
{\cal M}^{({\hat{u}})\,\rho_{3} \rho_{4}}_{\lambda_{a} \lambda_{b} \to \lambda_{1} \lambda_{2} \phi\phi} 
\label{2to4_reaction_SS_pompom}
\end{eqnarray}
with the $\hat{t}$- and $\hat{u}$-channel amplitudes:
\begin{equation}
\begin{split}
{\cal M}^{({\hat{t}})}_{\rho_{3} \rho_{4}} 
= 
& (-i) \bar{u}(p_{1}, \lambda_{1}) 
i\Gamma^{(\Pom pp)}_{\mu_{1} \nu_{1}}(p_{1},p_{a}) 
u(p_{a}, \lambda_{a})\,
i\Delta^{(\Pom)\, \mu_{1} \nu_{1}, \alpha_{1} \beta_{1}}(s_{13},t_{1})  \\
& \times 
i\Gamma^{(\Pom \phi\phi)}_{\rho_{1} \rho_{3} \alpha_{1} \beta_{1}}(\hat{p}_{t},-p_{3})\,
i\Delta^{(\phi)\,\rho_{1} \rho_{2}}(\hat{p}_{t})\,
i\Gamma^{(\Pom \phi\phi)}_{\rho_{4} \rho_{2} \alpha_{2} \beta_{2}}(p_{4},\hat{p}_{t})  \\
& \times 
i\Delta^{(\Pom)\, \alpha_{2} \beta_{2}, \mu_{2} \nu_{2}}(s_{24},t_{2}) \,
\bar{u}(p_{2}, \lambda_{2}) 
i\Gamma^{(\Pom pp)}_{\mu_{2} \nu_{2}}(p_{2},p_{b}) 
u(p_{b}, \lambda_{b}) \,,
\end{split}
\label{amplitude_t}
\end{equation}
\begin{equation} 
\begin{split}
{\cal M}^{({\hat{u}})}_{\rho_{3} \rho_{4}} 
= 
& (-i) \bar{u}(p_{1}, \lambda_{1}) 
i\Gamma^{(\Pom pp)}_{\mu_{1} \nu_{1}}(p_{1},p_{a}) 
u(p_{a}, \lambda_{a})\,
i\Delta^{(\Pom)\, \mu_{1} \nu_{1}, \alpha_{1} \beta_{1}}(s_{14},t_{1})  \\
& \times 
i\Gamma^{(\Pom \phi\phi)}_{\rho_{4} \rho_{1} \alpha_{1} \beta_{1}}(p_{4},\hat{p}_{u})\,
i\Delta^{(\phi)\,\rho_{1} \rho_{2}}(\hat{p}_{u})\,
i\Gamma^{(\Pom \phi\phi)}_{\rho_{2} \rho_{3} \alpha_{2} \beta_{2}}(\hat{p}_{u},-p_{3}) \\
& \times 
i\Delta^{(\Pom)\, \alpha_{2} \beta_{2}, \mu_{2} \nu_{2}}(s_{23},t_{2}) \,
\bar{u}(p_{2}, \lambda_{2}) 
i\Gamma^{(\Pom pp)}_{\mu_{2} \nu_{2}}(p_{2},p_{b}) 
u(p_{b}, \lambda_{b}) \,,
\end{split}
\label{amplitude_u}
\end{equation}
where $\hat{p}_{t} = p_{a} - p_{1} - p_{3}$,
$\hat{p}_{u} = p_{4} - p_{a} + p_{1}$, $s_{ij} = (p_{i} + p_{j})^{2}$,
$t_1 = (p_{1} - p_{a})^{2}$, and
$t_2 = (p_{2} - p_{b})^{2}$.
Here $\Delta^{(\Pom)}$ and $\Gamma^{(\Pom pp)}$ 
denote the effective propagator and proton vertex function,
respectively, for the tensorial pomeron.
The corresponding expressions, as given in Sec.~3 of \cite{Ewerz:2013kda}, are as follows:
\begin{eqnarray}
i \Delta^{(\Pom)}_{\mu \nu, \kappa \lambda}(s,t) &=& 
\frac{1}{4s} \left( g_{\mu \kappa} g_{\nu \lambda} 
                  + g_{\mu \lambda} g_{\nu \kappa}
                  - \frac{1}{2} g_{\mu \nu} g_{\kappa \lambda} \right)
(-i s \alpha'_{\Pom})^{\alpha_{\Pom}(t)-1}\,,
\label{A1}\\
i\Gamma_{\mu \nu}^{(\Pom pp)}(p',p)
&=&-i 3 \beta_{\Pom NN} F_{1}(t)
\left\lbrace 
\frac{1}{2} 
\left[ \gamma_{\mu}(p'+p)_{\nu} 
     + \gamma_{\nu}(p'+p)_{\mu} \right]
- \frac{1}{4} g_{\mu \nu} ( p\!\!\!/' + p\!\!\!/ )
\right\rbrace, \qquad \quad
\label{A4}
\end{eqnarray}
where $\beta_{\Pom NN} =1.87$~GeV$^{-1}$.
For extensive discussions of the properties of these terms we refer to \cite{Ewerz:2013kda}.
Here the pomeron trajectory $\alpha_{\Pom}(t)$
is assumed to be of standard linear form
(see, e.g., \cite{Donnachie:1992ny,Donnachie:2002en}):
\begin{eqnarray}
&&\alpha_{\Pom}(t) = \alpha_{\Pom}(0)+\alpha'_{\Pom}\,t, \nonumber \\
&&\alpha_{\Pom}(0) = 1.0808\,, \;\;
\alpha'_{\Pom} = 0.25 \; {\rm GeV}^{-2}\,.
\label{trajectory}
\end{eqnarray}
%

Our ansatz for the $\Pom \phi\phi$ vertex follows the one for the $\Pom \rho\rho$ 
in (3.47) of \cite{Ewerz:2013kda} 
with the replacements $a_{\Pom \rho \rho} \to a_{\Pom \phi \phi}$ and 
$b_{\Pom \rho \rho} \to b_{\Pom \phi \phi}$.
This was already used in Sec.~IV~B of \cite{Lebiedowicz:2018eui}.
The $\Pom \phi\phi$ vertex function is taken with the same Lorentz structure 
as for $f_{2} \gamma \gamma$ defined in (3.39) of \cite{Ewerz:2013kda}.
With $k', \mu$ and $k,\nu$ the momentum and vector index
of the outgoing and incoming $\phi$, respectively,
and $\kappa \lambda$ the pomeron indices,
the $\Pom \phi\phi$ vertex reads
\begin{eqnarray}
i\Gamma^{(\Pom \phi \phi)}_{\mu \nu \kappa \lambda}(k',k) =
i F_{M}((k'-k)^{2}) \left[
2a_{\Pom \phi \phi}\,  
\Gamma^{(0)}_{\mu \nu \kappa \lambda}(k',-k)\,
- b_{\Pom \phi \phi}\,\Gamma^{(2)}_{\mu \nu \kappa \lambda}(k',-k) \right] \,
\label{vertex_pomphiphi}
\end{eqnarray}  
%
with two rank-four tensor functions,
\begin{eqnarray}
\label{3.16}
&&\Gamma_{\mu\nu\kappa\lambda}^{(0)} (k_1,k_2) =
\Big[(k_1 \cdot k_2) g_{\mu\nu} - k_{2\mu} k_{1\nu}\Big] 
\Big[k_{1\kappa}k_{2\lambda} + k_{2\kappa}k_{1\lambda} - 
\frac{1}{2} (k_1 \cdot k_2) g_{\kappa\lambda}\Big] \,,\\
\label{3.17}
&&\Gamma_{\mu\nu\kappa\lambda}^{(2)} (k_1,k_2) = \,
 (k_1\cdot k_2) (g_{\mu\kappa} g_{\nu\lambda} + g_{\mu\lambda} g_{\nu\kappa} )
+ g_{\mu\nu} (k_{1\kappa} k_{2\lambda} + k_{2\kappa} k_{1\lambda} ) \nonumber \\
&& \qquad \qquad \qquad \quad - k_{1\nu} k_{2 \lambda} g_{\mu\kappa} - k_{1\nu} k_{2 \kappa} g_{\mu\lambda} 
- k_{2\mu} k_{1 \lambda} g_{\nu\kappa} - k_{2\mu} k_{1 \kappa} g_{\nu\lambda} 
\nonumber \\
&& \qquad \qquad \qquad \quad - [(k_1 \cdot k_2) g_{\mu\nu} - k_{2\mu} k_{1\nu} ] \,g_{\kappa\lambda} \,;
\end{eqnarray}
see Eqs.~(3.18) and (3.19) of \cite{Ewerz:2013kda}.
In (\ref{vertex_pomphiphi}) the coupling parameters 
$a_{\Pom \phi \phi}$ and $b_{\Pom \phi \phi}$ 
have dimensions GeV$^{-3}$ and GeV$^{-1}$, respectively.
In \cite{Lebiedowicz:2018eui} we have fixed the coupling parameters 
of the tensor pomeron to the $\phi$ meson based on the HERA experimental data for 
the $\gamma p \to \phi p$ reaction \cite{Derrick:1996af,Breitweg:1999jy}.
We take the coupling constants \mbox{$a_{\Pom \phi \phi} = 0.49$~GeV$^{-3}$} 
and \mbox{$b_{\Pom \phi \phi} = 4.27$~GeV$^{-1}$}
from Table~II of \cite{Lebiedowicz:2018eui} (see also Sec. IV~B there).

In the hadronic vertices we should take into account form factors
since the hadrons are extended objects.
The form factors $F_{1}(t)$ in (\ref{A4}) 
and $F_{M}(t)$ in (\ref{vertex_pomphiphi})
are chosen here as the electromagnetic form factors only for simplicity,
\begin{eqnarray}
&&F_{1}(t)= \frac{4 m_{p}^{2}-2.79\,t}{(4 m_{p}^{2}-t)(1-t/m_{D}^{2})^{2}}\,, 
\label{F1t}\\
&&F_{M}(t)= \frac{1}{1-t/\Lambda_{0}^{2}}\,;
\label{FMt}
\end{eqnarray}
see Eqs.~(3.29) and (3.34) of \cite{Ewerz:2013kda}, respectively.
In (\ref{F1t}) $m_{p}$ is the proton mass and \mbox{$m_{D}^{2} = 0.71$~GeV$^{2}$}
is the dipole mass squared.
As we discussed in Fig.~6 of \cite{Lebiedowicz:2018eui} 
we should take in (\ref{FMt}) $\Lambda_{0}^{2} = 1.0$~GeV$^{2}$ 
instead of $\Lambda_{0}^{2} = 0.5$~GeV$^{2}$ used for the $\Pom \rho \rho$ vertex in \cite{Ewerz:2013kda}.

Then, with the expressions for the propagators, vertices, and form factors,
from \cite{Ewerz:2013kda}
${\cal M}^{\rho_{3} \rho_{4}}$ can be written
in the high-energy approximation as
\begin{equation}
\begin{split}
& {\cal M}^{(\phi{\rm-exchange})\,\rho_{3} \rho_{4}}_{\lambda_{a}\lambda_{b}\to\lambda_{1}\lambda_{2}\phi\phi}
= \;
2 (p_1 + p_a)_{\mu_{1}} (p_1 + p_a)_{\nu_{1}}\, 
\delta_{\lambda_{1} \lambda_{a}} \, F_{1}(t_{1}) \,F_{M}(t_{1})\\
& \times 
\bigg\{
{V}^{\rho_{3} \rho_{1} \mu_{1} \nu_{1}}(s_{13}, t_{1}, \hat{p}_{t}, p_{3})\;
\Delta^{(\phi)}_{\rho_{1}\rho_{2}}(\hat{p}_{t})\;
{V}^{\rho_{4} \rho_{2} \mu_{2} \nu_{2}}(s_{24}, t_{2}, -\hat{p}_{t}, p_{4}) 
\, \left[ \hat{F}_{\phi}(\hat{p}_{t}^{2}) \right]^{2}\\
& \quad \; +
{V}^{\rho_{4} \rho_{1} \mu_{1} \nu_{1}}(s_{14}, t_{1}, -\hat{p}_{u}, p_{4})\;
\Delta^{(\phi)}_{\rho_{1}\rho_{2}}(\hat{p}_{u})\;
{V}^{\rho_{3} \rho_{2} \mu_{2} \nu_{2}}(s_{23}, t_{2}, \hat{p}_{u}, p_{3}) 
\, \left[ \hat{F}_{\phi}(\hat{p}_{u}^{2}) \right]^{2}
\bigg\}   \\
& \times 2 (p_2 + p_b)_{\mu_{2}} (p_2 + p_b)_{\nu_{2}}\, 
\delta_{\lambda_{2} \lambda_{b}} \, F_{1}(t_{2}) \,F_{M}(t_{2}) \,,
\end{split}
\label{amplitude_approx}
\end{equation}
where ${V}_{\mu \nu \kappa \lambda}$ reads as
\begin{equation}
\begin{split}
{V}_{\mu \nu \kappa \lambda}(s,t,k_{2},k_{1})= &
\frac{1}{4s} \,3 \beta_{\Pom NN} \, 
(- i s \alpha'_{\Pom})^{\alpha_{\Pom}(t)-1} \bigg[
 2 a_{\Pom \phi \phi} \Gamma_{\mu \nu \kappa \lambda}^{(0)}(k_{1},k_{2})
-b_{\Pom \phi \phi}  \Gamma_{\mu \nu \kappa \lambda}^{(2)}(k_{1},k_{2}) \bigg]\,.
\end{split}
\label{tensorial_function_aux2}
\end{equation}
The amplitude (\ref{amplitude_approx})
contains a form factor $\hat{F}_{\phi}(\hat{p}^{2})$
taking into account the off-shell dependencies of the intermediate $\phi$-mesons.
The form factor is normalised to unity at the on-shell point
$\hat{F}_{\phi}(m_{\phi}^{2}) = 1$ and parametrised here 
in the exponential form,
\begin{eqnarray} 
\hat{F}_{\phi}(\hat{p}^{2})=
\exp\left(\frac{\hat{p}^{2}-m_{\phi}^{2}}{\Lambda^{2}_{off,E}}\right) \,,
\label{off-shell_form_factors_exp} 
\end{eqnarray}
where the cutoff parameter $\Lambda_{off,E}$ could be adjusted to experimental data.

The relations (\ref{spinsum_aux}) are now easily checked from
(\ref{amplitude_approx}) and (\ref{tensorial_function_aux2})
using the properties of the tensorial functions (\ref{3.16}) and (\ref{3.17}); 
see (3.21) of \cite{Ewerz:2013kda}.
We can then make in (\ref{amplitude_approx}) the following replacement 
for the $\phi$-meson propagator:
\begin{eqnarray}
\Delta^{(\phi)}_{\rho_{1}\rho_{2}}(\hat{p}) \to -g_{\rho_{1}\rho_{2}} \, 
\Delta^{(\phi)}_{T}(\hat{p}^{2})\,,
\label{delta_phi}
\end{eqnarray}
where we take for $\hat{p}^{2} < 0$,
where $\Delta^{(\phi)}_{T}(\hat{p}^{2})$ must be real,
the simple lowest order expression
$(\Delta^{(\phi)}_{T}(\hat{p}^{2}))^{-1} = \hat{p}^{2}-m_{\phi}^{2}$.

We should take into account the fact that the exchanged 
intermediate object is not a simple spin-1 particle ($\phi$ meson) but may correspond 
to a Regge exchange, that is, 
the reggeization of the intermediate $\phi$ meson is necessary
(see, e.g., \cite{Lebiedowicz:2016zka}).
A simple way to include approximately the ``reggeization'' of the amplitude 
given in Eq.~(\ref{amplitude_approx}) is
by replacing the $\phi$-meson propagator
in both the $\hat{t}$- and $\hat{u}$-channel amplitudes by 
%
\begin{eqnarray}
\Delta^{(\phi)}_{\rho_{1}\rho_{2}}(\hat{p}) \to
\Delta^{(\phi)}_{\rho_{1}\rho_{2}}(\hat{p}) 
\left( \exp (i \phi(s_{34}))\, \frac{s_{34}}{s_{{\rm thr}}} \right)^{\alpha_{\phi}(\hat{p}^{2})-1} \,,
\label{reggeization}
\end{eqnarray}
where 
\begin{eqnarray}
&&s_{34} = (p_{3}+p_{4})^{2} = {\rm M}_{\phi \phi}^{2}\,, \nonumber \\
&&s_{\rm thr} = 4 m_{\phi}^{2}\,. 
\label{sthr}
\end{eqnarray}
Here we assume for the $\phi$ Regge trajectory
\begin{eqnarray}
&&\alpha_{\phi}(\hat{p}^{2}) = \alpha_{\phi}(0)+\alpha'_{\phi}\,\hat{p}^{2}, \nonumber\\
&&\alpha_{\phi}(0) = 0.1\,, \;\;
\alpha'_{\phi} = 0.9 \; {\rm GeV}^{-2}\,;
\label{trajectory_phi}
\end{eqnarray}
see Eq.~(5.3.1) of \cite{Collins:1977}.
In order to have the correct phase behaviour 
we introduced in (\ref{reggeization}) the function $\exp (i \phi(s_{34}))$ with
\begin{eqnarray}
\phi(s_{34}) =\frac{\pi}{2}\exp\left(\frac{s_{\rm thr}-s_{34}}{s_{{\rm thr}}}\right)-\frac{\pi}{2}\,.
\label{reggeization_aux}
\end{eqnarray}
This procedure of reggeization 
assures agreement with mesonic physics in the $\phi \phi$ system
close to threshold, $s_{34} = 4 m_{\phi}^2$ (no suppression), and 
it gives the Regge behaviour at large $s_{34}$.
However, some care is needed here, 
as the reggeization is only expected 
in general to hold in the $|\hat{p}^{2}|/s_{34} \ll 1$ regime.
In the reaction considered, 
both $\langle -\hat{p}_{t}^{2} \rangle$ and $\langle -\hat{p}_{u}^{2} \rangle$ 
are of order 1~GeV$^{2}$ (before reggeization)
with a cutoff for higher $|\hat{p}^{2}|$ provided in (\ref{amplitude_approx})
by the form factors $\hat{F}_{\phi}(\hat{p}^{2})$ (\ref{off-shell_form_factors_exp}).
Therefore, the propagator form in (\ref{reggeization}) and (\ref{reggeization_aux})
gives correct Regge behaviour for $s_{34} - 4 m_{\phi}^{2} \gg 1$~GeV$^{2}$
and $|\hat{p}^{2}|$ limited by the form factors,
whereas for smaller $s_{34}$ we have mesonic behaviour.

In Ref.~\cite{Harland-Lang:2013dia} it was argued that 
the reggeization should not be applied
when the rapidity distance between two centrally produced mesons, 
$\rm{Y_{diff}} = \rm{Y}_{3} - \rm{Y}_{4}$, tends to zero
\mbox{(i.e. for $|\hat{p}^{2}| \sim s_{34}$)}.
Indeed, for small $\rm{Y_{diff}}$ the two $\phi$ mesons may have large transverse
momenta leading to a large ${\rm M}_{\phi \phi}$.
Clearly this kinematic region has nothing to do with the Regge limit.
For large $\rm{Y_{diff}}$, on the other hand,
the form factors $\hat{F}_{\phi}(\hat{p}^{2})$ in (\ref{amplitude_approx})
limit the transverse momenta of the $\phi$'s but ${\rm M}_{\phi \phi}$ will be large.
That is, there we are in the Regge limit.
To take care of these two different regimes we propose
to use, as an alternative to (\ref{reggeization}),
a formula for the $\phi$ propagator which interpolates continuously
between the regions of low $\rm{Y_{diff}}$, where we use the standard
$\phi$ propagator, and of high $\rm{Y_{diff}}$ where we use the reggeized form (\ref{reggeization}):
\begin{eqnarray}
\Delta^{(\phi)}_{\rho_{1}\rho_{2}}(\hat{p})
\to \Delta^{(\phi)}_{\rho_{1}\rho_{2}}(\hat{p}) \,F({\rm Y_{diff}})
+ 
\Delta^{(\phi)}_{\rho_{1}\rho_{2}}(\hat{p})\,
\left[ 1 - F({\rm Y_{diff}}) \right]
\left( \exp (i \phi(s_{34}))\, \frac{s_{34}}{s_{\rm thr}} \right)^{\alpha_{\phi}(\hat{p}^{2})-1},\qquad
\label{second_procedure}
\end{eqnarray}
%
with a simple function 
\begin{eqnarray}
F({\rm Y_{diff}}) = \exp\left( -{\rm c_{y}} | {\rm Y_{diff}} | \right)\,.
\label{second_procedure_aux}
\end{eqnarray}
%
Here ${\rm c_{y}}$ is an unknown parameter 
which measures how fast one approaches to the Regge regime.

In Sec.~\ref{sec:results} below we shall compare 
the two prescriptions of reggeization,
(\ref{reggeization}) and (\ref{second_procedure}); 
see Figs.~\ref{fig:reggeization} and \ref{fig:3}.
Furthermore, we shall show in Fig.~\ref{fig:3b} that
a large size of the rapidity gap between the two $\phi$ mesons indeed
means automatically also large ${\rm M}_{\phi \phi}$ in our model.

\subsection{\boldmath{$f_{2}$} resonance production}
\label{sec:f2_phiphi}

Now we consider the amplitude for the reaction (\ref{2to4_reaction_phiphi})
through the $s$-channel $f_{2}$-meson exchange
as shown in Fig.~\ref{fig:4K_diagrams} (b).
The $f_{2}(2010)$, $f_{2}(2300)$, and $f_{2}(2340)$ mesons could be considered 
as potential candidates; see Table~\ref{table:table}.

The Born amplitude for the $\Pom \Pom$ fusion is given by
\begin{equation}
\begin{split}
{\cal M}^{(\Pom \Pom \to f_{2} \to \phi\phi)\,\rho_{3} \rho_{4}}_{\lambda_{a}\lambda_{b}\to\lambda_{1}\lambda_{2}\phi\phi} 
= & (-i)\,
\bar{u}(p_{1}, \lambda_{1}) 
i\Gamma^{(\Pom pp)\,\mu_{1} \nu_{1}}(p_{1},p_{a}) 
u(p_{a}, \lambda_{a})\;
i\Delta^{(\Pom)}_{\mu_{1} \nu_{1}, \alpha_{1} \beta_{1}}(s_{1},t_{1}) \\
& \times 
i\Gamma^{(\Pom \Pom f_{2})\,\alpha_{1} \beta_{1},\alpha_{2} \beta_{2}, \rho \sigma}(q_{1},q_{2}) \;
i\Delta^{(f_{2})}_{\rho \sigma, \alpha \beta}(p_{34})\;
i\Gamma^{(f_{2} \phi\phi)\,\alpha \beta \rho_{3} \rho_{4}}(p_{3},p_{4})\\
& \times 
i\Delta^{(\Pom)}_{\alpha_{2} \beta_{2}, \mu_{2} \nu_{2}}(s_{2},t_{2}) \;
\bar{u}(p_{2}, \lambda_{2}) 
i\Gamma^{(\Pom pp)\,\mu_{2} \nu_{2}}(p_{2},p_{b}) 
u(p_{b}, \lambda_{b}) \,,
\end{split}
\label{amplitude_f2_pomTpomT}
\end{equation}
where
$s_{1} = (p_{1} + p_{3} + p_{4})^{2}$,
$s_{2} = (p_{2} + p_{3} + p_{4})^{2}$, 
$q_{1} = p_{a} - p_{1}$, 
$q_{2} = p_{b} - p_{2}$, 
$t_{1} = q_{1}^{2}$, $t_{2} = q_{2}^{2}$,
and
$p_{34} = q_{1} + q_{2} = p_{3} + p_{4}$.

The $\Pom \Pom f_{2}$ vertex, including a form factor, can be written as 
\begin{eqnarray}
i\Gamma_{\mu \nu,\kappa \lambda,\rho \sigma}^{(\Pom \Pom f_{2})} (q_{1},q_{2}) =
\left( i\Gamma_{\mu \nu,\kappa \lambda,\rho \sigma}^{(\Pom \Pom f_{2})(1)} \mid_{\rm bare}
+ \sum_{j=2}^{7}i\Gamma_{\mu \nu,\kappa \lambda,\rho \sigma}^{(\Pom \Pom f_{2})(j)}(q_{1},q_{2}) \mid_{\rm bare} 
\right)
\tilde{F}^{(\Pom \Pom f_{2})}(q_{1}^{2},q_{2}^{2},p_{34}^{2}) \,.\nonumber\\
\label{vertex_pompomT}
\end{eqnarray}
Here and throughout our paper 
the label ``bare'' is used for a vertex,
as derived from a corresponding coupling Lagrangian \cite{Lebiedowicz:2016ioh},
without a form-factor function.
A possible choice for the 
$i\Gamma_{\mu \nu,\kappa \lambda,\rho \sigma}^{(\Pom \Pom f_{2})(j)}\mid_{\rm bare}$
coupling terms $j = 1, ..., 7$ is given in Appendix~A of \cite{Lebiedowicz:2016ioh}.
The~corresponding coupling constants $g_{\Pom \Pom f_{2}}^{(j)}$ are not known 
and should be fitted to existing and future experimental data.
In the following we shall, for the purpose of orientation,
assume that only $g_{\Pom \Pom f_{2}}^{(1)}$ is unequal to zero.
But we have checked that for the distributions studied here
the choice of $\Pom \Pom f_{2}$ coupling is not important;
see Sec.~\ref{sec:section_4} below.

In practical calculations, to describe the off-shell dependence in (\ref{vertex_pompomT}),
we take the factorized form for the $\Pom \Pom f_{2}$ form factor
%
\begin{eqnarray}
\tilde{F}^{(\Pom \Pom f_{2})}(q_{1}^{2},q_{2}^{2},p_{34}^{2}) = 
\tilde{F}_{M}(q_{1}^{2}) \tilde{F}_{M}(q_{2}^{2}) F^{(\Pom \Pom f_{2})}(p_{34}^{2})\,
\label{Fpompommeson}
\end{eqnarray}
normalised to
$\tilde{F}^{(\Pom \Pom f_{2})}(0,0,m_{f_{2}}^{2}) = 1$.
%
We will further set
\begin{eqnarray}
&&\tilde{F}_{M}(t)=\frac{1}{1-t/\tilde{\Lambda}_{0}^{2}}\,,
\quad \tilde{\Lambda}_{0}^{2} = 1\;{\rm GeV}^{2}\,; 
\label{FM_t}\\
&&F^{(\Pom \Pom f_{2})}(p_{34}^{2}) = 
\exp{ \left( \frac{-(p_{34}^{2}-m_{f_{2}}^{2})^{2}}{\Lambda_{f_{2}}^{4}} \right)}\,,
\quad \Lambda_{f_{2}} = 1\;{\rm GeV}\,.
\label{Fpompommeson_ff}
\end{eqnarray}

For the $f_{2} \phi \phi$ vertex we take the following ansatz
(in analogy to the $f_{2}\gamma\gamma$ vertex; see (3.39) of \cite{Ewerz:2013kda}):
\begin{eqnarray}
i\Gamma^{(f_{2} \phi \phi)}_{\mu \nu \kappa \lambda}(p_{3},p_{4}) &=&
i\dfrac{2}{M_{0}^{3}} g'_{f_{2} \phi \phi}\,  
\Gamma^{(0)}_{\mu \nu \kappa \lambda}(p_{3},p_{4})\,
F'^{(f_{2} \phi \phi)}(p_{34}^{2}) \nonumber \\
&&- i \dfrac{1}{M_{0}} g''_{f_{2} \phi \phi}\,\Gamma^{(2)}_{\mu \nu \kappa \lambda}(p_{3},p_{4})\,
F''^{(f_{2} \phi \phi)}(p_{34}^{2})\,,
\label{vertex_f2phiphi}
\end{eqnarray}  
with $M_{0} = 1$~GeV and dimensionless coupling constants 
$g'_{f_{2} \phi \phi}$ and $g''_{f_{2} \phi \phi}$ 
being free parameters.
The explicit tensorial functions 
$\Gamma_{\mu \nu \kappa \lambda}^{(i)}(p_{3},p_{4})$, 
$i = 0, 2$, are given by (\ref{3.16}) and (\ref{3.17}), respectively.
The relations (\ref{spinsum_aux}) can now be checked from (\ref{amplitude_f2_pomTpomT}) 
and (\ref{vertex_f2phiphi}) using again (3.21) of~\cite{Ewerz:2013kda}.
Different form factors $F'$ and $F''$ are allowed \textit{a priori} in~(\ref{vertex_f2phiphi}).
We assume that
\begin{eqnarray}
F'^{(f_{2} \phi \phi)}(p_{34}^{2}) = 
F''^{(f_{2} \phi \phi)}(p_{34}^{2}) = 
F^{(\Pom \Pom f_{2})}(p_{34}^{2}) \,.
\label{Fpompommeson_ff_tensor}
\end{eqnarray}
%
%
%

In the high-energy approximation we can write the amplitude 
for the $\Pom \Pom$ fusion as
\begin{equation}
\begin{split}
& {\cal M}^{(\Pom \Pom \to f_{2} \to \phi\phi)\,\rho_{3} \rho_{4}}_{\lambda_{a}\lambda_{b}\to\lambda_{1}\lambda_{2}\phi\phi}
= 3 \beta_{\Pom NN}  \, 2(p_1 + p_a)_{\mu_{1}} (p_1 + p_a)_{\nu_{1}}\, 
\delta_{\lambda_{1} \lambda_{a}}\, F_1(t_1)  \;
\frac{1}{4 s_{1}} (- i s_{1} \alpha'_{\Pom})^{\alpha_{\Pom}(t_{1})-1} \\ 
& \quad \quad
\times 
\Gamma^{(\Pom \Pom f_{2})\,\mu_{1} \nu_{1}, \mu_{2} \nu_{2}, \alpha \beta}(q_{1},q_{2})\,
\Delta^{(f_{2})}_{\alpha \beta, \kappa \lambda}(p_{34})\,
\Gamma^{(f_{2} \phi \phi)\, \kappa \lambda \rho_{3} \rho_{4}}(p_{3},p_{4})\\
& \quad \quad\times 
\frac{1}{4 s_{2}} (- i s_{2} \alpha'_{\Pom})^{\alpha_{\Pom}(t_{2})-1}\,
3 \beta_{\Pom NN}  \, 2 (p_2 + p_b)_{\mu_{2}} (p_2 + p_b)_{\nu_{2}}\, 
\delta_{\lambda_{2} \lambda_{b}}\, F_1(t_2) \,.
\end{split}
\label{amplitude_approx_rhorho}
\end{equation}
%
We use in (\ref{amplitude_approx_rhorho}) the tensor-meson propagator with the simple Breit-Wigner form
\begin{eqnarray}
\Delta_{\mu \nu, \kappa \lambda}^{(f_{2})}(p_{34})&=&
\frac{1}{p_{34}^{2}-m_{f_{2}}^2+i m_{f_{2}} \Gamma_{f_{2}}}
\left[ 
\frac{1}{2} 
( \hat{g}_{\mu \kappa} \hat{g}_{\nu \lambda}  + \hat{g}_{\mu \lambda} \hat{g}_{\nu \kappa} )
-\frac{1}{3} 
\hat{g}_{\mu \nu} \hat{g}_{\kappa \lambda}
\right] \,, 
\label{prop_f2}
\end{eqnarray}
where $\hat{g}_{\mu \nu} = -g_{\mu \nu} + p_{34 \mu} p_{34 \nu} / p_{34}^2$,
$\Gamma_{f_{2}}$ is the total decay width of the $f_{2}$ resonance,
and $m_{f_{2}}$ is its mass. 
We take their numerical values from PDG~\cite{Tanabashi:2018oca};
see Table~\ref{table:table} in Sec.~\ref{sec:section_two_meson}.

\subsection{Pseudoscalar and scalar resonance production}
\label{sec:eta_phiphi}

As was mentioned in Sec.~\ref{sec:intro}, the scalar $f_{0}(2100)$
and the pseudoscalar $\eta(2100)$, $\eta(2225)$, and $X(2500)$ states 
were seen in $J/\psi \to \gamma \phi \phi$ \cite{Ablikim:2016hlu}.
In \cite{Ablikim:2016hlu} the authors found that the most significant contribution 
to $\phi \phi$ comes from the $\eta(2225)$ resonance.

The above resonances can also contribute to $\phi\phi$ CEP in addition to
the continuum and the $f_{2}(2340)$ processes discussed 
in Secs.~\ref{sec:pompom_phiphi} and \ref{sec:f2_phiphi}, respectively.
Therefore, in our analysis we should consider these possibilities.
But for simplicity we will limit our discussion to 
the CEP of the $f_{0}(2100)$ and the $\eta(2225)$ mesons
with subsequent decay to $\phi \phi$.

The Born amplitude for the $\Pom \Pom$ fusion to $\phi \phi$
through an $s$-channel $\eta$-like resonance $\widetilde{M}$ is given by
\begin{equation}
\begin{split}
{\cal M}^{(\Pom \Pom \to \widetilde{M} \to \phi\phi)\,\rho_{3} \rho_{4}}_{\lambda_{a}\lambda_{b}\to\lambda_{1}\lambda_{2}\phi\phi} 
= & (-i)\,
\bar{u}(p_{1}, \lambda_{1}) 
i\Gamma^{(\Pom pp)\,\mu_{1} \nu_{1}}(p_{1},p_{a}) 
u(p_{a}, \lambda_{a})\;
i\Delta^{(\Pom)}_{\mu_{1} \nu_{1}, \alpha_{1} \beta_{1}}(s_{1},t_{1}) \\
& \times 
i\Gamma^{(\Pom \Pom \widetilde{M})\,\alpha_{1} \beta_{1},\alpha_{2} \beta_{2}}(q_{1},q_{2}) \;
i\Delta^{(\widetilde{M})}(p_{34})\;
i\Gamma^{(\widetilde{M} \phi\phi)\,\rho_{3} \rho_{4}}(p_{3},p_{4})\\
& \times 
i\Delta^{(\Pom)}_{\alpha_{2} \beta_{2}, \mu_{2} \nu_{2}}(s_{2},t_{2}) \;
\bar{u}(p_{2}, \lambda_{2}) 
i\Gamma^{(\Pom pp)\,\mu_{2} \nu_{2}}(p_{2},p_{b}) 
u(p_{b}, \lambda_{b}) \,.
\end{split}
\label{amplitude_eta_pomTpomT}
\end{equation}
The effective $\Pom \Pom \widetilde{M}$ vertex was discussed in Sec.~2.2 of \cite{Lebiedowicz:2013ika}.
As was shown there, in general more than one coupling structure $\Pom \Pom \widetilde{M}$ is possible.
The general $\Pom \Pom \widetilde{M}$ vertex constructed in Sec.~2.2 of \cite{Lebiedowicz:2013ika}
corresponds to the sum of the values
$(l,S) = (1,1)$ and $(3,3)$
with the dimensionless coupling parameters
$g'_{\Pom \Pom \widetilde{M}}$ and $g''_{\Pom \Pom \widetilde{M}}$, respectively.
The~resulting $\Pom \Pom \widetilde{M}$ vertex, including a form factor, is given as follows
\begin{eqnarray}
i\Gamma_{\mu \nu,\kappa \lambda}^{(\Pom \Pom \widetilde{M})} (q_{1},q_{2}) &=&
\left( 
  i\Gamma_{\mu \nu,\kappa \lambda}'^{(\Pom \Pom \widetilde{M})}(q_{1},q_{2}) \mid_{\rm bare}
+ i\Gamma_{\mu \nu,\kappa \lambda}''^{(\Pom \Pom \widetilde{M})}(q_{1},q_{2}) \mid_{\rm bare} 
\right)
\tilde{F}^{(\Pom \Pom \widetilde{M})}(q_{1}^{2},q_{2}^{2},p_{34}^{2})\,,\nonumber \\
\label{vertex_pompomPS}\\
%
%
i\Gamma_{\mu \nu,\kappa \lambda}'^{(\Pom \Pom \widetilde{M})}(q_{1}, q_{2})\mid_{\rm bare} &=&
i \, \frac{g_{\Pom \Pom \widetilde{M}}'}{2 M_{0}} \,
\left( g_{\mu \kappa} \varepsilon_{\nu \lambda \rho \sigma}
      +g_{\nu \kappa} \varepsilon_{\mu \lambda \rho \sigma}
      +g_{\mu \lambda}\varepsilon_{\nu \kappa \rho \sigma}
      +g_{\nu \lambda}\varepsilon_{\mu \kappa \rho \sigma} \right) \nonumber \\
&&\times 
(q_{1}-q_{2})^{\rho} p_{34}^{\sigma}\,,
\label{vertex_pompomPS_11}\\
%
%
i\Gamma_{\mu \nu,\kappa \lambda}''^{(\Pom \Pom \widetilde{M})}(q_{1}, q_{2})\mid_{\rm bare} &=&
i \, \frac{g_{\Pom \Pom \widetilde{M}}''}{M_{0}^{3}} \, 
\lbrace \varepsilon_{\nu \lambda \rho \sigma} \left[ q_{1 \kappa} q_{2 \mu} - (q_{1} \cdot q_{2}) g_{\mu \kappa} \right] +
\varepsilon_{\mu \lambda \rho \sigma} \left[ q_{1 \kappa} q_{2 \nu} - (q_{1} \cdot q_{2}) g_{\nu \kappa} \right]  \nonumber \\
&&+
\varepsilon_{\nu \kappa \rho \sigma}  \left[ q_{1 \lambda} q_{2 \mu} - (q_{1} \cdot q_{2}) g_{\mu \lambda} \right] +
\varepsilon_{\mu \kappa \rho \sigma}  \left[ q_{1 \lambda} q_{2 \nu} - (q_{1} \cdot q_{2}) g_{\nu \lambda} \right] 
\rbrace \nonumber \\
&&\times 
(q_{1}-q_{2})^{\rho} p_{34}^{\sigma}\,;
\label{vertex_pompomPS_33}
\end{eqnarray}
see (2.4) and (2.6) of \cite{Lebiedowicz:2013ika}.
For $\widetilde{M} = \eta$ and $\widetilde{M} = \eta'(958)$,
the corresponding coupling constants were fixed in \cite{Lebiedowicz:2013ika} (see Table~4 there)
to differential distributions of the WA102 Collaboration \cite{Barberis:1998ax,Kirk:2000ws}.
For the $\Pom \Pom \eta(2225)$ coupling, relevant for CEP of $\phi \phi$,
there are no data to determine it.
Therefore, we consider, for simplicity, only the term $(l,S) = (1,1)$ in~(\ref{vertex_pompomPS}).
That is, we set $g''_{\Pom\Pom\eta(2225)}=0$.
We take the same factorized form for the pomeron-pomeron-$\eta(2225)$ form factor 
as in (\ref{Fpompommeson})--(\ref{Fpompommeson_ff}).

For the $\eta \phi \phi$ vertex we make the following ansatz:
%
\begin{eqnarray}
i\Gamma^{(\eta \phi \phi)}_{\mu \nu}(p_{3},p_{4}) =
i\dfrac{1}{2 M_{0}}g_{\eta \phi \phi}\,  
\varepsilon_{\mu \nu \kappa \lambda} p_{3}^{\kappa} p_{4}^{\lambda}\,
F^{(\eta \phi \phi)}(p_{34}^{2})\,,
\label{vertex_etaphiphi}
\end{eqnarray}  
with $M_{0} = 1$~GeV and $g_{\eta \phi \phi}$ being a free parameter.

The amplitude for $\phi \phi$ CEP through the scalar $f_{0}(2100)$ meson 
is as for $\eta(2225)$ in (\ref{amplitude_eta_pomTpomT}) but
with $\Gamma^{(\Pom \Pom \eta)}$, $\Gamma^{(\eta \phi \phi)}$, and $\Delta^{(\eta)}$
replaced by $\Gamma^{(\Pom \Pom f_{0})}$, $\Gamma^{(f_{0} \phi \phi)}$, and $\Delta^{(f_{0})}$,
respectively.
\mbox{In Appendix A of \cite{Lebiedowicz:2016zka}}, 
a similar amplitude for the reaction
$pp \to pp (f_{0} \to \rho^{0}\rho^{0})$ is written.
The effective $\Pom \Pom f_{0}$ vertex is discussed 
in detail in Appendix~A of \cite{Lebiedowicz:2013ika}.
As was shown there, the $\Pom \Pom f_{0}$ vertex corresponds to the sum of two $(l,S)$ couplings,
$(l,S) = (0,0)$ and $(2,2)$, with corresponding coupling parameters
$g'_{\Pom \Pom f_{0}}$ and $g''_{\Pom \Pom f_{0}}$, respectively.
The vertex is written as follows:
\begin{eqnarray}
i\Gamma_{\mu \nu,\kappa \lambda}^{(\Pom \Pom f_{0})} (q_{1},q_{2}) =
\left( 
  i\Gamma_{\mu \nu,\kappa \lambda}'^{(\Pom \Pom f_{0})} \mid_{\rm bare}
+ i\Gamma_{\mu \nu,\kappa \lambda}''^{(\Pom \Pom f_{0})}(q_{1},q_{2}) \mid_{\rm bare} 
\right)
\tilde{F}^{(\Pom \Pom f_{0})}(q_{1}^{2},q_{2}^{2},p_{34}^{2})\,;
\label{vertex_pompomS}
\end{eqnarray}
see (A.17)--(A.21) of \cite{Lebiedowicz:2013ika}.
Due to the same reason as for the $\eta(2225)$ meson, 
we restrict in (\ref{vertex_pompomS}) to one term $(l,S) = (0,0)$.
We take the same form for the pomeron-pomeron-$f_{0}(2100)$ form factor
as in (\ref{Fpompommeson})--(\ref{Fpompommeson_ff}).

In Appendix~A of \cite{Lebiedowicz:2016zka} we discussed our ansatz for the $f_{0} \rho \rho$ vertex;
see (A.7) there. 
For the $f_{0} \phi \phi$ vertex, of interest to us here,
we make the same ansatz but with coupling parameters $g'_{f_{0} \phi \phi}$ and $g''_{f_{0} \phi \phi}$
instead of $g'_{f_{0} \rho \rho}$ and $g''_{f_{0} \rho \rho}$, respectively.
For simplicity, we assume in the following $g'_{f_{0} \phi \phi} = 0$.
We get then
\begin{eqnarray}
i\Gamma^{(f_{0} \phi \phi)}_{\mu \nu}(p_{3},p_{4}) =
i\dfrac{2}{M_{0}}g''_{f_{0} \phi \phi}\,  
\left[ p_{4\,\mu} p_{3\,\nu} - (p_{3} \cdot p_{4}) g_{\mu \nu} \right]\,
F''^{(f_{0} \phi \phi)}(p_{34}^{2})\,,
\label{vertex_f0phiphi}
\end{eqnarray}  
where $g''_{f_{0} \phi \phi}$ is a parameter to be determined from experiment.
Here the $\Pom \Pom f_{0}(2100)$ and $f_{0}(2100) \phi \phi$ coupling parameters 
are essentially unknown at present.

\textit{A priori} different form factors 
$F^{(\eta \phi \phi)}$ and $F''^{(f_{0} \phi \phi)}$ are allowed
in (\ref{vertex_etaphiphi}) and (\ref{vertex_f0phiphi}), respectively.
We assume $F^{(\eta \phi \phi)} = F''^{(f_{0} \phi \phi)} = F^{(\Pom \Pom f_{2})}$; 
see Eq.~(\ref{Fpompommeson_ff}).

\subsection{Diffractive production of \boldmath{$\phi \phi$} continuum with odderon exchanges}
\label{sec:triple_diff}
\begin{figure}
(a)\includegraphics[width=7.cm]{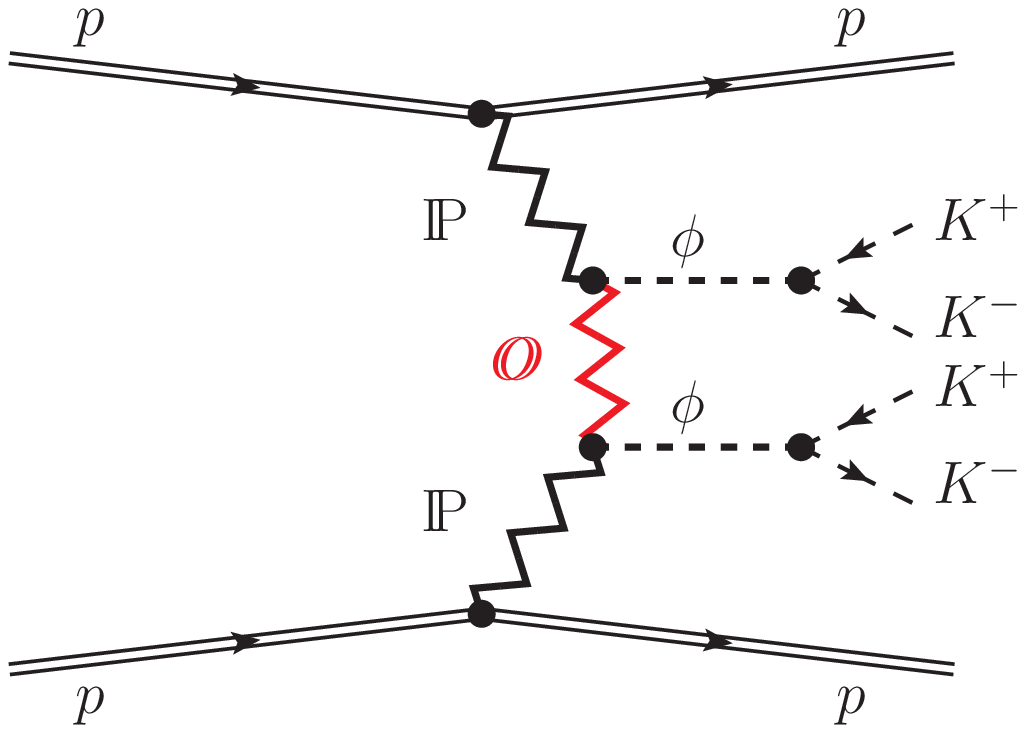}    
(b)\includegraphics[width=7.cm]{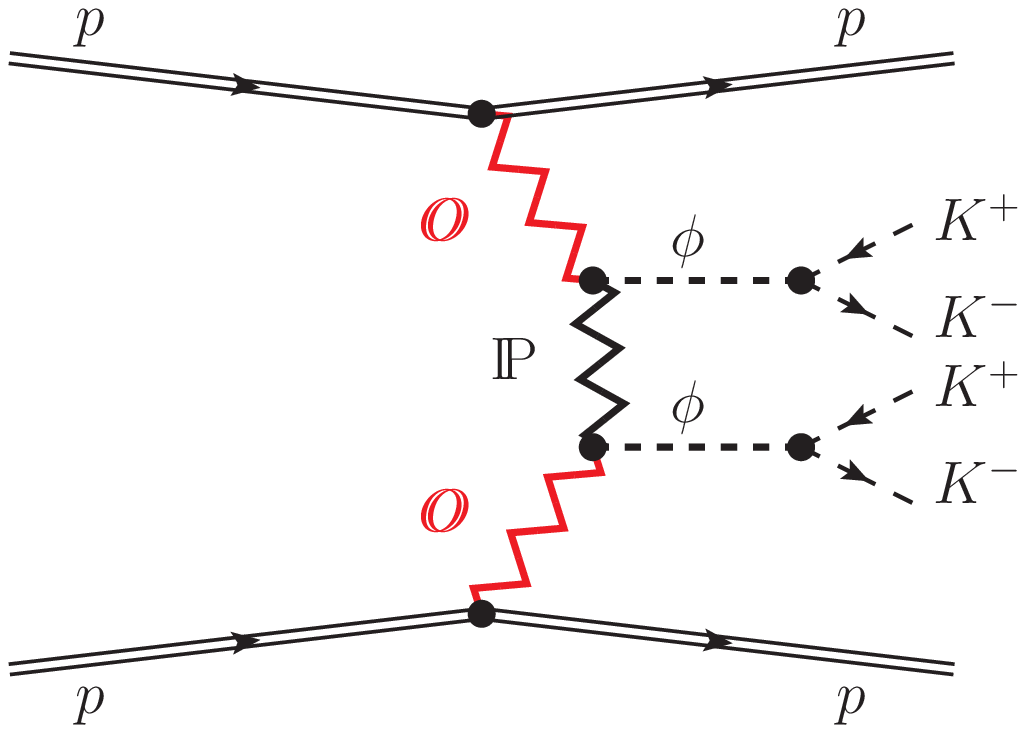}  
  \caption{\label{fig:odderon}
The Born-level diagrams for diffractive production of a $\phi$-meson pair
with one and two odderon exchanges.}
\end{figure}
The diffractive production of two $\phi$ mesons seems to offer a good possibility
to identify and/or study the odderon exchanges \cite{Ewerz:2003xi}.
At high energy there are two types of processes represented by the diagrams in Fig.~\ref{fig:odderon}.
So far these processes have not yet been calculated or even estimated.
A particularly important case worthy of attention is diagram (a) in Fig.~\ref{fig:odderon}.
The advantage of this process compared to that in diagram (b) 
is that in diagram (a) no odderon-proton vertex is involved.
Because the coupling of the odderon to the proton is probably small,
one could expect $\sigma^{(\Ode-\Pom-\Ode)} \ll \sigma^{(\Pom-\Ode-\Pom)}$.
Therefore, in the following we neglect the contribution 
with two odderon exchanges in the calculation.

The amplitude for the process shown by diagram (a) in Fig.~\ref{fig:odderon}
has the same form as the amplitude with the $\phi$-meson exchange
discussed in Sec.~\ref{sec:pompom_phiphi}; 
see Eqs.~(\ref{2to4_reaction_SS_pompom})--(\ref{amplitude_u}).
But here we have to make the following replacements:
\begin{eqnarray}
&&i\Delta^{(\phi)}_{\mu \nu}(\hat{p}) 
\to i\Delta^{(\Ode)}_{\mu \nu}(s_{34},\hat{p}^{2})\,,
\\
&&i\Gamma^{(\Pom \phi\phi)}_{\mu \nu \kappa \lambda}(k',k)
\to i\Gamma^{(\Pom \Ode \phi)}_{\mu \nu \kappa \lambda}(k',k)\,.
\end{eqnarray}
Our ansatz for the effective propagator of the $C = -1$ odderon 
follows (3.16) and (3.17) of~\cite{Ewerz:2013kda},
\begin{eqnarray}
&&i \Delta^{(\Ode)}_{\mu \nu}(s,t) = 
-i g_{\mu \nu} \frac{\eta_{\Ode}}{M_{0}^{2}} (-i s \alpha'_{\Ode})^{\alpha_{\Ode}(t)-1}\,,
\label{A12} \\
&&\alpha_{\Ode}(t) = \alpha_{\Ode}(0)+\alpha'_{\Ode}\,t\,,
\label{A13}
\end{eqnarray}
where in (\ref{A12}) we have $M_{0}^{-2} = 1$~(GeV)$^{-2}$ for dimensional reasons.
Furthermore, $\eta_{\Ode}$ is a parameter with value $\pm 1$ and $\alpha_{\Ode}(t)$
is the odderon trajectory, assumed to be linear in~$t$.
We choose, as an example, the slope parameter for the odderon
the same as for the pomeron in (\ref{trajectory}).
For the odderon intercept we choose a number of representative values.
That is, we shall show results for
\begin{eqnarray}
\eta_{\Ode} = \pm 1\,, \;\;
\alpha'_{\Ode} = 0.25 \; {\rm GeV}^{-2}\,,\;\;
\alpha_{\Ode}(0) = 1.05,\, 1.00,\, 0.95\,.
\label{A14}
\end{eqnarray}

The odderon-exchange diagram presented in Fig.~\ref{fig:odderon}~(a),
due to the Regge-based parametrisation with the odderon intercept $\alpha_{\Ode}(0) \sim 1.0$,
should be especially relevant in the region of large rapidity separation of the $\phi$ mesons 
and large $\phi \phi$ invariant masses.
This will be discussed further in Sec.~\ref{sec:section_6}.

\begin{figure}
(a)\includegraphics[width=6.cm]{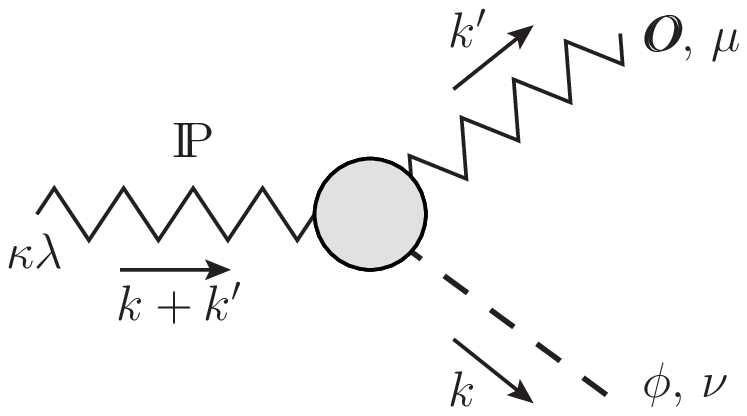} 
(b)\includegraphics[width=6.cm]{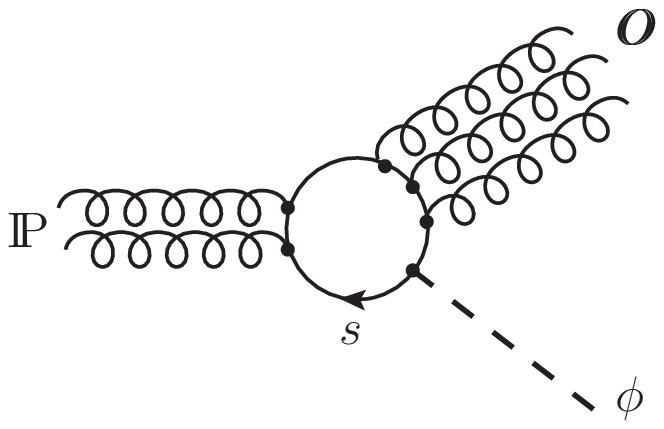} 
  \caption{\label{QCD_POphi_coupling}
(a) Generic diagram for the $\Pom \Ode \phi$ vertex 
with momentum and Lorentz-indices assignments.
(b) A QCD diagram contributing to the $\Pom \Ode \phi$ vertex.}
\end{figure}

For the $\Pom \Ode \phi$ vertex we use an ansatz analogous to the $\Pom \rho \rho$ vertex;
see (3.47) of \cite{Ewerz:2013kda}. 
We get then,
orienting the momenta of the $\Ode$ and the $\phi$ outwards
as shown in Fig.~\ref{QCD_POphi_coupling}~(a), the following formula:
\begin{eqnarray}
i\Gamma^{(\Pom \Ode \phi)}_{\mu \nu \kappa \lambda}(k',k) =
i F^{(\Pom \Ode \phi)}((k+k')^{2},k'^{2},k^{2})
\left[ 2\,a_{\Pom \Ode \phi}\, \Gamma^{(0)}_{\mu \nu \kappa \lambda}(k',k)
- b_{\Pom \Ode \phi}\,\Gamma^{(2)}_{\mu \nu \kappa \lambda}(k',k) \right].\qquad
\label{A15}
\end{eqnarray}  
Here $k',\mu$ and $k,\nu$ are the momentum and vector index of the odderon
and the $\phi$, respectively;
$a_{\Pom \Ode \phi}$ and $b_{\Pom \Ode \phi}$ are (unknown) coupling constants;
and $F^{(\Pom \Ode \phi)} \left( (k+k')^{2},k'^{2},k^{2} \right)$ is a form factor.
In~practical calculations we take the factorized form
for the $\Pom \Ode \phi$ form factor,
%
\begin{eqnarray}
F^{(\Pom \Ode \phi)} ( (k+k')^{2},k'^{2},k^{2} ) = 
F((k+k')^{2})\, F(k'^{2})\, F^{(\Pom \Ode \phi)}(k^{2})\,, 
\label{Fpomodephi}
\end{eqnarray}
where we adopt the monopole form
\begin{eqnarray}
F(k^{2})= \frac{1}{1-k^{2}/\Lambda^{2}}\,,
\label{Fpomodephi_ff}
\end{eqnarray}
and $F^{(\Pom \Ode \phi)}(k^{2})$ is a form factor normalised to
$F^{(\Pom \Ode \phi)}(m_{\phi}^{2}) = 1$.
The coupling parameters $a_{\Pom \Ode \phi}$, $b_{\Pom \Ode \phi}$ in (\ref{A15})
and the cutoff parameter $\Lambda^{2}$ in the form factor (\ref{Fpomodephi_ff})
could be adjusted to experimental data.

In Fig.~\ref{QCD_POphi_coupling}~(b) we show a QCD diagram 
which will contribute to the $\Pom \Ode \phi$ vertex.
The ``normal'' decay of a $\phi$ meson from the QCD point of view
is to three gluons produced in the annihilation of the $s \bar{s}$ quarks.
A higher order correction can involve a five-gluon decay.
Turning such a diagram around we arrive at the $\Pom \Ode \phi$ coupling
shown in Fig.~\ref{QCD_POphi_coupling}~(b).

For the considered reaction $p p \to p p \phi \phi$,
the $\phi \phi$ subsystem energy $\sqrt{s_{34}} = {\rm M}_{\phi \phi}$
is not very high and at threshold starts from $\sqrt{s_{34}} = 2 m_{\phi}$.
The odderon-exchange amplitude 
applies for larger, certainly not too small, $\sqrt{s_{34}}$.
At low energies the Regge type of interaction is not realistic
and should be switched off. To achieve this requirement
we shall multiply the odderon-exchange amplitude
by a simple, purely phenomenological factor:
%
\begin{eqnarray}
F_{{\rm thr}}(s_{34}) = 1 - \exp\left(\frac{s_{{\rm thr}}-s_{34}}{s_{{\rm thr}}}\right)\,,
\label{odderon_interpolation}
\end{eqnarray}
with $s_{{\rm thr}} = 4 m_{\phi}^{2}$.
Our prescription leads to ${\cal M}^{(\Ode{\rm-exchange})}_{pp \to pp \phi \phi} \to 0$ 
when $s_{34} \to s_{{\rm thr}}$.
The form factors of Eqs.~(\ref{Fpomodephi}) and (\ref{Fpomodephi_ff})
then guarantee that in our calculation the odderon only
contributes in the Regge regime $|\hat{p}^{2}| \ll s_{34}$.

\subsection{\boldmath{$\gamma$}-exchange mechanism}
\label{sec:gamma_exch}

\begin{figure}
\includegraphics[width=7.cm]{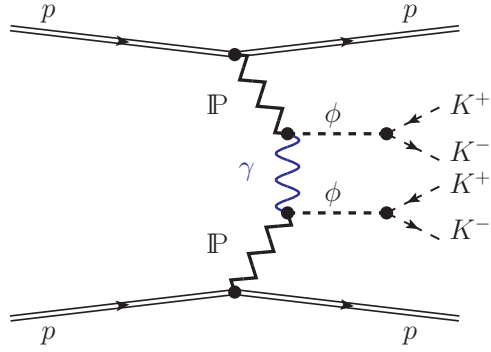} 
  \caption{\label{fig:gamma}
The Born-level diagram for diffractive production of a $\phi$-meson pair
with an intermediate photon exchange.}
\end{figure}

The amplitude for the process shown by the diagram in Fig.~\ref{fig:gamma}
has the same form as the amplitude with the $\phi$-meson exchange
discussed in Sec.~\ref{sec:pompom_phiphi}; 
see Eqs.~(\ref{amplitude_approx}) and (\ref{tensorial_function_aux2}).
But we have to make the following replacements:
\begin{eqnarray}
&&\Delta^{(\phi)}_{\rho_{1}\rho_{2}}(\hat{p}) \to \Delta^{(\gamma)}_{\rho_{1}\rho_{2}}(\hat{p})
= -\frac{g_{\rho_{1}\rho_{2}}}{\hat{p}^{2}}\,,
\\
&&\hat{F}_{\phi}(\hat{p}^{2}) \to \hat{F}_{\gamma}(\hat{p}^{2})\,,
\end{eqnarray}
where we assume that $\hat{F}_{\gamma}(\hat{p}^{2}) = F_{M}(\hat{p}^{2})$ (\ref{FMt})
and $\Lambda_{0}^{2} = 1.0$~GeV$^{2}$, and
\begin{eqnarray}
&&a_{\Pom \phi \phi} \to a_{\Pom \gamma \phi} = \frac{e}{\gamma_{\phi}} \,a_{\Pom \phi \phi}\,,\\
&&b_{\Pom \phi \phi} \to b_{\Pom \gamma \phi} = \frac{e}{\gamma_{\phi}} \,b_{\Pom \phi \phi}\,,
\label{replacements}
\end{eqnarray}
where $e>0, \gamma_{\phi} < 0$, and $\gamma_{\phi}^{2} = 4 \pi/0.0716$ 
[see Eq.~(5.3) of \cite{Donnachie:2002en} 
and Eqs. (3.23)--(3.25) and Sec.~4 of \cite{Ewerz:2013kda}].

\section{Results}
\label{sec:results}

In this section we wish to present first results 
for the $pp \to pp K^{+}K^{-}K^{+}K^{-}$ reaction 
via the intermediate $\phi(1020) \phi(1020)$ state
corresponding to the diagrams shown in Figs.~\ref{fig:4K_diagrams}--\ref{fig:gamma}.
In practice we work with
the amplitudes in high-energy approximation;
see~(\ref{amplitude_approx})~and~(\ref{amplitude_approx_rhorho}).

\subsection{Comparison with the WA102 data}
\label{sec:section_4}

It was noticed in \cite{Barberis:1998bq} that
the cross section for the production of a $\phi \phi$ system, 
for the same interval of $|x_{F,\phi \phi}| \leqslant 0.2$,
is almost independent of the center-of-mass energy.
The~experimental results are
$\sigma_{{\rm exp}}^{(\phi \phi)} = 42 \pm 9$~nb at $\sqrt{s} = 12.7$~GeV \cite{Armstrong:1986ky},
$\sigma_{{\rm exp}}^{(\phi \phi)} = 36 \pm 6$~nb at $\sqrt{s} = 23.8$~GeV \cite{Armstrong:1989hz},
and $\sigma_{{\rm exp}}^{(\phi \phi)} = 41.0 \pm 3.7$~nb at $\sqrt{s} = 29.1$~GeV \cite{Barberis:1998bq}.
This suggests that the double-pomeron-exchange mechanism
shown in Fig.~\ref{fig:4K_diagrams} is the dominant one for the $pp \to pp \phi \phi$ reaction
in the above energy range.
In the following we neglect, therefore, secondary reggeon exchanges.

\begin{figure}[!ht]
(a)\includegraphics[width=0.46\textwidth]{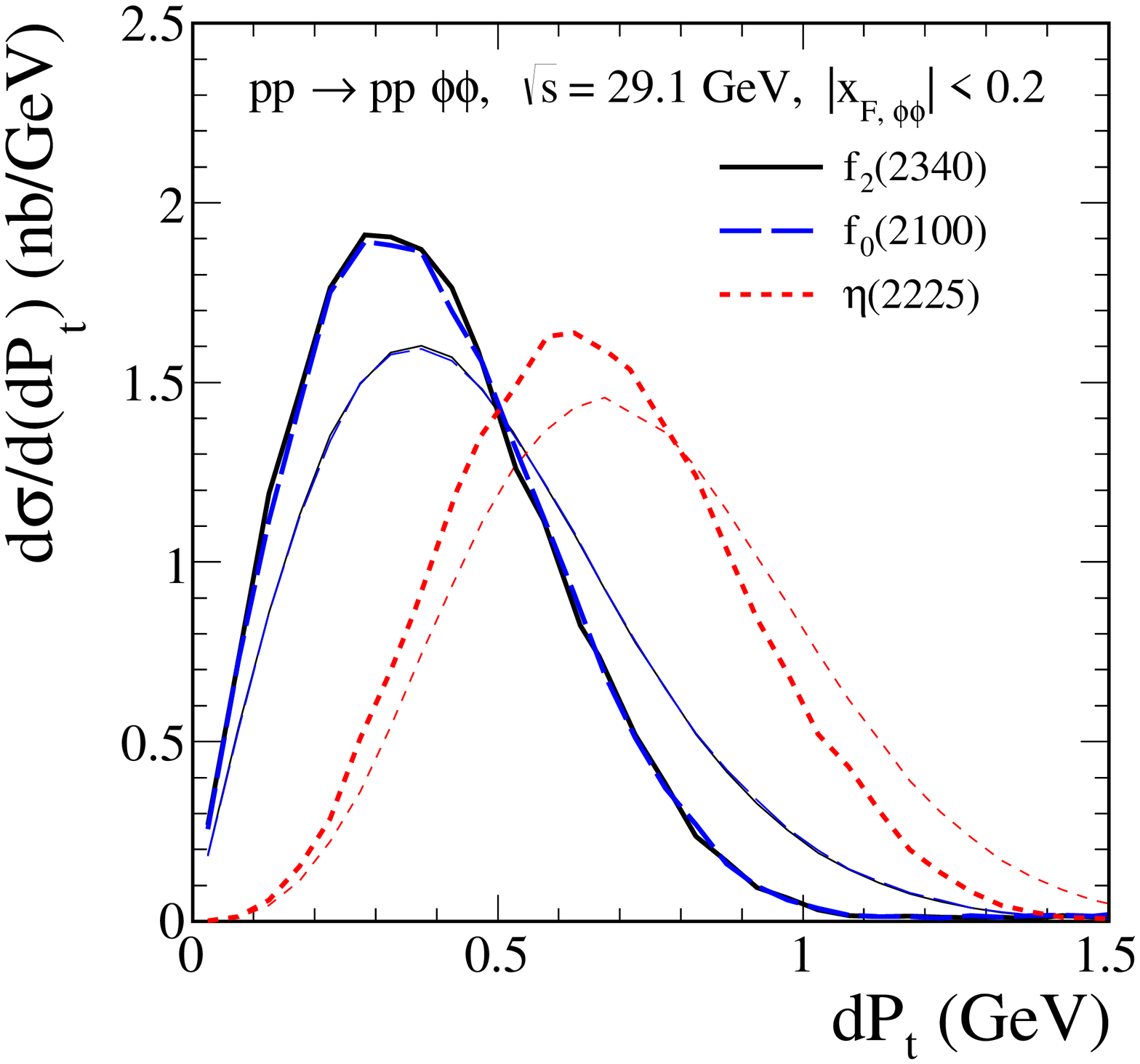}
(b)\includegraphics[width=0.46\textwidth]{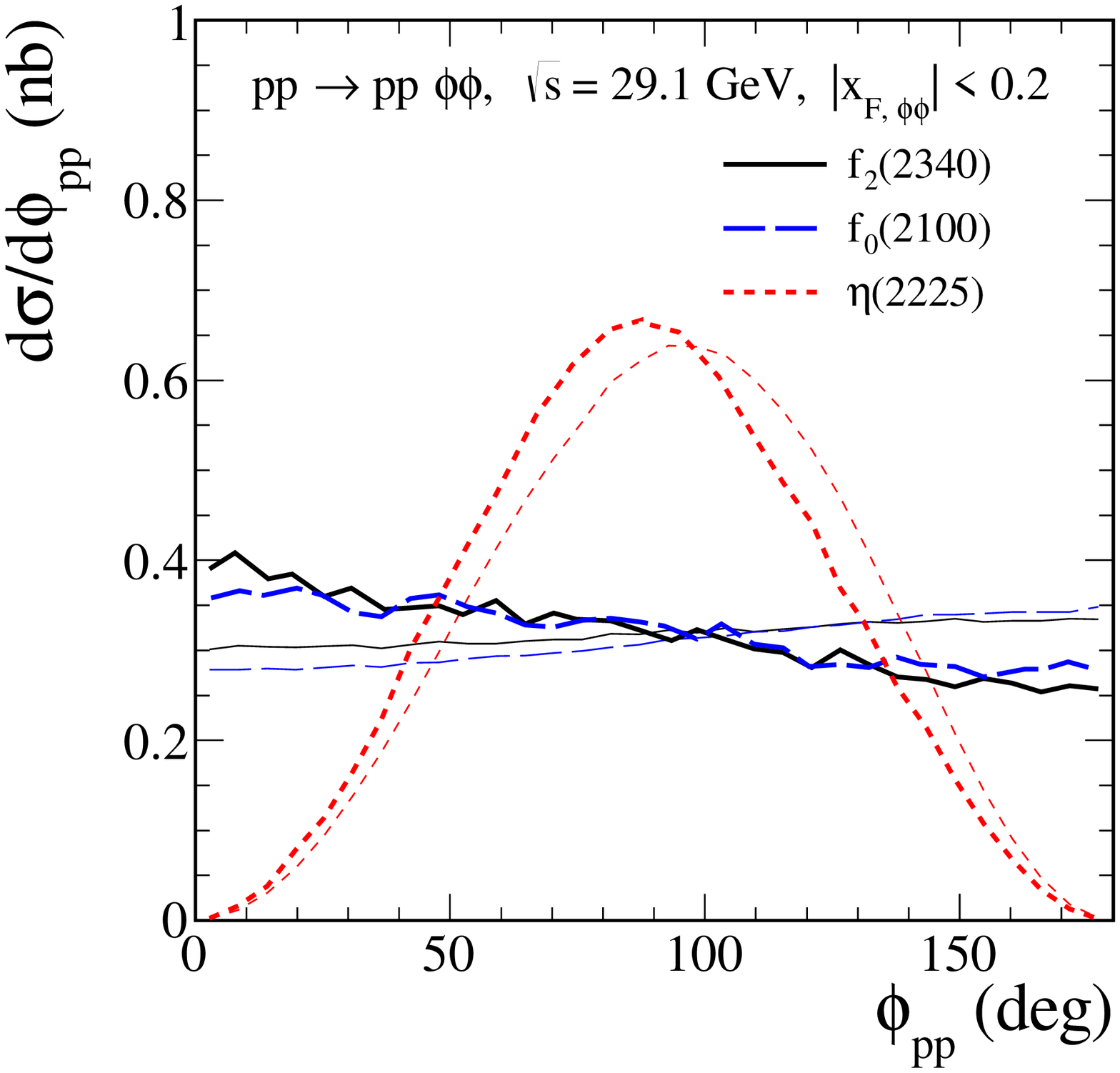}
\caption{\label{fig:0}
The distribution in $\rm{dP_{t}}$ (\ref{dPt_variable}) and in $\phi_{pp}$
for the central exclusive $\phi \phi$ production \mbox{at $\sqrt{s} = 29.1$~GeV}
and $|x_{F,\phi \phi}| \leqslant 0.2$.
The results for scalar, pseudoscalar, and tensor resonances
without (the thin lines) and with (the thick lines) 
absorptive corrections are shown.
Because here we are interested only in the shape of the distributions,
we normalised the differential distributions arbitrarily to 1~nb
for both cases, with and without absorption corrections.}
\end{figure}

In principle, there are many possible resonances 
with $J^{PC} = 0^{++}, 0^{-+}, 2^{++}$ that may contribute 
to the $pp \to pp \phi \phi$ reaction represented by the diagram (b) in Fig.~\ref{fig:4K_diagrams}; 
see the fifth column in Table~\ref{table:table}.
Therefore, before comparing with the experimental data, 
let us first concentrate on the general characteristics of resonant production
via the pomeron-pomeron fusion.
We shall consider only three resonances as representative examples:
$f_{0}(2100)$, $\eta(2225)$, and $f_{2}(2340)$. 
For illustration, in Fig.~\ref{fig:0} 
we present the shape of distributions in $\rm{dP_{t}}$ and $\phi_{pp}$
for the experimental conditions as in the WA102 experiment \cite{Barberis:1998bq},
that is, for $\sqrt{s} = 29.1$~GeV and 
$|x_{F,\phi \phi}| \leqslant 0.2$.
Here $\rm{dP_{t}}$ is the ``glueball-filter variable'' 
\begin{eqnarray}
\bdPt = \bqta - \bqtb = \bptb - \bpta \,, \quad \rm{dP_{t}} = |\bdPt|\,,
\label{dPt_variable}
\end{eqnarray}
and $\phi_{pp}$ is the azimuthal angle between the transverse momentum vectors 
$\bpta$, $\bptb$ of the outgoing protons.
The results without (the thin lines) and with (the thick lines) 
absorptive corrections are shown in Fig.~\ref{fig:0}. 
The differential distributions have been normalised to 1~nb
for both cases, with and without absorptive corrections.
We can conclude that only the scalar and tensor resonances
have similar characteristics as the WA102 experimental distributions \cite{Barberis:1998bq}
shown in Fig.~\ref{fig:2} below.
In the following we will assume that the $f_{2}(2340)$ resonance dominates.

\begin{figure}[!ht]
\includegraphics[width=0.48\textwidth]{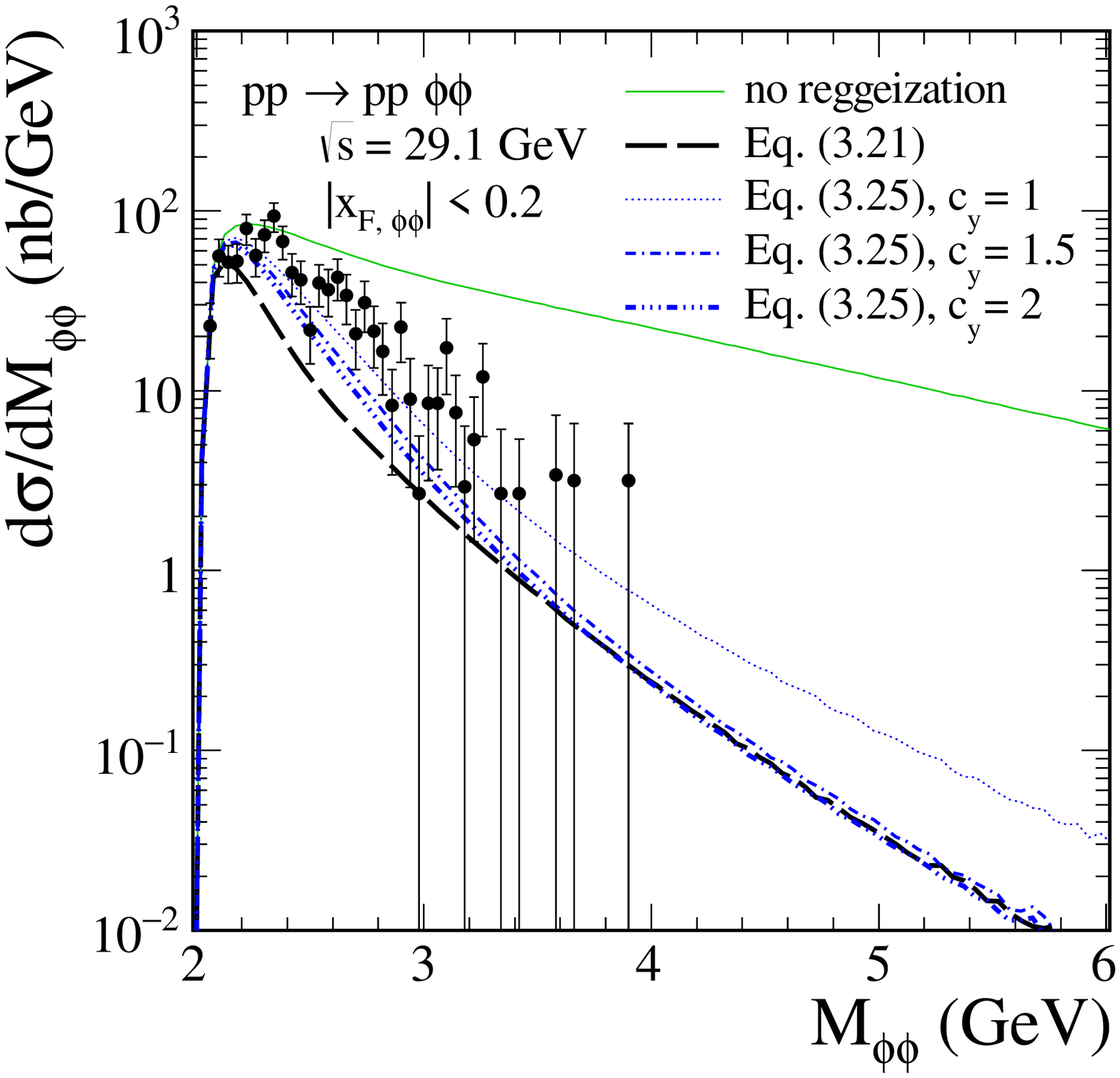}
\includegraphics[width=0.48\textwidth]{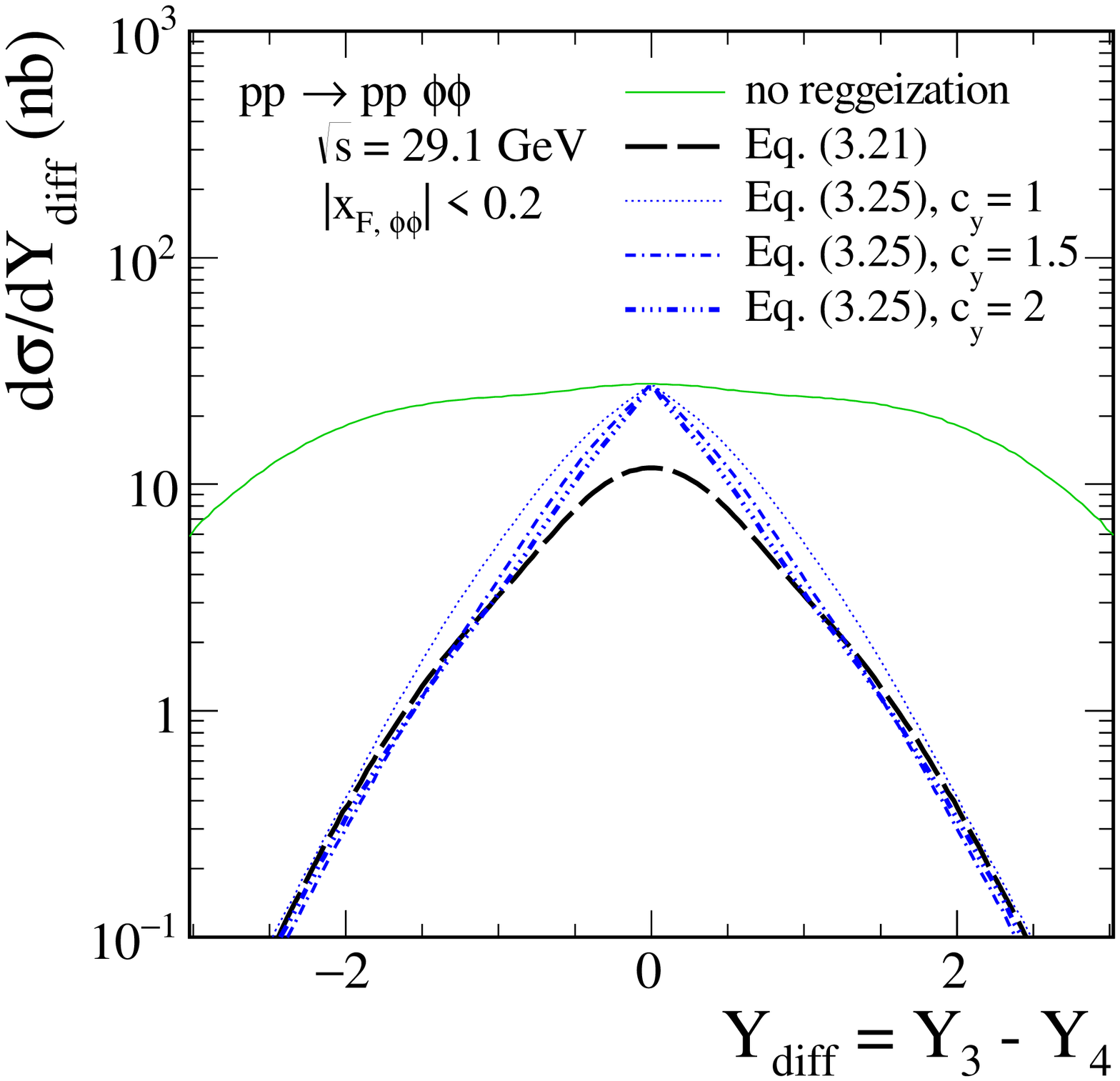}
\caption{\label{fig:reggeization}
The distributions in $\phi\phi$ invariant mass (the left panel)
and in $\rm{Y_{diff}}$, the rapidity distance between the two $\phi$ mesons 
(the right panel), for the $\phi$-exchange continuum contribution.
The calculations were done for $\sqrt{s} = 29.1$~GeV and $|x_{F,\phi \phi}| \leqslant 0.2$.
In the left panel we show the WA102 experimental data \cite{Barberis:2000em} 
normalised to the total cross section $\sigma_{{\rm exp}}^{(\phi\phi)} = 41$~nb 
from \cite{Barberis:1998bq}.
The green solid line corresponds to the non-reggeized contribution.
The results for the two prescriptions of reggeization,
(\ref{reggeization}) and (\ref{second_procedure}), are shown
by the black and blue lines, respectively.
The absorption effects are included here.}
\end{figure}

In Fig.~\ref{fig:reggeization} we show the results for the $\phi \phi$ continuum process
via the $\phi$-meson exchange mechanism represented 
by diagram (a) in Fig.~\ref{fig:4K_diagrams}.
In the left panel we present the $\phi \phi$ invariant mass distributions 
and in the right panel the distributions in \mbox{$\rm{Y_{diff}} = \rm{Y}_{3} - \rm{Y}_{4}$}.
In our calculation we take $\Lambda_{off,E} = 1.6$~GeV in (\ref{off-shell_form_factors_exp})
and the $\Pom \phi \phi$ coupling parameters from \cite{Lebiedowicz:2018eui}.
It is clearly seen from the left panel that the result without reggeization
(see the green solid line) is well above the WA102 experimental data 
\cite{Barberis:1998bq,Barberis:2000em} normalised to the total cross section 
$\sigma_{{\rm exp}}^{(\phi\phi)} = 41$~nb from \cite{Barberis:1998bq}.
The reggeization effect that leads to the suppression of the cross section 
should be applied here, but the way it should be included is less obvious.
We show results for two prescriptions of reggeization
given by Eqs.~(\ref{reggeization}) and (\ref{second_procedure}).
We have checked that for the considered reaction 
$\langle -\hat{p}_{t}^{2} \rangle$, $\langle -\hat{p}_{u}^{2} \rangle
\simeq 1$~GeV$^2$ (before reggeization). 
We have $s_{34} - 4 m_{\phi}^2 > 2$~GeV$^2$ 
for $\sqrt{s_{34}} = {\rm M}_{\phi \phi} > 2.5$~GeV.
It can therefore be expected that the prescription (\ref{reggeization}) 
is relevant near threshold and
especially for ${\rm M}_{\phi \phi} \gtrsim 3$~GeV.
However, in the light of the discussion after Eq.~(\ref{reggeization})
how to treat the low-${\rm M}_{\phi \phi}$ region,
we consider also the alternative prescription (\ref{second_procedure})
combined with (\ref{second_procedure_aux}).
In this case we present predictions for ${\rm c_{y}} = 1, 1.5$, and 2 in (\ref{second_procedure_aux}).
One can clearly see no effect of the reggeization at $\rm{Y_{diff}} = 0$.
The reggeization becomes more important 
when ${\rm M}_{\phi \phi}$ and $|{\rm Y_{diff}}|$ increase.
For ${\rm c_{y}} = 2$ and ${\rm M}_{\phi \phi} \gtrsim 3.5$~GeV
we get similar results from (\ref{second_procedure})
as from the first prescription (\ref{reggeization}).

\begin{figure}[!ht]
(a)\includegraphics[width=0.46\textwidth]{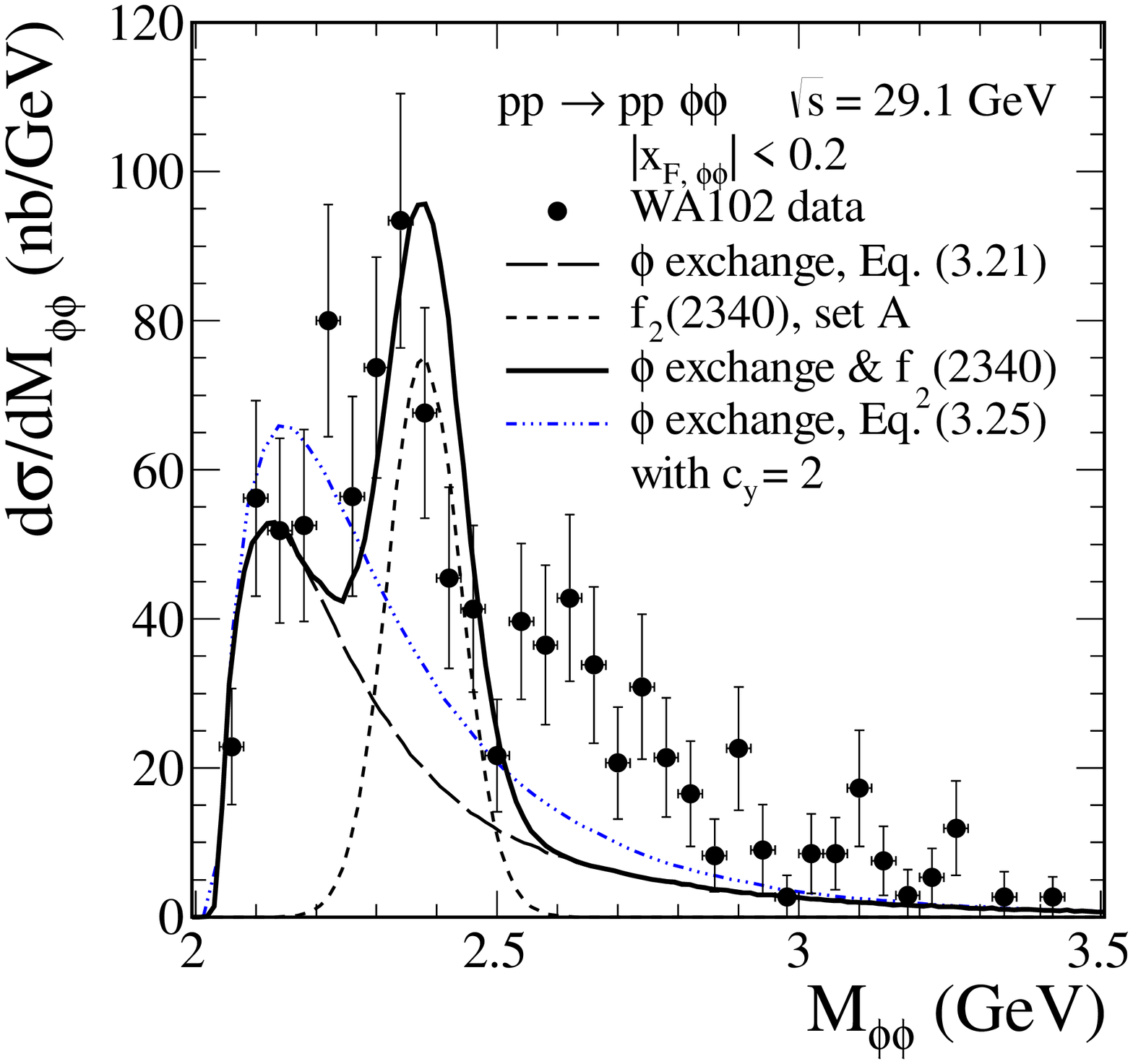}
(b)\includegraphics[width=0.46\textwidth]{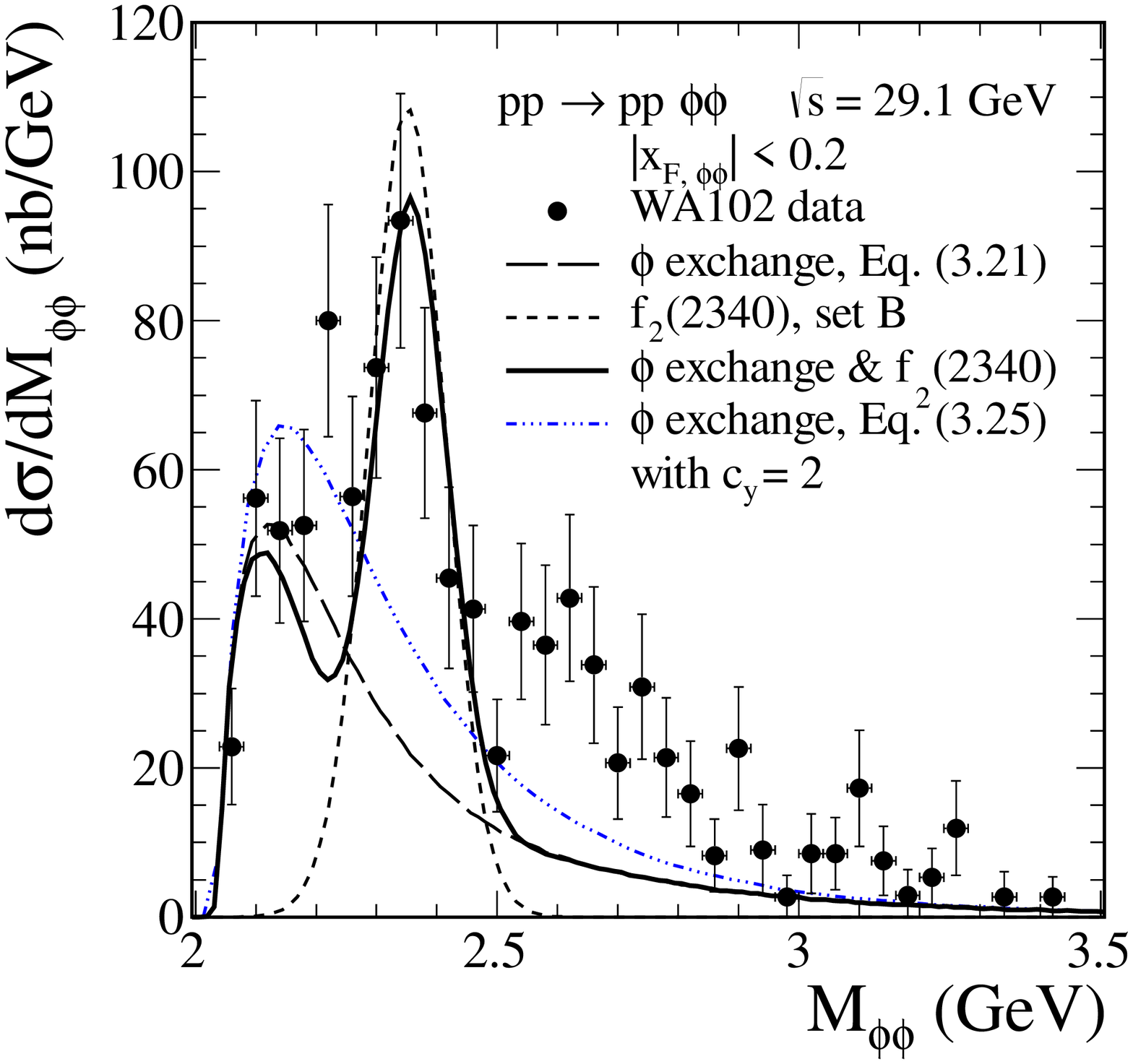}
\caption{\label{fig:1}
Invariant mass distributions for the central $\phi \phi$ system compared to the WA102 data
\cite{Barberis:2000em} at $\sqrt{s} = 29.1$~GeV and $|x_{F,\phi \phi}| \leqslant 0.2$.
The data points have been normalised to the total cross section 
$\sigma_{{\rm exp}}^{(\phi\phi)} = 41$~nb from \cite{Barberis:1998bq}.
We show results for two sets of the parameters from Table~\ref{table:parameters},
set~A [see panel~(a)] and set~B [see panel~(b)].
The black long-dashed line corresponds to the reggeized $\phi$-exchange contribution
[Eq.~(\ref{reggeization})],
while the black short-dashed line corresponds to the $f_{2}(2340)$ resonance term,
and the black solid line represents the coherent sum of both contributions.
For comparison, we show also the blue dashed-dotted line that
corresponds to the reggeized $\phi$-exchange contribution using Eq.~(\ref{second_procedure}).
The absorption effects are included here.}
\end{figure}

In Fig.~\ref{fig:1} we compare our predictions 
including now the two mechanisms shown in Fig.~\ref{fig:4K_diagrams} 
to the WA102 data \cite{Barberis:1998bq,Barberis:2000em}
for the $\phi \phi$ invariant mass distribution  
from the $pp \to pp \phi \phi$ reaction.
With our choice to keep only one $\Pom \Pom f_{2}$ coupling
from (\ref{vertex_pompomT}), namely $g^{(1)}_{\Pom \Pom f_{2}}$,
the distributions depend on the product of the couplings
$g^{(1)}_{\Pom \Pom f_{2}} g'_{f_{2} \phi \phi}$ and
$g^{(1)}_{\Pom \Pom f_{2}} g''_{f_{2} \phi \phi}$
with $g'_{f_{2} \phi \phi}$ and $g''_{f_{2} \phi \phi}$
given in (\ref{vertex_f2phiphi}).
Again, for orientation purposes,
we shall assume here and in the following that only either
the first or the second of the above products of couplings is nonzero.
In the parameter set~A we choose 
$g^{(1)}_{\Pom \Pom f_{2}} g'_{f_{2} \phi \phi} \neq 0$,
and in set~B we choose
$g^{(1)}_{\Pom \Pom f_{2}} g''_{f_{2} \phi \phi} \neq 0$;
see Table~\ref{table:parameters}.
Of course, once good measurements of all the relevant distributions
of our reaction are available, one can try -- as will be correct --
to fit a linear combination of the above two coupling terms to the data.
Thus, we show in Fig.~\ref{fig:1}
results
for the two sets of parameters given in Table~\ref{table:parameters},
set~A [see panel~(a)] and set~B [see panel~(b)].
The long-dashed lines represent results for the reggeized $\phi$-exchange contribution.
The short-dashed lines represent results 
for the $f_{2}(2340) \to \phi \phi$ resonance contribution.
The solid lines represent the coherent sum of both contributions.
We found a rather good agreement near ${\rm M}_{\phi \phi} = 2.3$~GeV, 
taking into account only the continuum and $f_{2}(2340)$ meson, 
although the possibility of an $f_{2}(2300)$ meson contribution cannot be ruled out.
Our predictions indicate therefore that in such a case we are dealing 
rather with an upper limit of the cross section for the $f_{2}$-resonance term.
We wish to point out here that the interference of the continuum
and resonance contributions depends on subtle details
(choice of the couplings for resonant term, phase interpolation for the continuum term). 

By comparing the theoretical results and the differential cross sections
obtained by the WA102 Collaboration
we fixed the parameters of the off-shell $\hat{t}/\hat{u}$-channel $\phi$-meson form factor 
($\Lambda_{off,E}$ in (\ref{off-shell_form_factors_exp})) 
and the $\Pom \Pom f_{2}$ and $f_{2} \phi \phi$ couplings.
For the convenience of the reader 
we have collected in Table~\ref{table:parameters}
the default numerical values of the parameters of our model used in the calculations.
\begin{table}[!h]
\caption{Some parameters of our model.
The columns indicate the equation numbers where
the parameter is defined and their numerical values used in the calculations.}
\begin{tabular}{|l|l|l|l|}
\hline
Parameters for			& Equation 	& Value (Set~A) & Value (Set~B)\\ \hline
\textbf{\boldmath{$\phi$}-exchange mechanism}	& &  &\\ \hline  	 
$a_{\Pom \phi \phi}$ & (\ref{tensorial_function_aux2}); 
Sec. IV~B of \cite{Lebiedowicz:2018eui} & 0.49~GeV$^{-3}$ & 0.49~GeV$^{-3}$   \\
$b_{\Pom \phi \phi}$ & (\ref{tensorial_function_aux2}); 
Sec. IV~B of \cite{Lebiedowicz:2018eui} & 4.27~GeV$^{-1}$ & 4.27~GeV$^{-1}$    \\
$\Lambda_{0}^{2}$ & (\ref{FMt});
Sec. IV~B of \cite{Lebiedowicz:2018eui} & 1.0~GeV$^{2}$ & 1.0~GeV$^{2}$ 	 	\\ 
$\Lambda_{off,E}$ &	 (\ref{off-shell_form_factors_exp}) & 1.6~GeV & 1.6~GeV  \\ \hline
\textbf{\boldmath{$\Pom\Pom \to f_{2}(2340) \to \phi \phi$} mechanism}& & & \\ \hline 
$g_{\Pom \Pom f_{2}}^{(1)}$ $g_{f_{2} \phi \phi}'$  & (\ref{vertex_pompomT}) \textit{et seq}.; 
(\ref{vertex_f2phiphi}) & 12.0 & 0.0		\\
$g_{\Pom \Pom f_{2}}^{(1)}$ $g_{f_{2} \phi \phi}''$ & (\ref{vertex_pompomT}) \textit{et seq}.; 
(\ref{vertex_f2phiphi}) & 0.0 	& 7.0 \\
$\tilde{\Lambda}_{0}^{2}$ & (\ref{FM_t}) & 1.0~GeV$^{2}$ & 1.0~GeV$^{2}$ \\
$\Lambda_{f_{2}}$ &	(\ref{Fpompommeson_ff})--(\ref{Fpompommeson_ff_tensor}) 
& 1.0~GeV  	& 1.0~GeV \\ \hline
\end{tabular}
\label{table:parameters}
\end{table}

It can be observed that the WA102 experimental point 
at ${\rm M}_{\phi\phi} \approx 2.2$~GeV is well above our 
theoretical result ($\phi$-exchange contribution) and
it may signal the presence of the $f_{J}(2220)$ resonance.
As was shown in Fig.~\ref{fig:0},
mesons with $J = 0$ and $J = 2$ have similar characteristics. 
Therefore, the answer to the question 
about the spin of $f_{J}(2220)$ cannot be easily given
by studying the $\phi\phi$ decay channel.
Our model calculation, including only two contributions,
the reggeized $\phi(1020)$-meson exchange and the production 
via the intermediate $f_{2}(2340)$, describes the WA102 experimental data 
up to ${\rm M}_{\phi \phi} = 2.5$~GeV reasonably well; see Fig.~\ref{fig:1}.
We cannot exclude a small contribution of the $X(2500)$ meson 
which was seen in $J/\psi \to \gamma \phi \phi$ \cite{Ablikim:2016hlu}.
Including the other resonances will only be meaningful once experiments with
better statistics become available.
Hopefully this will be the case at the LHC.
The behaviour at higher values of ${\rm M}_{\phi \phi} \gtrsim 2.5$~GeV
will be further discussed in Sec.~\ref{sec:section_6}.

\begin{figure}[!ht]
(a)\includegraphics[width=0.46\textwidth]{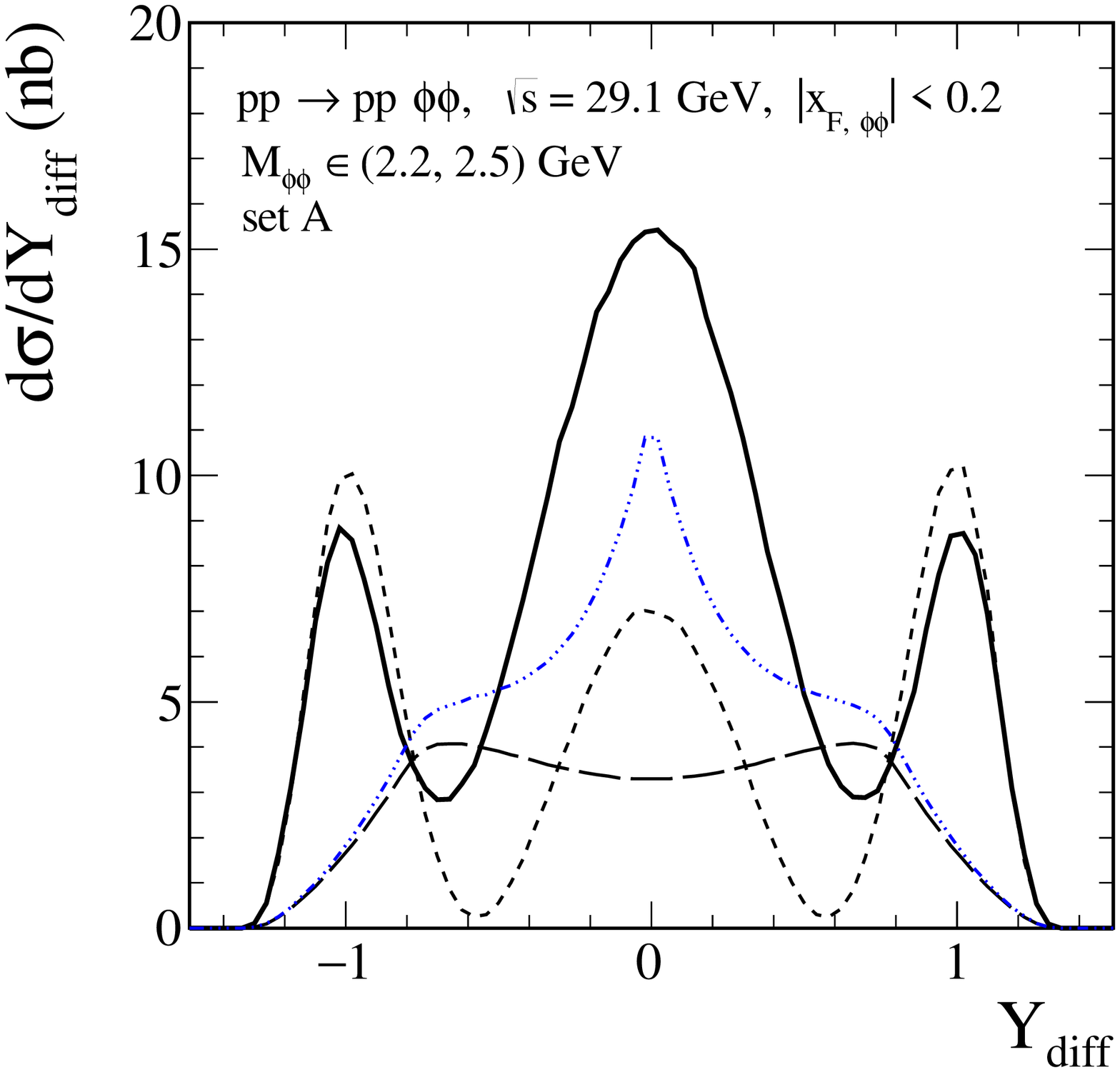}
(b)\includegraphics[width=0.46\textwidth]{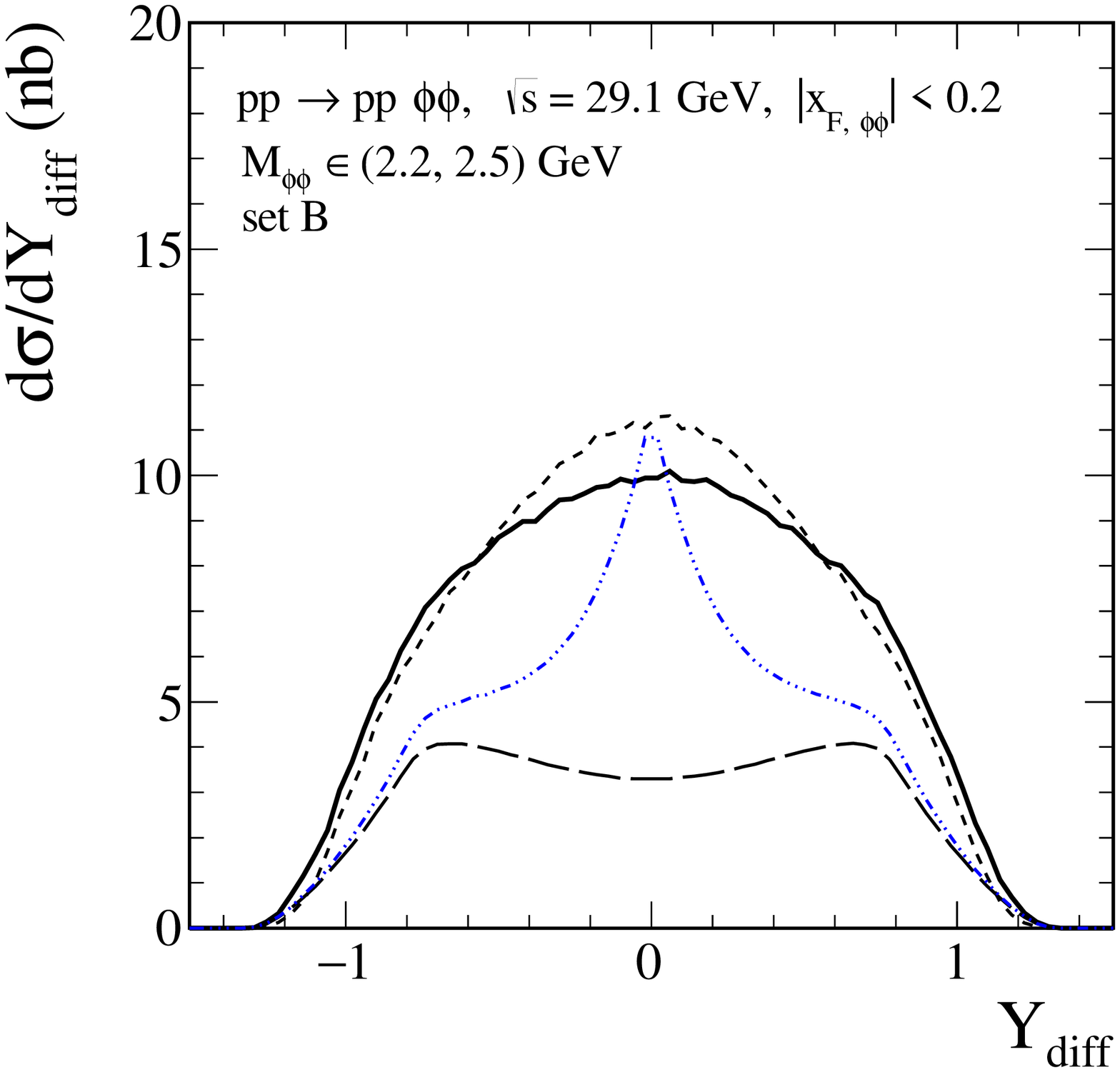}
\caption{\label{fig:1a}
The distributions in rapidity distance between two centrally produced $\phi(1020)$ mesons 
$\rm{Y_{diff}} = \rm{Y}_{3} - \rm{Y}_{4}$
at $\sqrt{s} = 29.1$~GeV for $|x_{F,\phi \phi}| \leqslant 0.2$
and ${\rm M}_{\phi \phi} \in (2.2,2.5)$~GeV.
The meaning of the lines is the same as in Fig.~\ref{fig:1}.
Here we show results for the two sets, A and B, of the parameters; see Table~\ref{table:parameters}.
The absorption effects are included here.}
\end{figure}

From Fig.~\ref{fig:1a} it is clearly seen that the shape of the $\rm{Y_{diff}}$ distribution
is sensitive to the choice of the $f_{2} \phi \phi$ coupling (\ref{vertex_f2phiphi})
and of the reggeization ansatz.
For the $\phi$ continuum process we show the results obtained 
for the two reggeization prescriptions, (\ref{reggeization}) and (\ref{second_procedure}).
Here $\rm{Y}_{3}$, $\rm{Y}_{4}$ are the rapidities of the two $\phi$ mesons.
We show results in the $\phi \phi$ invariant mass window, 
${\rm M}_{\phi\phi} \in (2.2,2.5)$~GeV, where
tensor glueball candidates with masses around $2.3$~GeV are expected.
Two sets of the parameters, set~A and set~B, from Table~\ref{table:parameters} 
give different results.
It can, therefore, be expected that the $\rm{Y_{diff}}$ variable 
will be very helpful in determining the $f_{2}\phi\phi$ coupling using results 
expected from LHC measurements, in particular, if they cover a wider range of rapidities.
This will be presented further in Figs.~\ref{fig:4} and \ref{fig:4aux}.
We have checked that for the reaction $pp \to pp (\Pom \Pom \to f_{2}(2340) \to \phi \phi)$
discussed here the shapes of the $\rm{Y_{diff}}$ distributions 
do not depend significantly
on the choice of the $\Pom \Pom f_{2}$ vertex coupling (\ref{vertex_pompomT}).
This is a different situation compared to the one observed 
by us for the $pp \to pp (\Pom \Pom \to f_{2}(1270) \to \pi^{+} \pi^{-})$ reaction; 
see Figs.~7 and 8 of \cite{Lebiedowicz:2016ioh} and \cite{Lebiedowicz:2019por}. 

In Fig.~\ref{fig:2} in the panels (a), (b), and (c)
we compare our model results to the WA102 data on the differential distributions
$d\sigma/d(\rm{dP_{t}})$, $d\sigma/d\phi_{pp}$, and $d\sigma/d|t|$ 
(that is $d\sigma/d|t_{1}|$ or $d\sigma/d|t_{2}|$), respectively.
Here we used in the calculations the parameter set~B \mbox{of Table~\ref{table:parameters}}.
We have checked that for these three observables
the results obtained with the parameter set~A 
of Table~\ref{table:parameters} are similar.
The theoretical results correspond to the calculations including absorptive effects 
calculated at the amplitude level and related to the $pp$ nonperturbative interactions.
Note that in the panels (a), (b), and (c) we also show
the Born result for the $\phi$-exchange contribution.
The ratio of full and Born cross sections $\langle S^{2} \rangle$ (the gap survival factor)
at $\sqrt{s} = 29.1$~GeV is $\langle S^{2} \rangle \cong 0.4$.
From Figs.~\ref{fig:0} and \ref{fig:2} we see the influence of absorption effects on
the shape of distributions in $\phi_{pp}$ and $\rm{dP_{t}}$.
\begin{figure}[!ht]
(a)\includegraphics[width=0.46\textwidth]{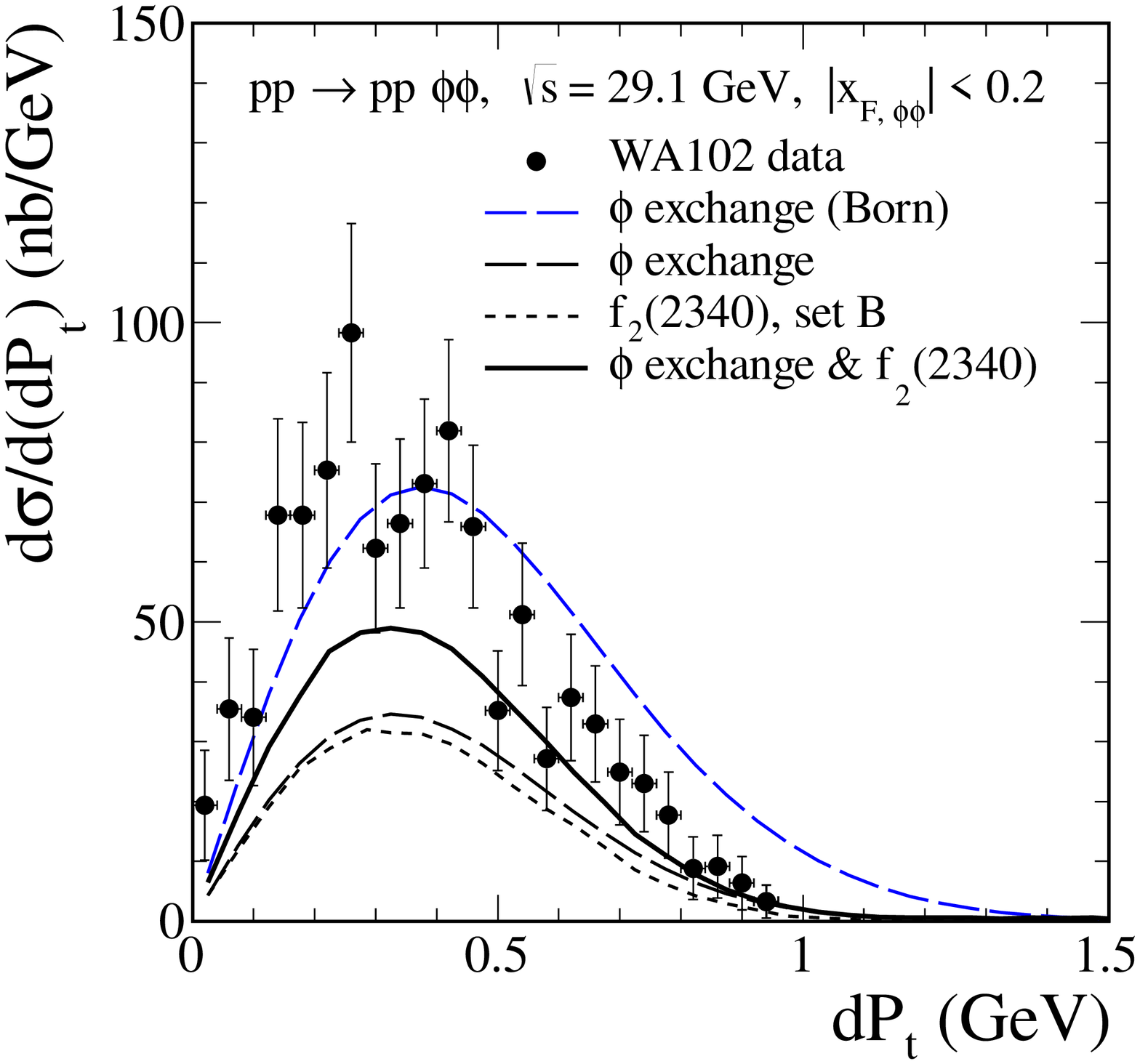}
(b)\includegraphics[width=0.46\textwidth]{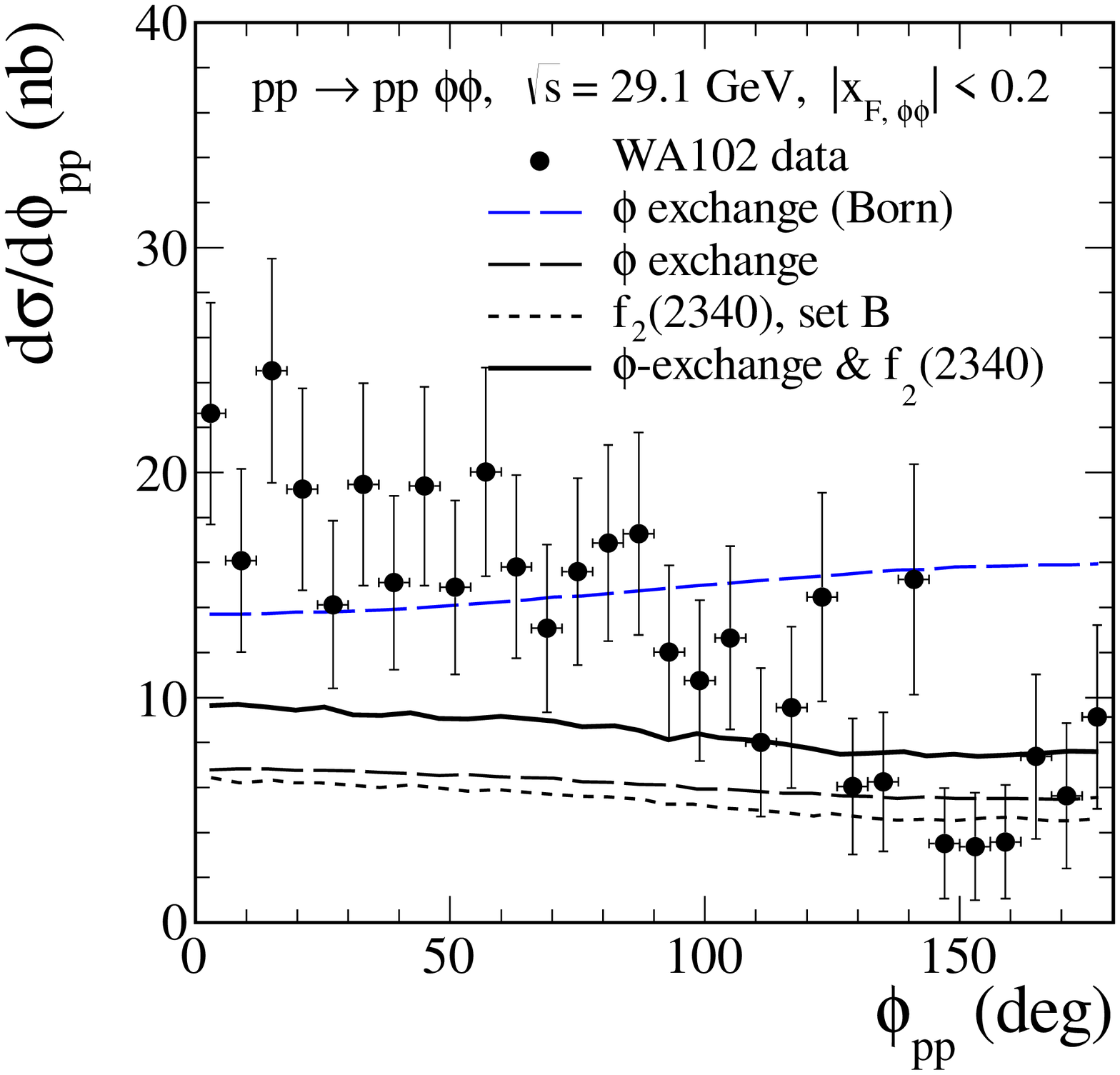}
(c)\includegraphics[width=0.46\textwidth]{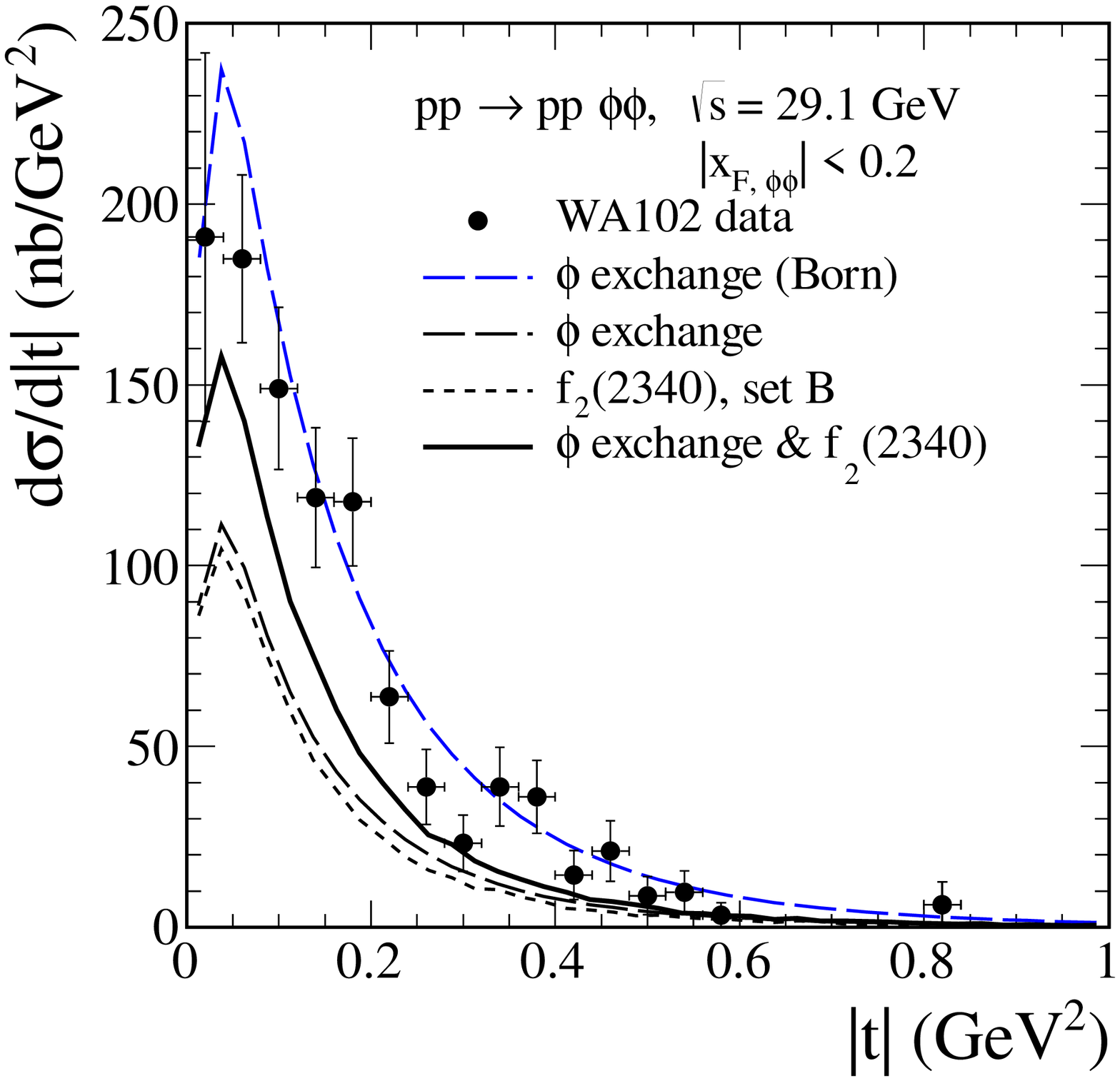}
\caption{\label{fig:2}
Differential cross sections for the central exclusive $\phi \phi$ production at $\sqrt{s} = 29.1$~GeV
and $|x_{F,\phi \phi}| \leqslant 0.2$.
The data points from \cite{Barberis:1998bq} have been normalised to the total cross section 
$\sigma_{{\rm exp}}^{(\phi\phi)} = 41$~nb given there.
The meaning of the lines is the same as in Fig.~\ref{fig:1}~(b) (set B).
Here we show results for the $\phi$-exchange contribution using Eq.~(\ref{reggeization}).
The absorption effects are included,
but, for comparison,  we also show the reggeized $\phi$-exchange contribution
in the Born approximation (without absorption effects)
corresponding to the upper blue long-dashed line.
}
\end{figure}

So far we have tried to adjust parameters of the continuum
and the $f_{2}(2340)$ resonance terms in order not to exceed
the WA102 experimental data for the $\phi \phi$ invariant mass distribution.
We see that limiting to these mechanisms 
we cannot describe the data for ${\rm M}_{\phi \phi} > 2.5$~GeV.
In consequence we underestimate experimental distributions also in Fig.~\ref{fig:2}.
Clearly, an additional mechanism is needed to resolve this problem.
We shall discuss a possible solution of this problem in Sec.~\ref{sec:section_6}.

\subsection{Predictions for the LHC experiments}
\label{sec:section_5}

\begin{figure}[!ht]
\includegraphics[width=0.48\textwidth]{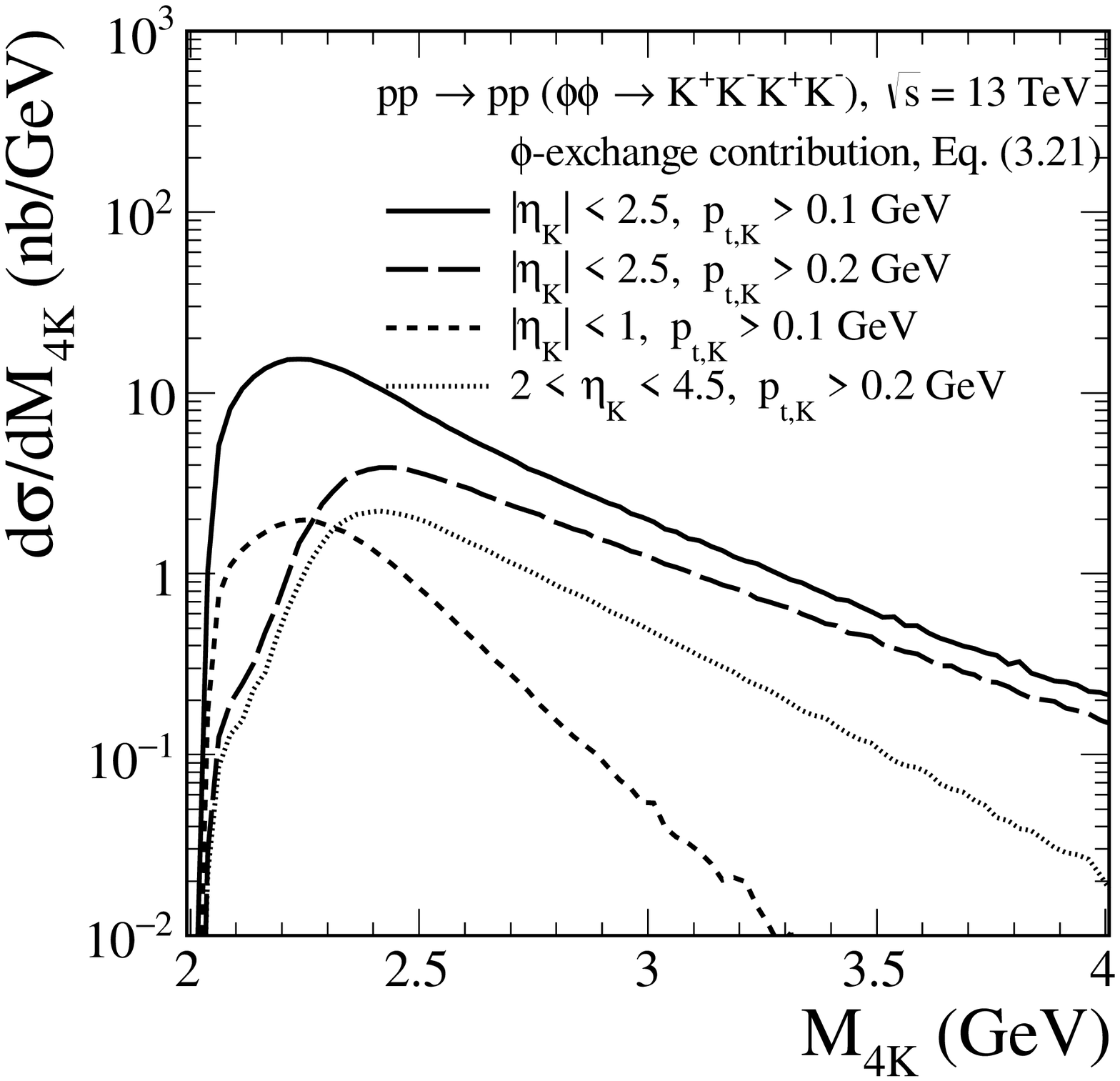}
\includegraphics[width=0.48\textwidth]{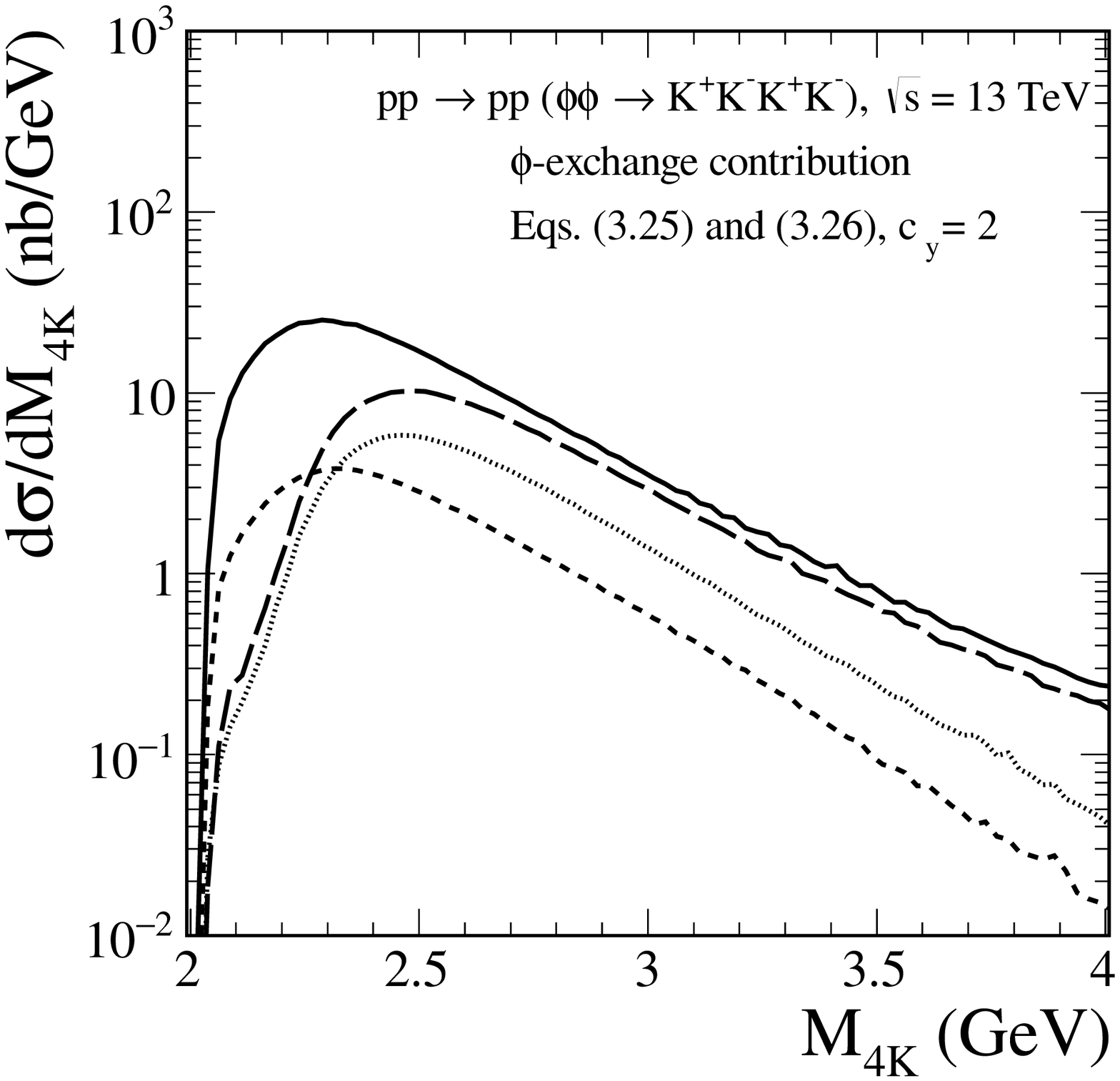}
\includegraphics[width=0.48\textwidth]{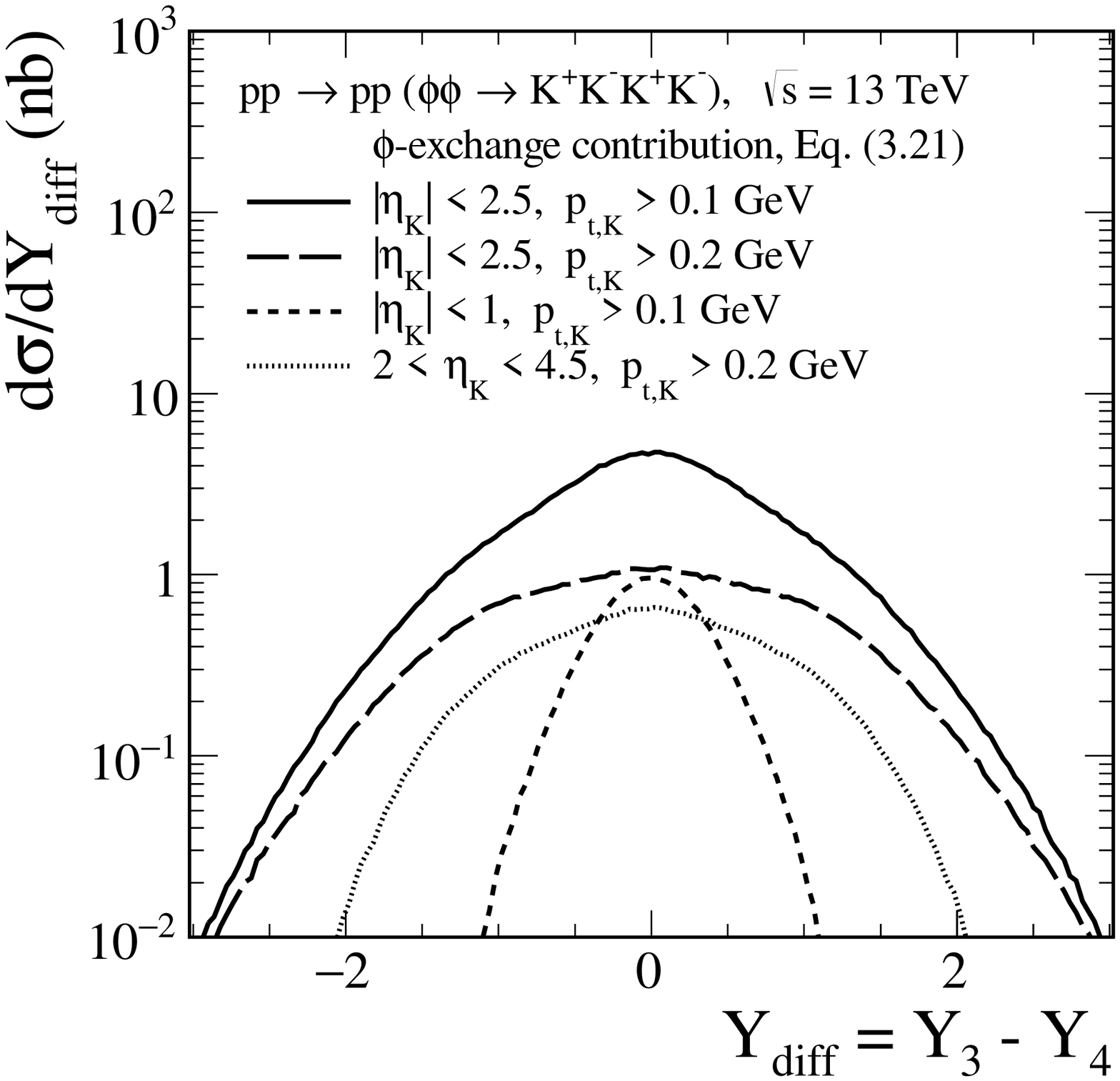}
\includegraphics[width=0.48\textwidth]{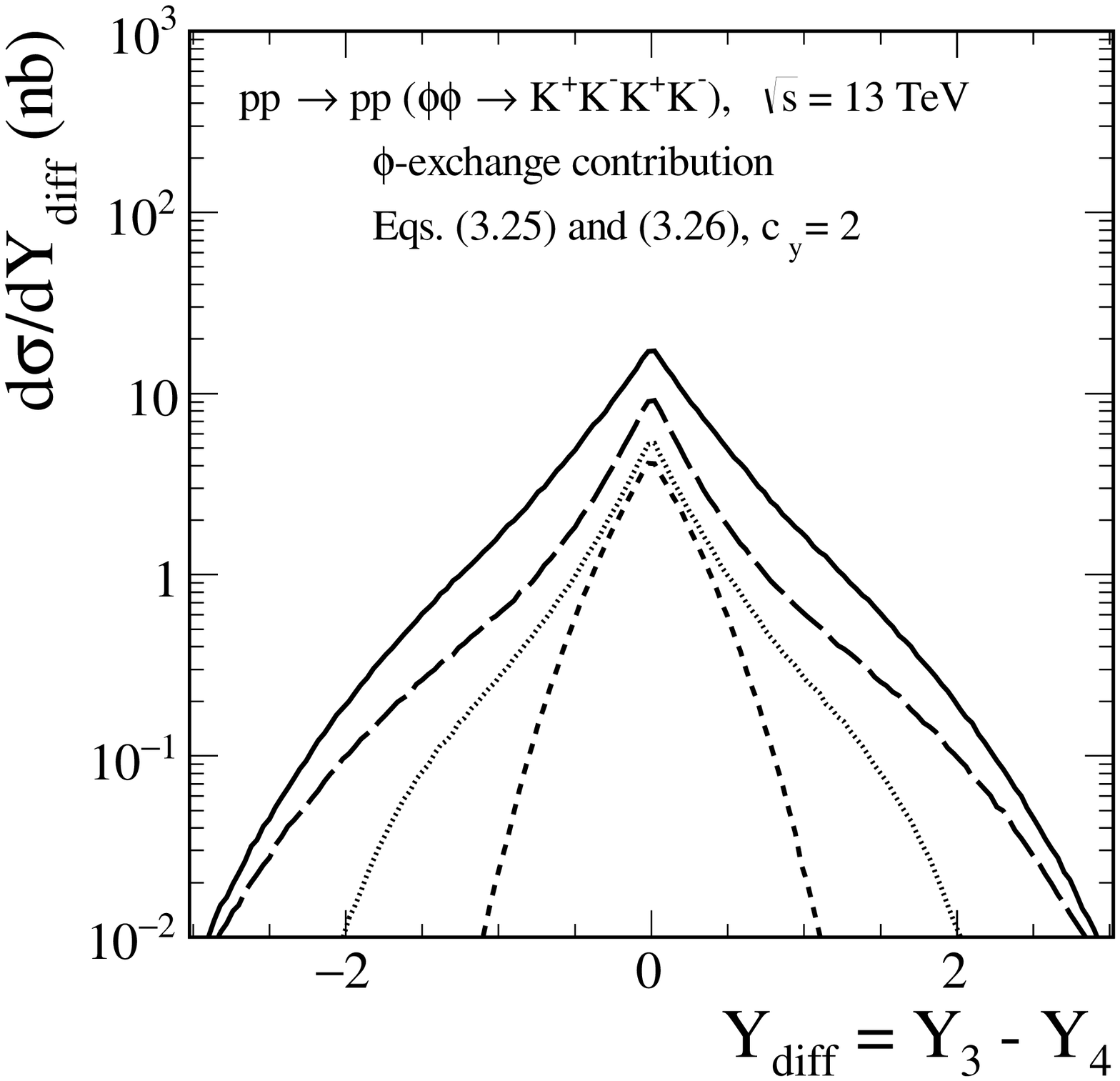}
\caption{\label{fig:3}
Differential cross sections as a function of the four-kaon invariant mass
(top panels) and as function of $\rm{Y_{diff}}$ (bottom panels)
for the $\phi$-exchange mechanism
calculated for $\sqrt{s} = 13$~TeV 
with the kinematical cuts specified in the figure legends.
The results for the two prescriptions of reggeization 
(\ref{reggeization}) and (\ref{second_procedure}) are presented.
The absorption effects are included here.}
\end{figure}

We start from a discussion of the results for the $pp \to ppK^{+} K^{-}K^{+} K^{-}$ reaction 
obtained from the $\phi(1020)$-exchange mechanism discussed in Sec.~\ref{sec:pompom_phiphi}. 
The calculations were done for $\sqrt{s} = 13$~TeV
with typical experimental cuts on pseudorapidities 
and transverse momenta of centrally produced kaons.
The ratio of the full and Born cross sections
at $\sqrt{s} = 13$~TeV is approximately $\langle S^{2} \rangle \cong 0.2$.
In Fig.~\ref{fig:3} 
we present the $K^{+} K^{-} K^{+} K^{-}$ invariant mass distributions
(see the top panels)
and the distributions in $\rm{Y_{diff}} = \rm{Y}_{3} - \rm{Y}_{4}$
(see the bottom panels) 
calculated for $\sqrt{s} = 13$~TeV with the kinematical cuts specified in the figure legends.
Here $\rm{Y}_{3}$ and $\rm{Y}_{4}$ mean $\rm{Y}_{K^{+}K^{-}}$ 
where the kaons are produced from the same $\phi$ meson decay.
Of course, the larger the detector coverage in $\eta_{K}$, 
the larger becomes $|\rm{Y_{diff}}|$.
In the calculations we take into account the intermediate $\phi$-meson reggeization.
We show results for the two prescriptions of reggeization,
(\ref{reggeization}) and (\ref{second_procedure}),
see the left and right panels, respectively.
The results shown in the right panel 
were calculated with ${\rm c_{y}} = 2$ in (\ref{second_procedure_aux}).
We see that the choice of reggeization has a large impact on the results.
The reggeization effect leads to a damping of the four-kaon invariant mass distributions.
From the top panels, we see that
increasing the $p_{t,K}$ cut from 0.1~GeV to 0.2~GeV
significantly suppresses the cross section at small ${\rm M}_{4K}$.
The first scenario of reggeization, Eq.~(\ref{reggeization}),
also significantly suppresses the region
when $\rm{Y}_3 \approx \rm{Y}_4$, that is, for $\rm{Y_{diff}} \simeq 0$.
This is slightly different for the second reggeization scenario (\ref{second_procedure});
see the bottom right panel.
The cross section for the $\phi \phi$-continuum contribution
is about 2 orders of magnitude smaller than
the cross section for the $\rho \rho$-continuum contribution
discussed in \cite{Lebiedowicz:2016zka}.

\begin{figure}[!ht]
\includegraphics[width=0.48\textwidth]{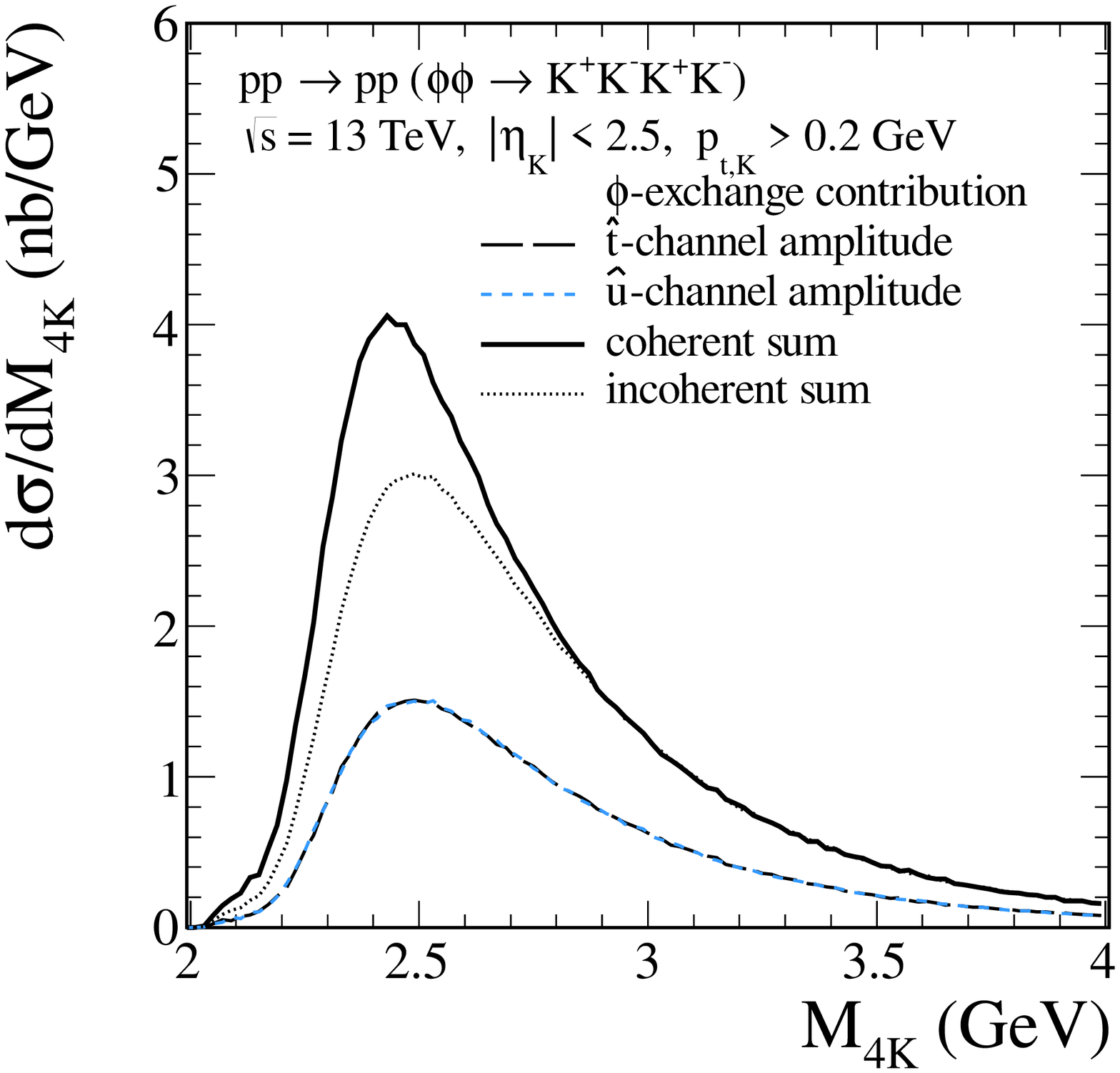}
\includegraphics[width=0.48\textwidth]{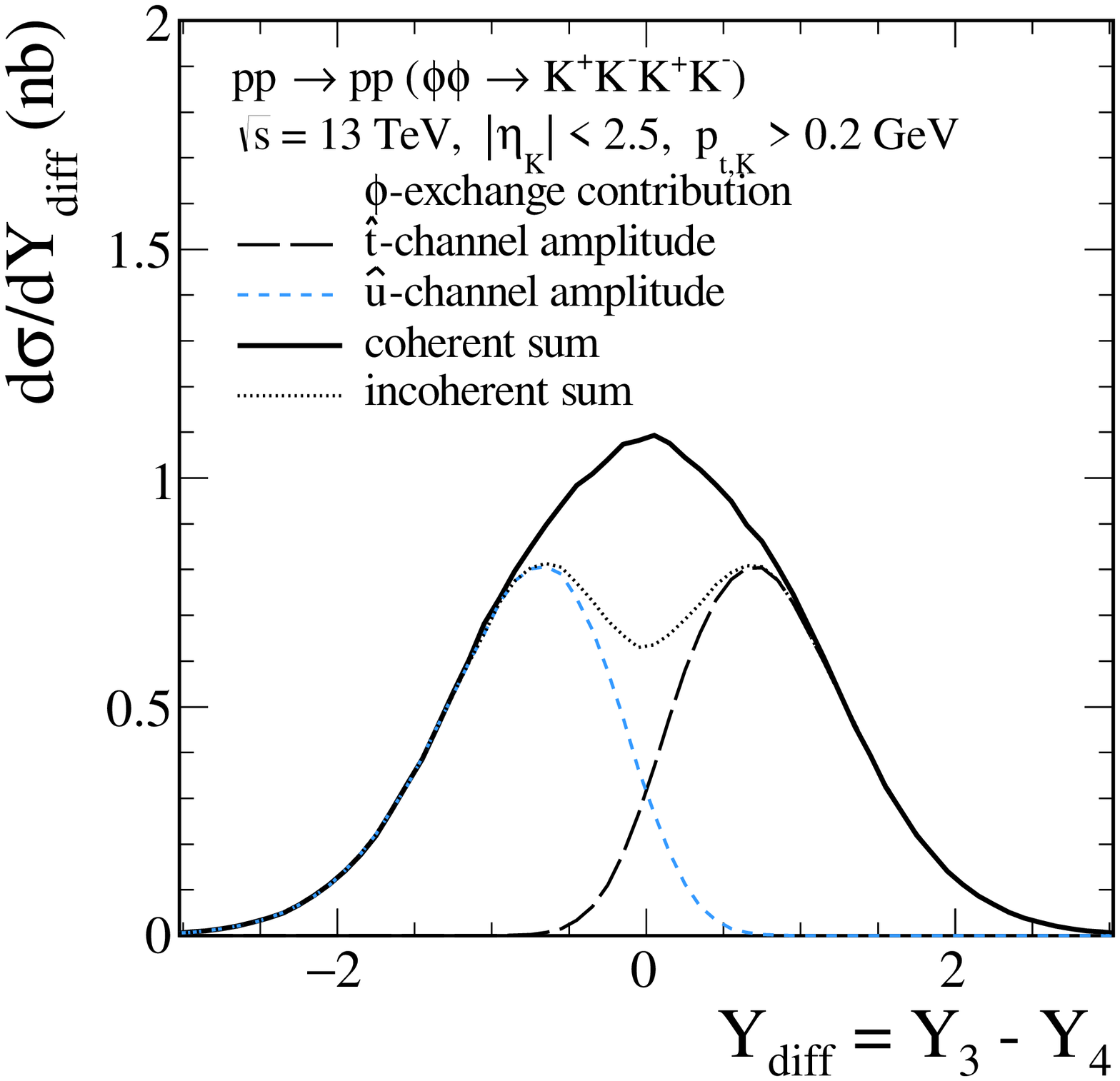}
\caption{\label{fig:3a}
Differential cross sections as a function of the four-kaon invariant mass
(left panel) and as a function of $\rm{Y_{diff}}$ (right panel)
for the $pp \to pp (\phi \phi \to K^{+}K^{-}K^{+}K^{-})$ reaction 
calculated for $\sqrt{s} = 13$~TeV and $|\eta_{K}| < 2.5$, $p_{t, K} > 0.2$~GeV.
The results for the $\phi$-exchange contribution are presented.
The black solid lines correspond to the coherent sum of the $\hat{t}$- and $\hat{u}$-channel amplitudes.
Their incoherent sum is shown by the black dotted lines for comparison.
The~black long-dashed and blue short-dashed lines correspond to the results 
for the individual $\hat{t}$ and $\hat{u}$ terms, respectively.
The absorption effects are included here.}
\end{figure}

In Fig.~\ref{fig:3a} we show further features of the ${\rm M}_{4K}$ and 
the $\rm{Y_{diff}}$ distributions for \mbox{the $\phi$-exchange} contribution.
We show results for the reggeization prescription (\ref{reggeization}).
The black solid line represents the complete result 
with the coherent sum of the $\hat{t}$- and $\hat{u}$-channel amplitudes;
see Eqs.~(\ref{amplitude_t}) and (\ref{amplitude_u}), respectively.
The black long-dashed and blue short-dashed lines represent the results
for their individual contributions, respectively.
The black dotted line corresponds to the incoherent sum of $\hat{t}$ and $\hat{u}$ contributions.
We can see that the complete result indicates a large interference effect
between the $\hat{t}$- and $\hat{u}$-channel diagrams.
This effect occurs in the region at low ${\rm M}_{4K}$ and $|\rm{Y_{diff}}| < 1$.
It can, therefore, be expected that the identification of diffractively produced 
high-mass resonances that decay into $\phi \phi$ pairs 
(e.g., $\eta_{c}$, $\chi_{c0}$, $\chi_{c2}$) should be possible at the LHC.
For this purpose, one could study the distribution 
$d^{2}\sigma/d{\rm M}_{4K}d\mathrm{Y_{diff}}$
for the $pp \to pp K^{+}K^{-}K^{+}K^{-}$ reaction;
see the discussion in \cite{Lebiedowicz:2018sdt} for the $pp \to pp p\bar{p}$ reaction.

\begin{figure}[!ht]
\includegraphics[width=0.42\textwidth]{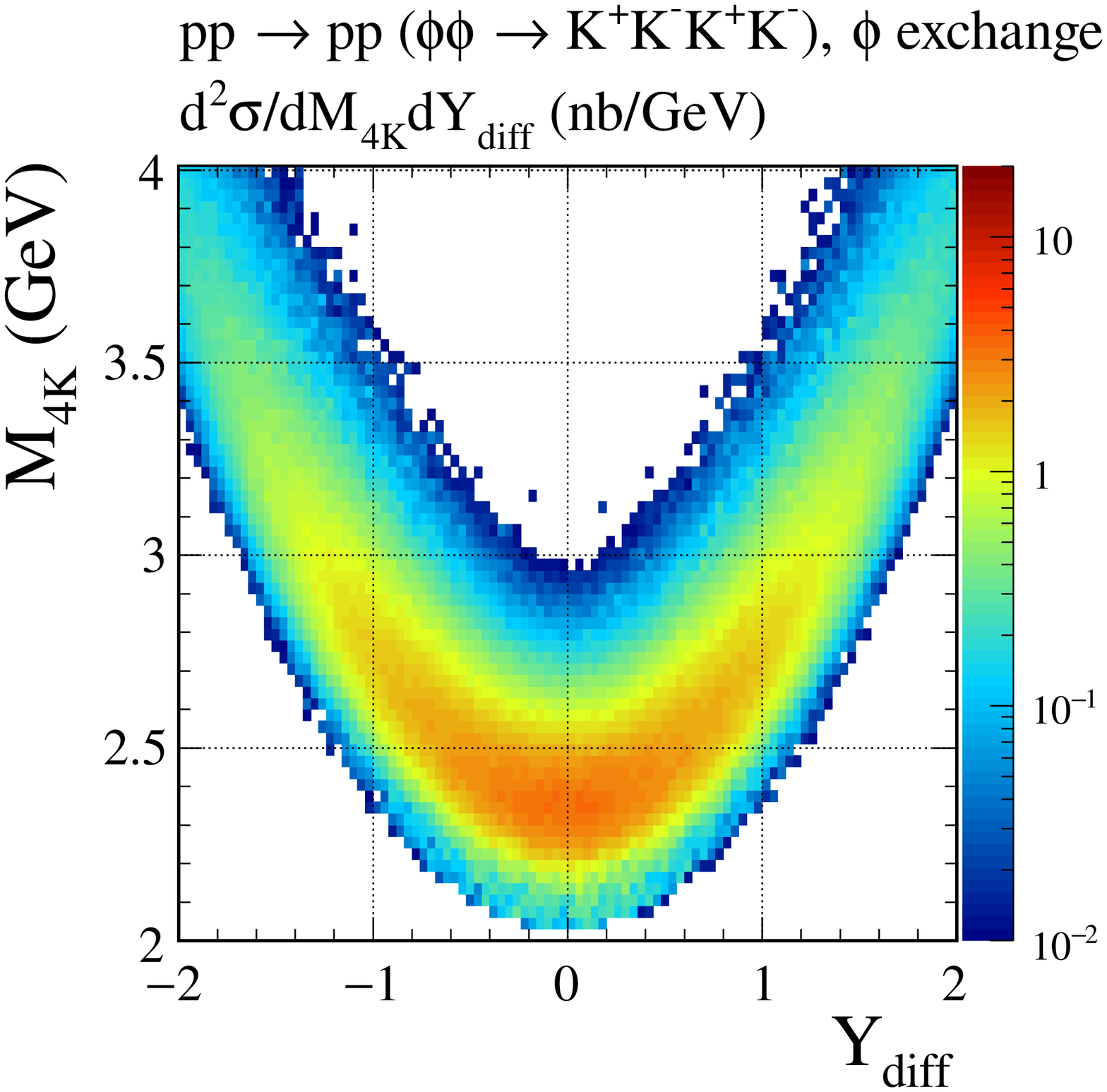}
\includegraphics[width=0.42\textwidth]{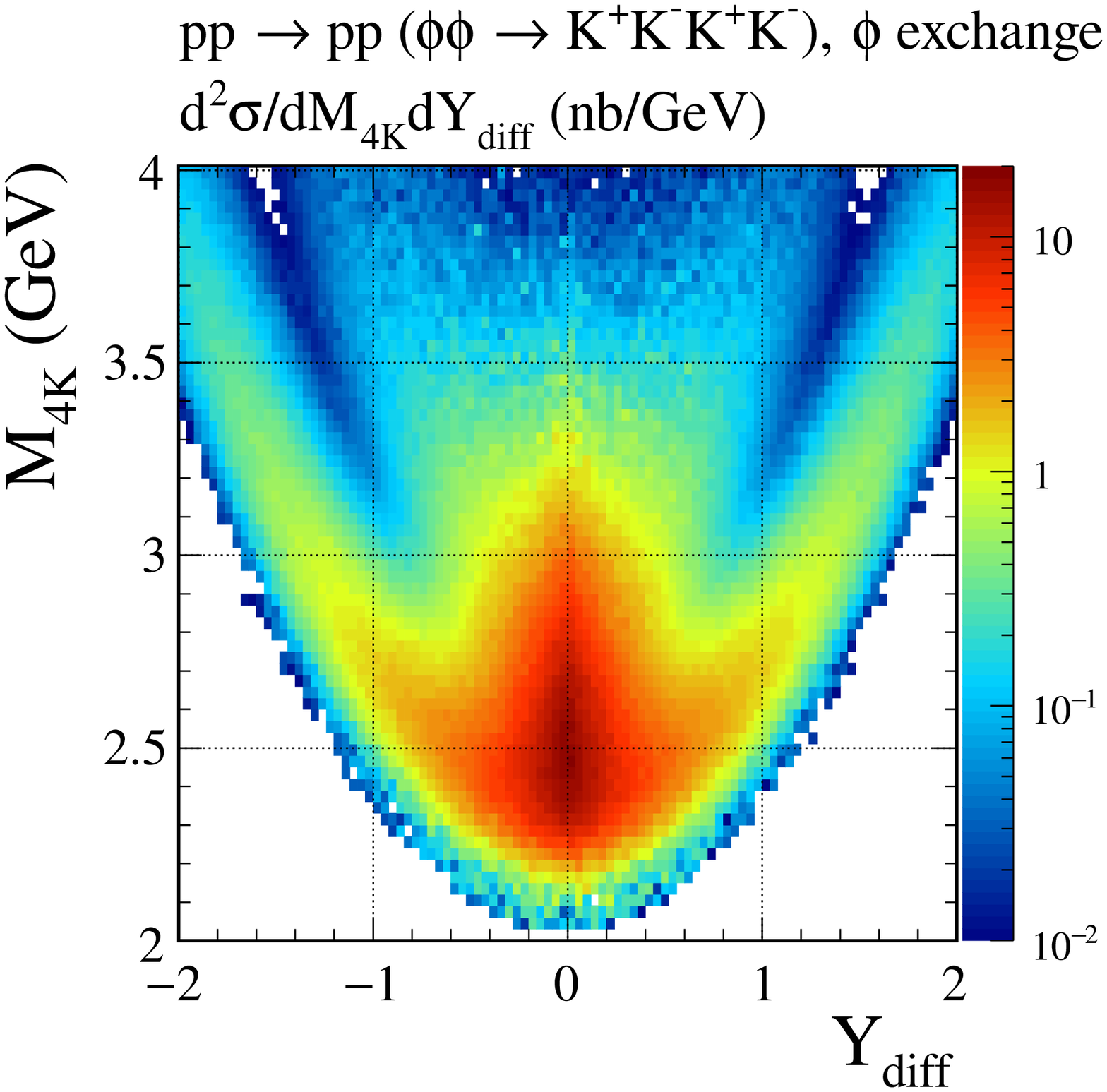}
\caption{\label{fig:3b}
The two-dimensional distributions in $(\rm{Y_{diff}},{\rm M}_{4K})$ 
for the diffractive continuum four-kaon production 
for $\sqrt{s} = 13$~TeV and $|\eta_{K}| < 2.5$, $p_{t, K} > 0.2$~GeV.
The results for two prescriptions of reggeization are presented.
The result in the left panel corresponds to the prescription (\ref{reggeization}),
and the result in the right panel corresponds to Eqs.~(\ref{second_procedure})
and (\ref{second_procedure_aux}) for ${\rm c_{y}} = 2$. 
The absorption effects are included here.}
\end{figure}

In Fig.~\ref{fig:3b} we show the distribution in $(\rm{Y_{diff}}, {\rm M}_{4K})$ for the continuum
$4K$ production via the reggeized $\phi$-exchange mechanism.
In the left panel we show the results for (\ref{reggeization})
and in the right panel for (\ref{second_procedure}) and (\ref{second_procedure_aux}) 
with ${\rm c_{y}} = 2$.
We note that with our prescriptions of reggeization
and taking into account the kinematic cuts we have a clear correlation:
large $|\rm{Y_{diff}}|$ automatically means large ${\rm M}_{4K}$.
Basically this is due to the fact that the transverse momenta of the outgoing
$\phi$ mesons stay rather small due to the form factors in (\ref{amplitude_approx}).
The behaviour of these distributions for large ${\rm M}_{4K} = {\rm M}_{\phi\phi}$
can be understood as follows. 
We are in essence studying here the reaction
$\Pom \Pom \to \phi \phi$ through $\phi$, respectively,
for large ${\rm M}_{\phi\phi}$ $\phi_{\Reg}$ ($\phi$ reggeon) exchange.
We expect then the maximum of this differential cross section for one $\phi$ forward
and the other backward.
This configuration corresponds to large ${\rm M}_{4K}$ and $|\rm{Y_{diff}}|$,
giving the ``ridge'' in Fig.~\ref{fig:3b}.
In contrast, for ${\rm M}_{4K}$ near threshold the contributions from
the $\hat{t}$ and $\hat{u}$ exchange diagrams overlap and interfere constructively;
see Fig.~\ref{fig:3a}.
This effect gives the enhancement at small ${\rm M}_{4K}$ and small $|\rm{Y_{diff}}|$
in Fig.~\ref{fig:3b}.
Because of kinematic separation of the $\hat{t}$- and $\hat{u}$-channel 
continuum contributions for ${\rm M}_{4K} > 3$~GeV
the $\eta_{c}$ and $\chi_{c}$ mesons 
could be searched for preferentially at $\rm{Y_{diff}} = 0$.
If the reggeization ansatz (\ref{reggeization}) 
is close to what is realised in nature these resonances $\eta_{c}$, $\chi_{c}$
should be clearly visible at small $|\rm{Y_{diff}}|$.
However, the reggeization ansatz (\ref{second_procedure}) gives
a larger continuum contribution at small $|\rm{Y_{diff}}|$;
see also the lower panels of Fig.~\ref{fig:3}.
Thus, if (\ref{second_procedure}) is close to the truth,
the identification of the above resonance contributions would be more difficult.

\begin{figure}[!ht]
\includegraphics[width=0.48\textwidth]{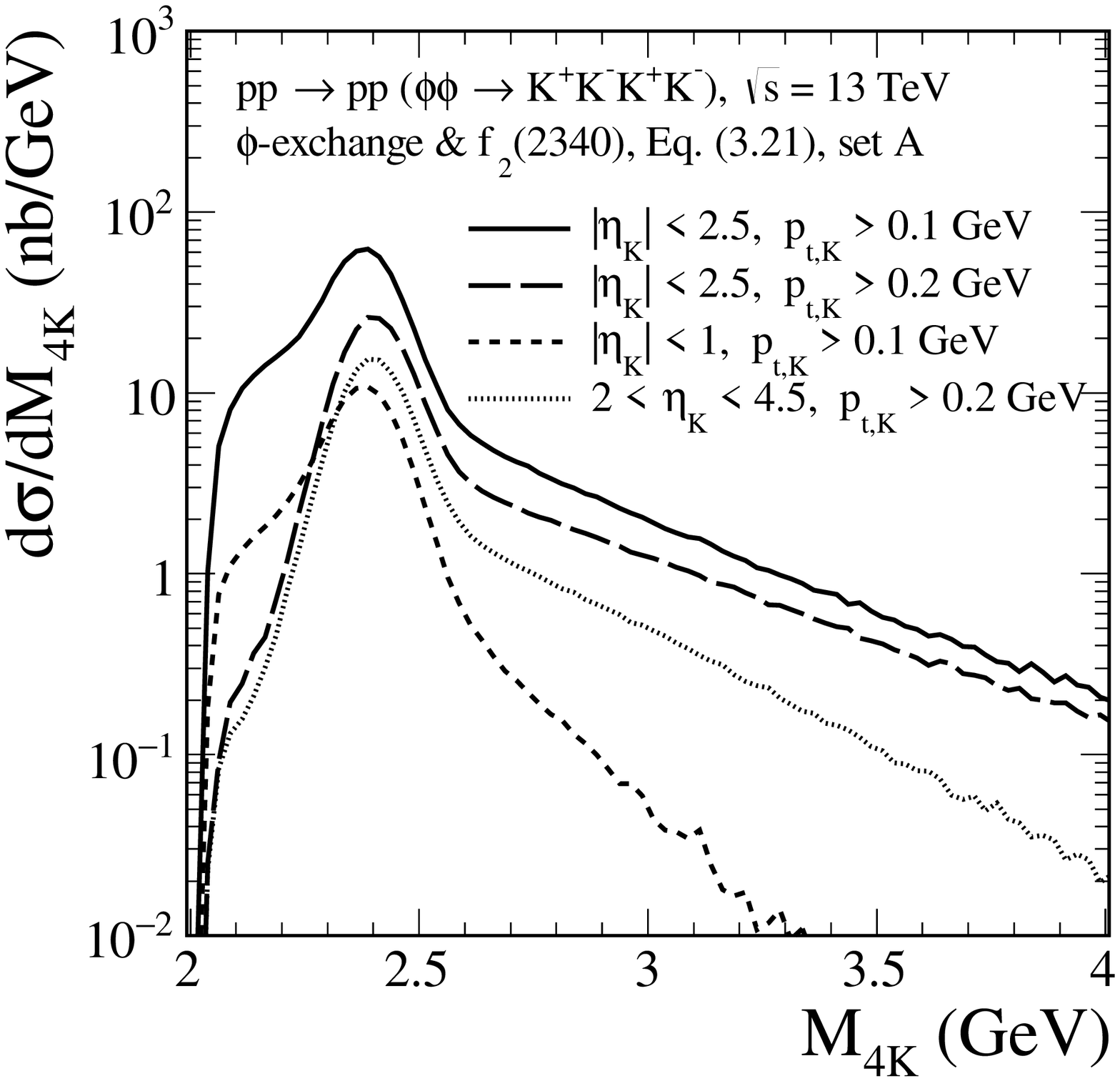}
\includegraphics[width=0.48\textwidth]{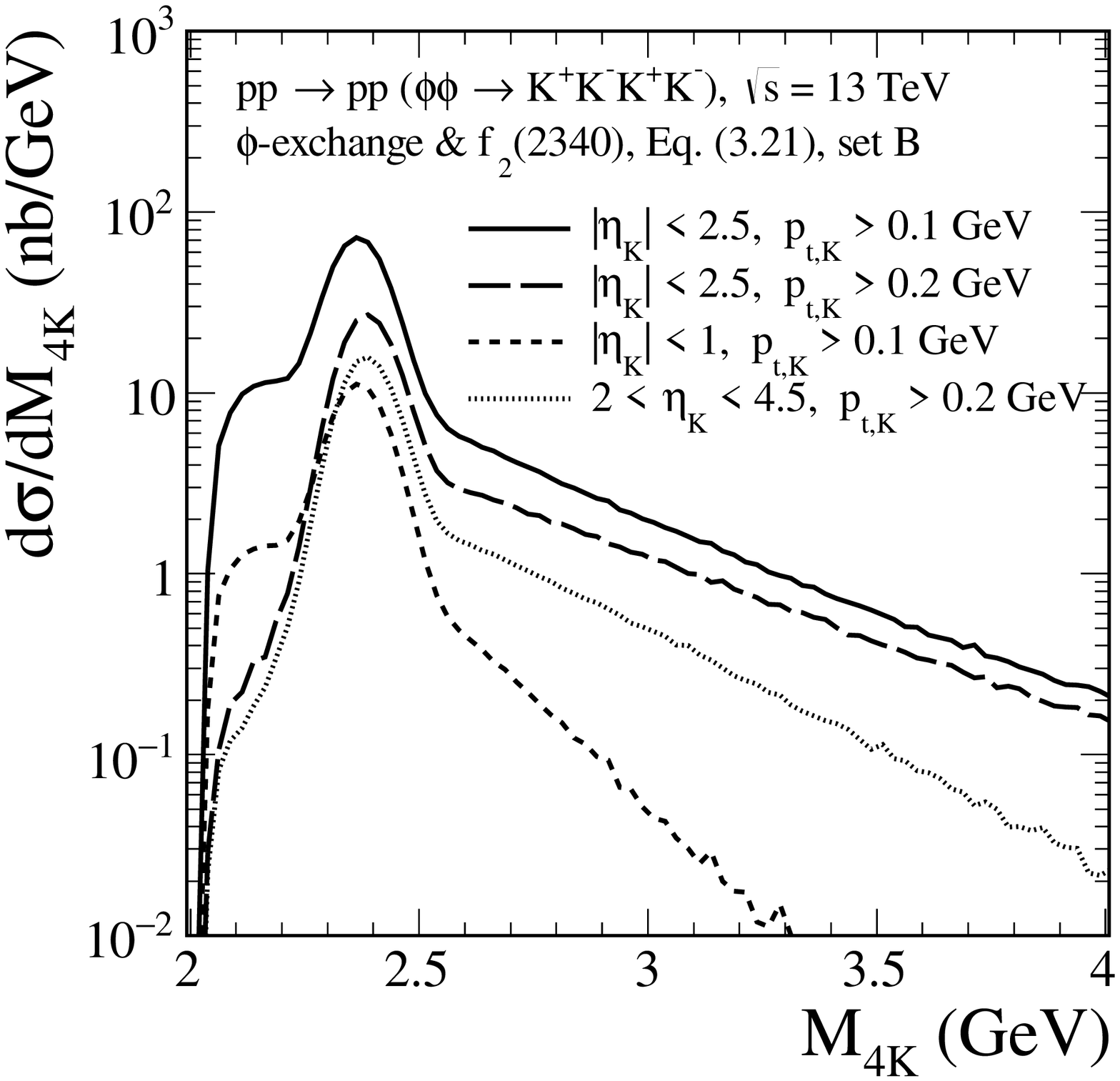}\\
\includegraphics[width=0.48\textwidth]{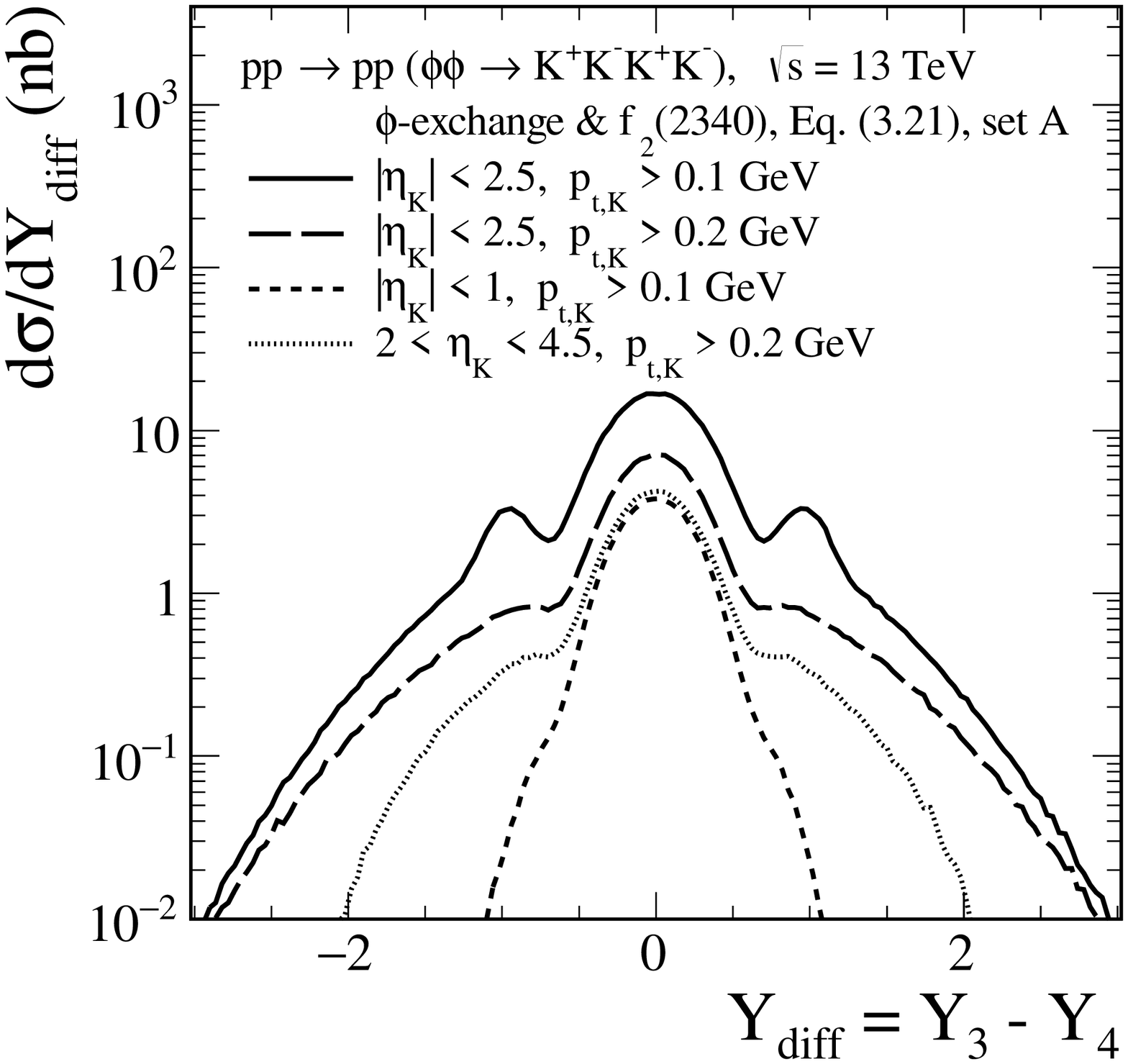}
\includegraphics[width=0.48\textwidth]{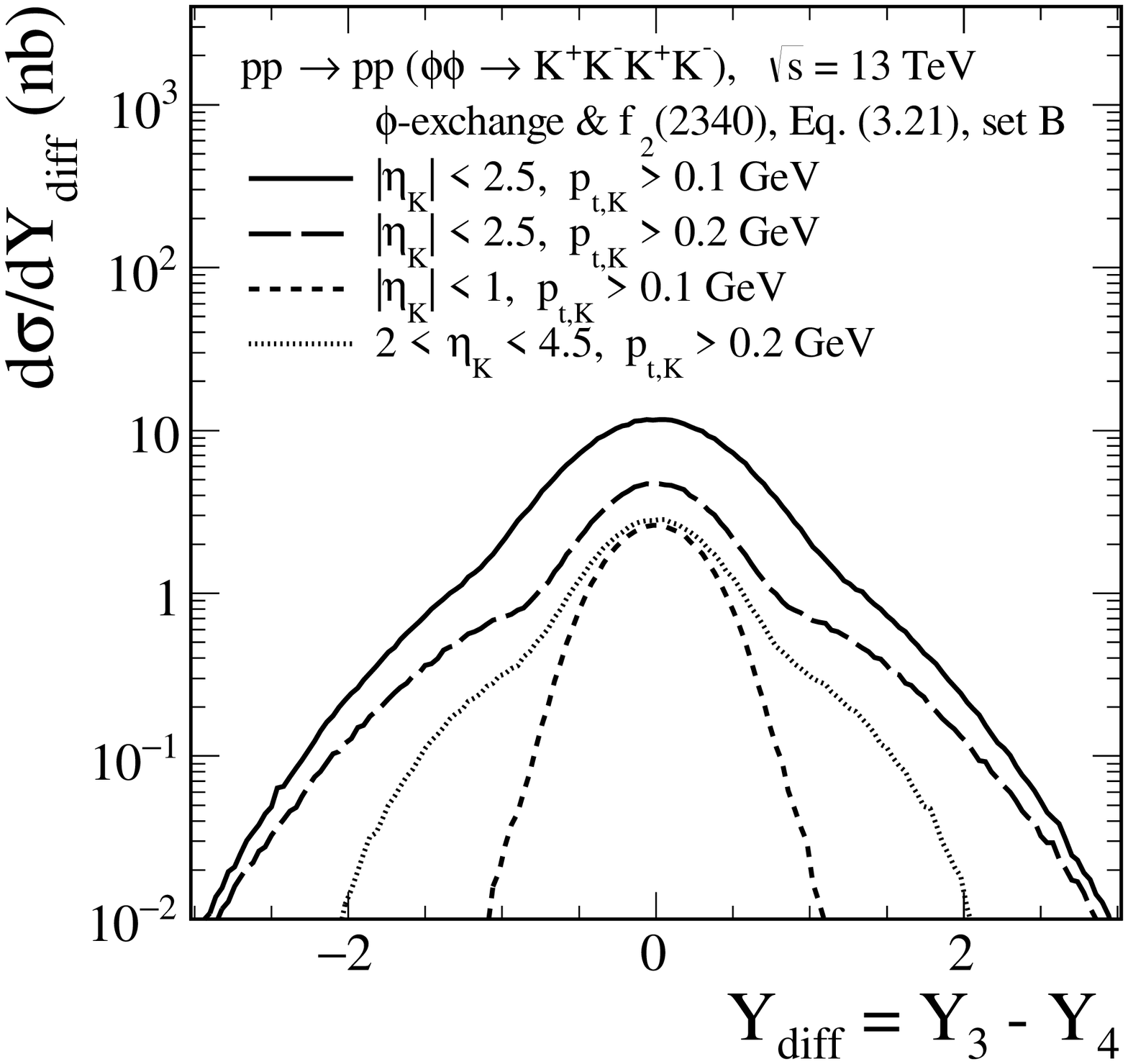}
\caption{\label{fig:4}
Differential cross sections as a function of the four-kaon invariant mass
(top panels) and as a function of $\rm{Y_{diff}}$ (bottom panels)
at $\sqrt{s} = 13$~TeV for typical experimental cuts.
The lines represent a coherent sum of the $\phi(1020)$-exchange
and the $f_{2}(2340)$ terms.
We show results for two sets of the parameters from Table~\ref{table:parameters},
set~A (see the left panels) and set~B (see the right panels).
The absorption effects are included here.}
\end{figure}

In Fig.~\ref{fig:4} we present predictions
for the $pp \to ppK^{+} K^{-}K^{+} K^{-}$ reaction
including both the continuum $\phi$-exchange contribution
and the $f_{2}(2340)$ contribution
for two sets of the parameters fixed from the WA102 data; 
see Fig.~\ref{fig:1} and Table~\ref{table:parameters}.
As can be clearly seen from Fig.~\ref{fig:4}, 
the resonance contribution generates, 
in both the ${\rm M}_{4K}$ and the $\rm{Y_{diff}}$ distributions,
patterns with a complicated structure. 
In the calculations we include the $\phi$-exchange contribution
using the reggeization prescription (\ref{reggeization})
and the dominant tensor $f_{2}(2340)$ resonance decaying into the $\phi \phi$ pair 
leading finally to the $K^{+} K^{-}K^{+} K^{-}$ final state. 
The resonance $f_{2}(2340)$ contribution is visible on top of 
the $\phi$-exchange continuum contribution.
We can see that the complete result indicates a large interference
effect of both terms.
In principle, there may also be contributions from
other tensor mesons and from $\eta$- and $f_{0}$-type mesons; 
see the fifth column in Table~\ref{table:table}. 

\begin{figure}[!ht]
\includegraphics[width=0.45\textwidth]{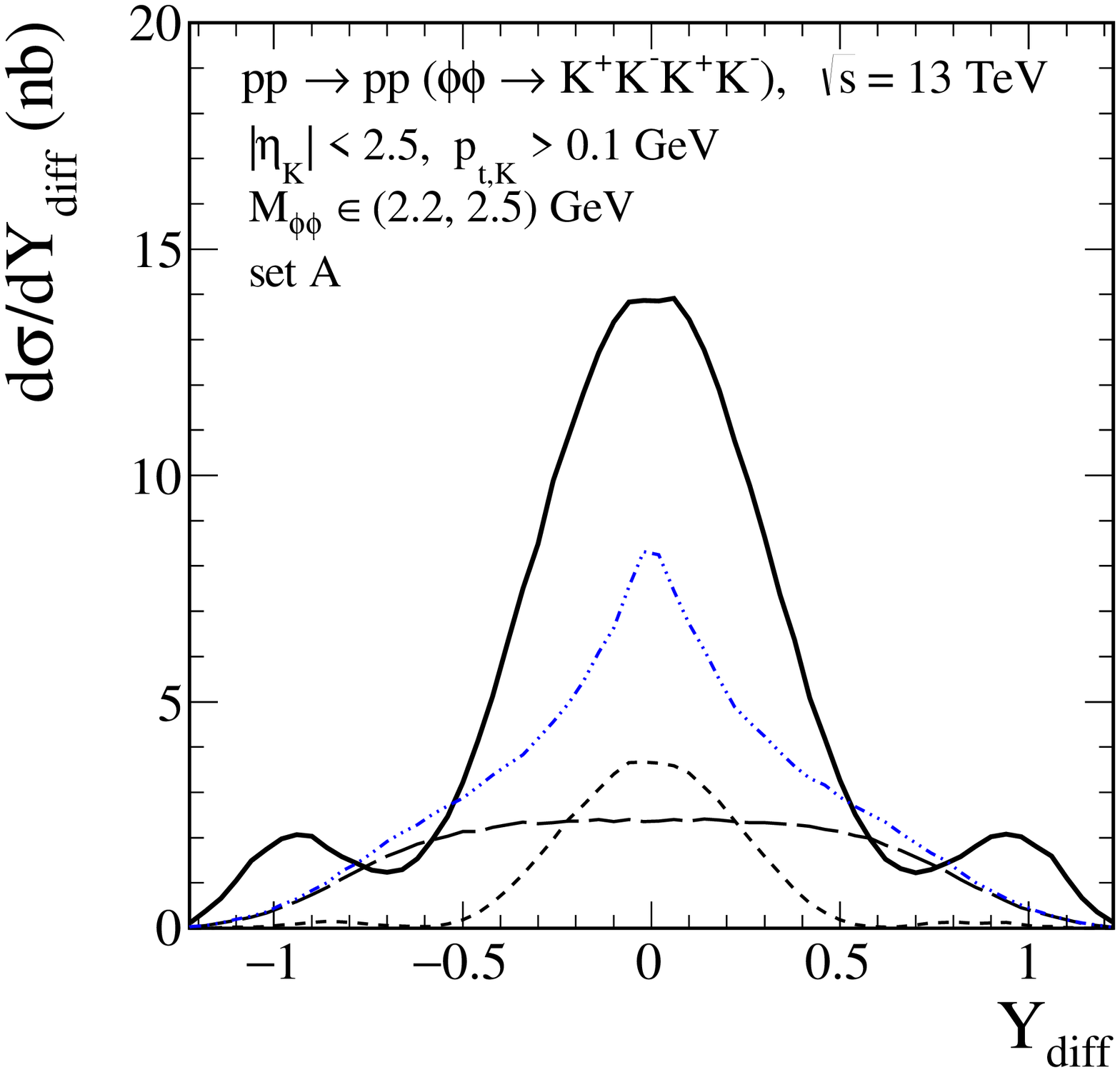}
\includegraphics[width=0.45\textwidth]{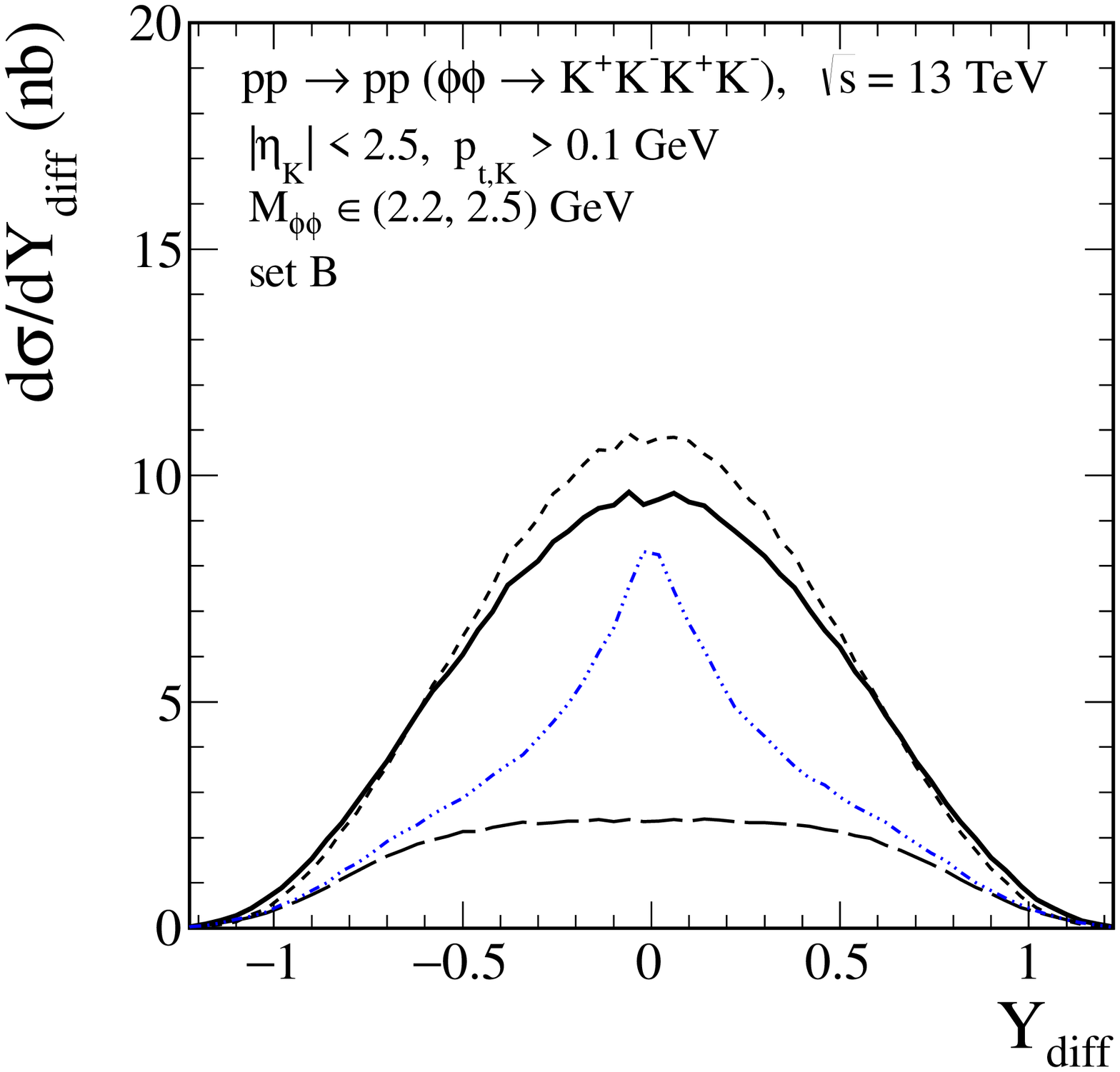}
\includegraphics[width=0.45\textwidth]{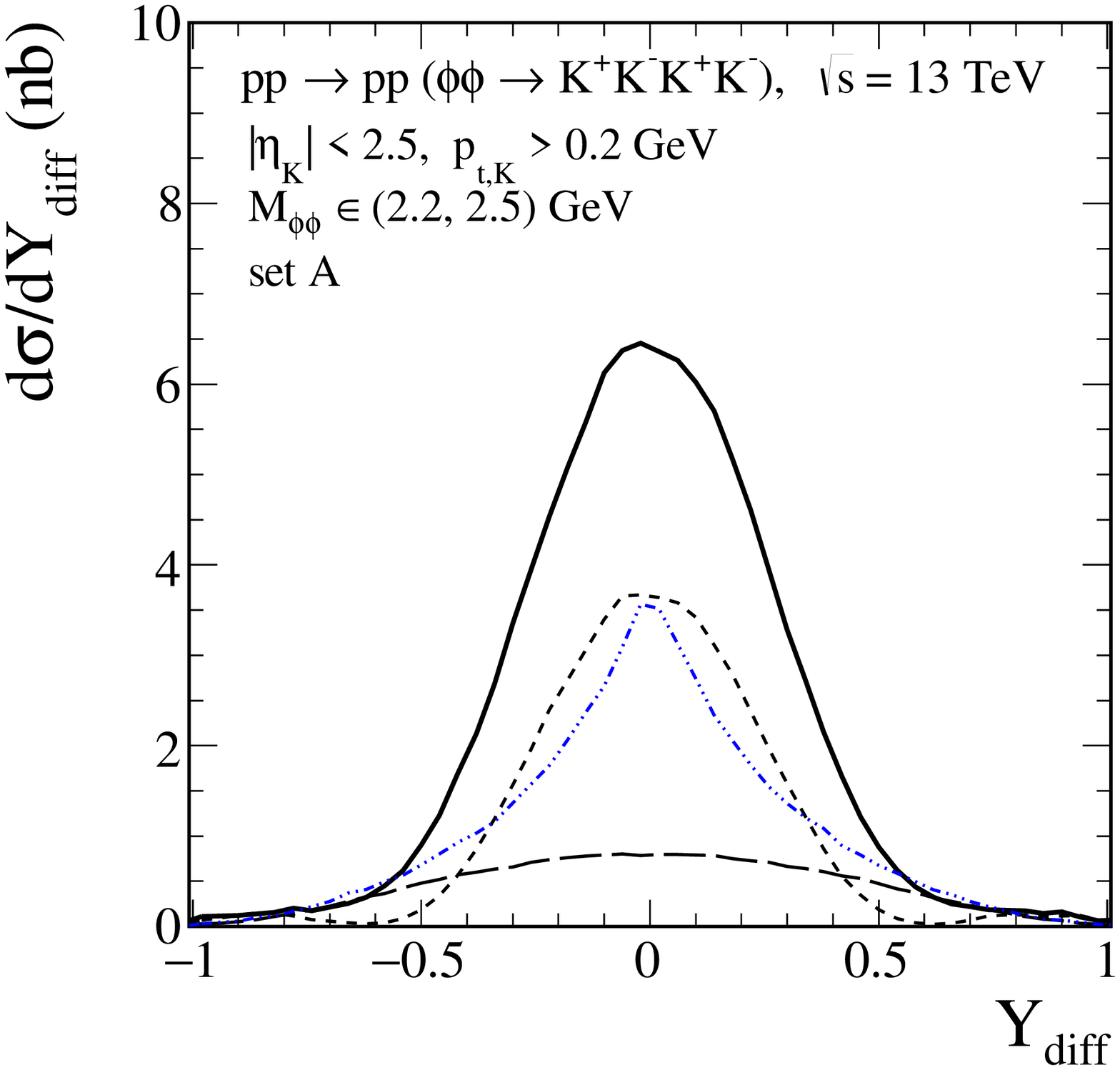}
\includegraphics[width=0.45\textwidth]{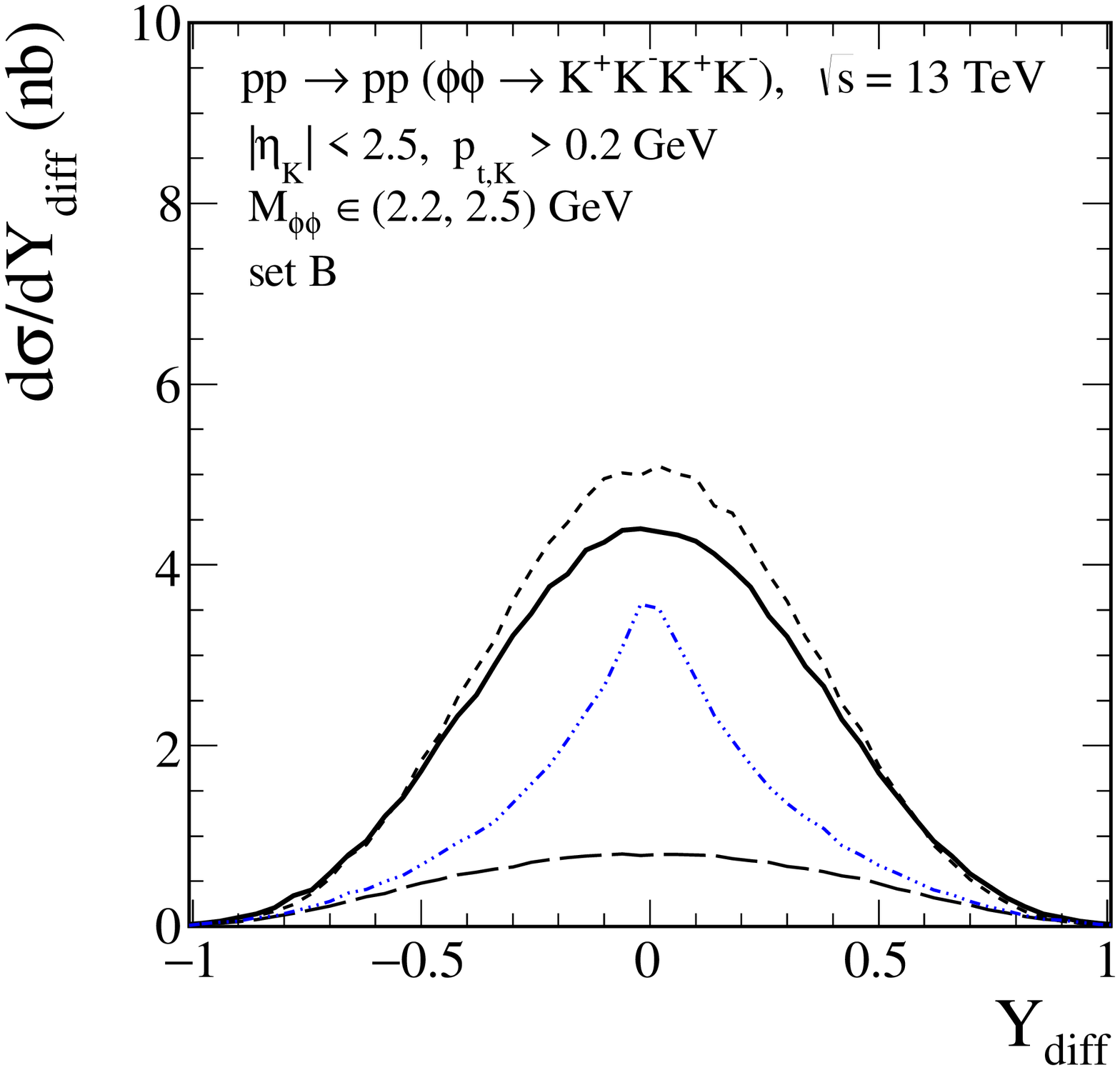}
\includegraphics[width=0.45\textwidth]{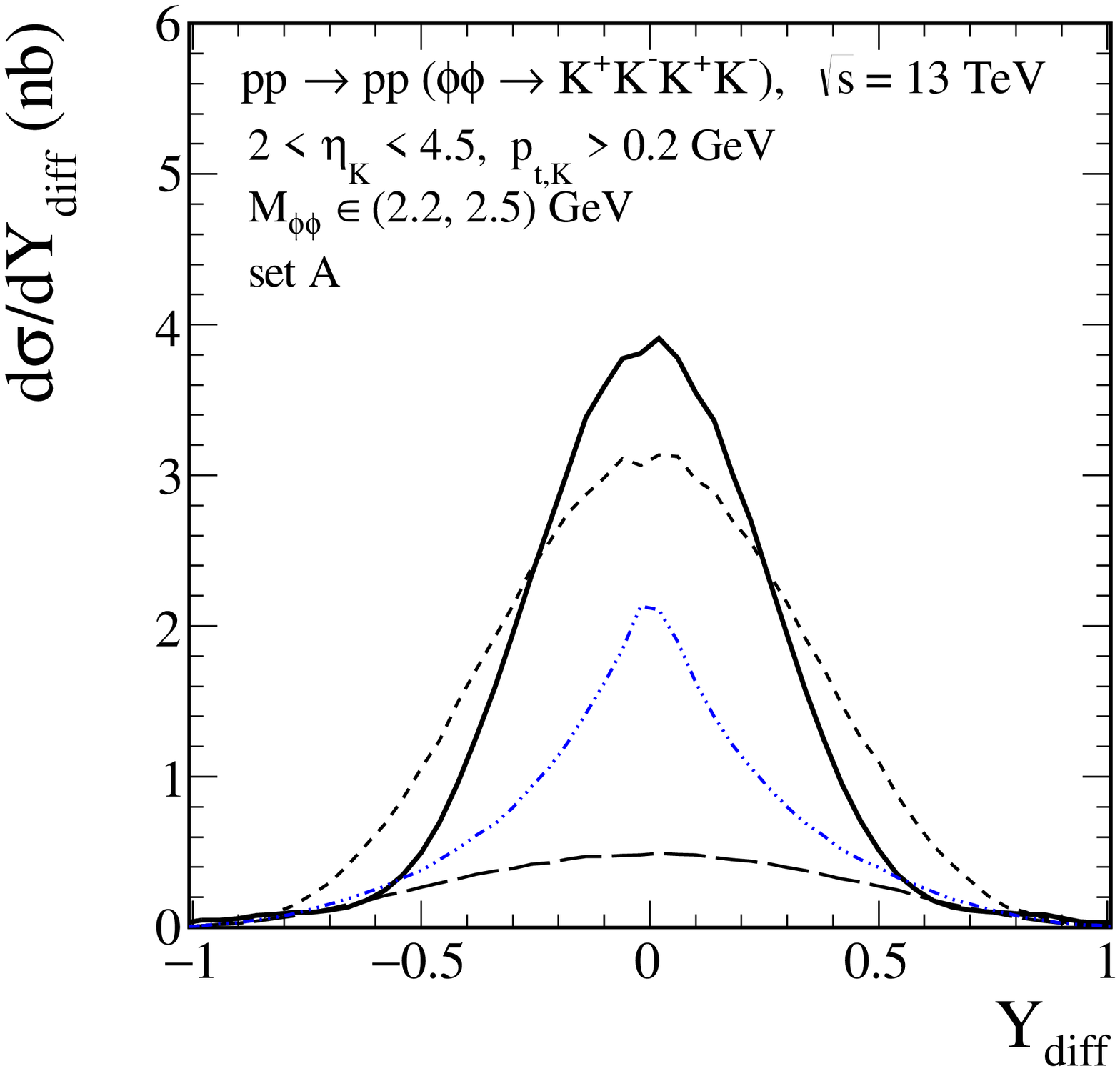}
\includegraphics[width=0.45\textwidth]{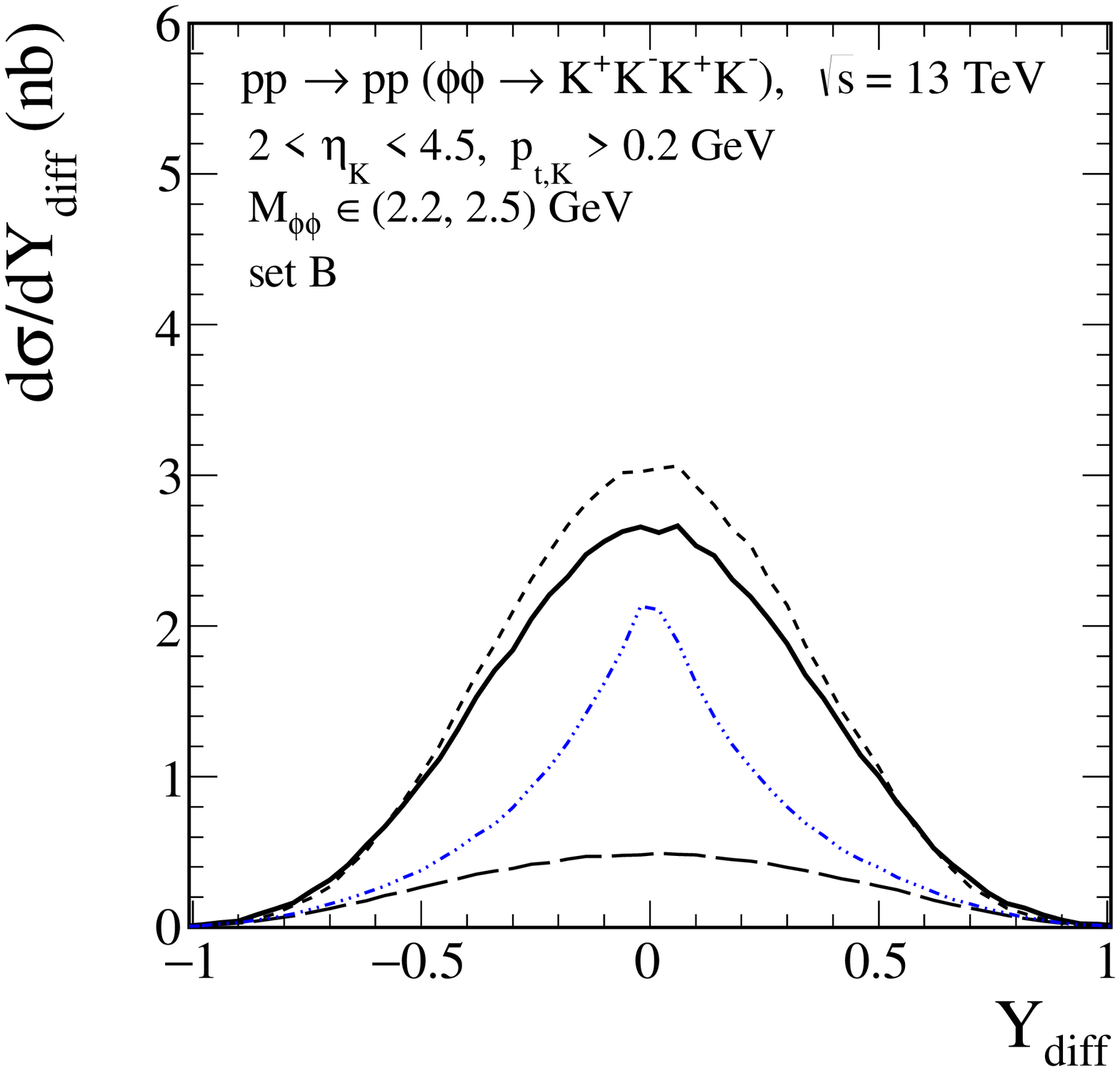}
\caption{\label{fig:4aux}
The distributions in $\rm{Y_{diff}}$ 
at $\sqrt{s} = 13$~TeV for different experimental cuts
on $\eta_{K}$ and $p_{t, K}$, and for \mbox{${\rm M}_{\phi \phi} \in (2.2,2.5)$~GeV}.
The meaning of the lines is the same as in Fig.~\ref{fig:1}.
The absorption effects are included here.}
\end{figure}

In Fig.~\ref{fig:4aux} we show the distributions in $\rm{Y_{diff}}$
for different experimental conditions, 
$|\eta_{K}| < 2.5$, $p_{t, K} > 0.2$~GeV,
$|\eta_{K}| < 2.5$, $p_{t, K} > 0.1$~GeV,
$2.0 < \eta_{K} < 4.5$, $p_{t, K} > 0.2$~GeV, from the top to bottom panels, respectively,
and in the mass range \mbox{${\rm M}_{\phi \phi} \in (2.2,2.5)$~GeV}.
We show results for the two sets, A and B, of the parameters corresponding to the left and right panels.
For the $\phi$-exchange contribution we show also results for
the alternative prescription (\ref{second_procedure}) 
and for ${\rm c_{y}} = 2$ in (\ref{second_procedure_aux}).
From Figs.~\ref{fig:4}~(bottom panels) and \ref{fig:4aux}
we can see that
the distribution in $\rm{Y_{diff}}$ can be used to determine 
the $f_{2}(2340) \to \phi \phi$ coupling (\ref{vertex_f2phiphi}),
in particular, if low $p_{t,K}$ will be available.

\begin{figure}[!ht]
\includegraphics[width=0.48\textwidth]{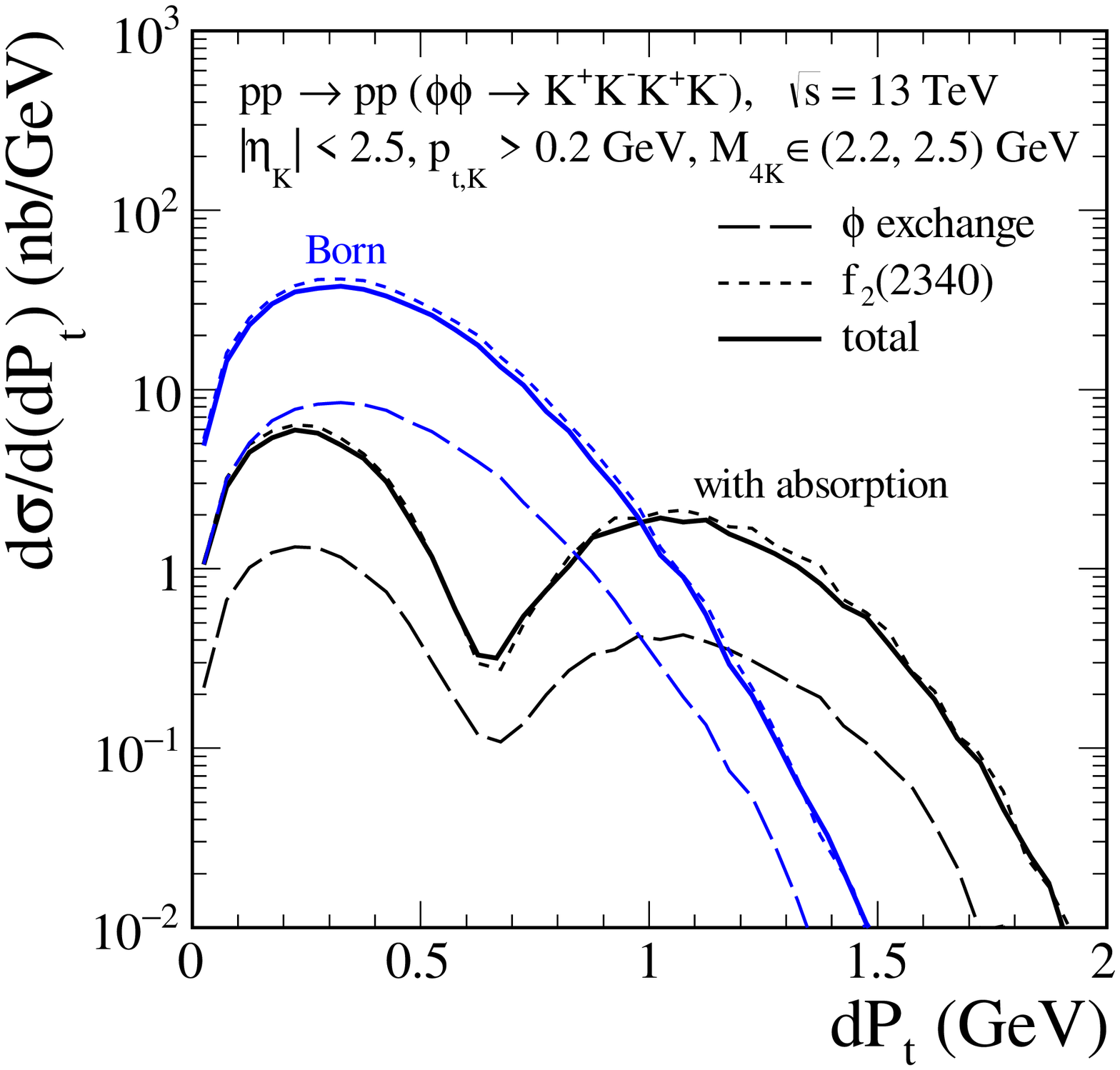}
\includegraphics[width=0.48\textwidth]{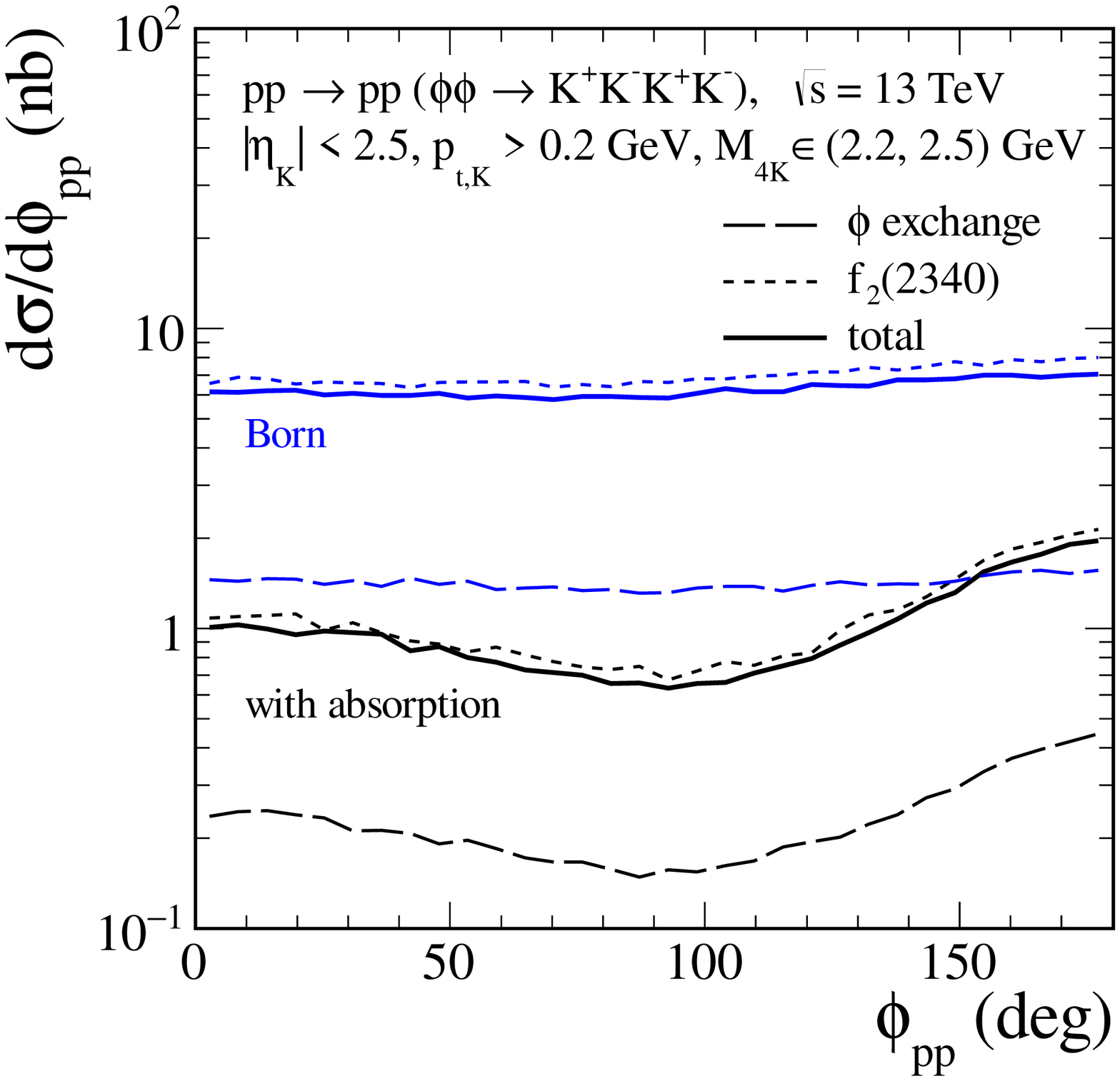}
\caption{\label{fig:5}
Distributions in $\rm{dP_{t}}$, the ``glueball filter'' variable (left panel),
and in proton-proton relative azimuthal angle $\phi_{pp}$
(right panel) for the $pp \to pp (\phi \phi \to K^{+}K^{-}K^{+}K^{-})$ reaction
through the $\phi$-exchange and $f_{2}(2340)$ mechanisms.
Here the parameter set~B from Table~\ref{table:parameters} 
and the reggeization formula (\ref{reggeization}) were used.
The predictions shown correspond to $\sqrt{s} = 13$~TeV and include cuts
for $|\eta_{K}| < 2.5$, $p_{t,K} >0.2$~GeV, and ${\rm M}_{4K} \in (2.2,2.5)$~GeV.
The black lines correspond to the results with the absorption effects included.
For comparison, the blue lines, marked ``Born'', correspond 
to the results without absorption.}
\end{figure}

In Fig.~\ref{fig:5} we discuss the observables $\rm{dP_{t}}$ (\ref{dPt_variable})
and $\phi_{pp}$ for which the distributions 
are very sensitive to the absorption effects.
The results shown correspond to $\sqrt{s} = 13$~TeV and include cuts
for $|\eta_{K}| < 2.5$, $p_{t,K} >0.2$~GeV, and ${\rm M}_{4K} \in (2.2,2.5)$~GeV.
Quite a different pattern can be seen for the Born case and for the case with absorption included.
The absorptive corrections lead to significant modification
of the shape of the $\phi_{pp}$ distribution and lead to an increase of the cross section
for large $\rm{dP_{t}}$. 
This effect could be verified in future experiments when both protons are measured, 
e.g., by the CMS-TOTEM and the ATLAS-ALFA experimental groups.

In Table~\ref{tab:table2} we have collected integrated cross sections 
in nb for different experimental cuts
for the exclusive $K^{+}K^{-}K^{+}K^{-}$ production
including only the contributions shown in Fig.~\ref{fig:4K_diagrams}.
The results were obtained in the calculations with 
the tensor pomeron exchanges.
The absorption effects are included in the calculations.
\begin{table}[]
\centering
\caption{The integrated cross sections in nb 
for the central exclusive $K^{+}K^{-}K^{+}K^{-}$ production
in proton-proton collisions via the intermediate $\phi\phi$ system
due to the mechanisms shown in Fig.~\ref{fig:4K_diagrams}.
The results have been calculated for $\sqrt{s}=13$~TeV
and some typical experimental cuts
using the parameter set~B from Table~\ref{table:parameters}.
The calculations for the $\phi$-exchange contribution were made using (\ref{reggeization}).
The absorption effects are included here.}
\label{tab:table2}
\begin{tabular}{ll|c|c|c|c}
\cline{3-5}
&  &  \multicolumn{3}{c|}{Cross sections (nb)} \\ \cline{1-5}
\multicolumn{1}{|c|}{$\sqrt{s}$, TeV} & Cuts 
&  Total &   $\phi$ exchange      &    $f_{2}(2340)$       &  \\ 
\cline{1-5}
\multicolumn{1}{|l|}{13} &  $|\eta_{K}| < 1$, $p_{t, K} > 0.1$~GeV
& \;\;2.11 & 0.83 & \;\;2.00 &  \\ 
\multicolumn{1}{|l|}{13} &  $|\eta_{K}| < 2.5$, $p_{t, K} > 0.1$~GeV
& 16.16 & 8.30 & 12.80 &  \\ 
\multicolumn{1}{|l|}{13} &  $|\eta_{K}| < 2.5$, $p_{t, K} > 0.2$~GeV
& \;\;5.75 & 2.67 & \;\;4.47 &  \\ 
\multicolumn{1}{|l|}{13} &  $2 < \eta_{K} < 4.5$, $p_{t, K} > 0.2$~GeV
& \;\;3.06 & 1.26 & \;\;2.62 &  \\ 
\cline{1-5}
\end{tabular}
\end{table}

\subsection{Results including odderon exchange}
\label{sec:section_6}

In this section we shall discuss possibilities to observe odderon-exchange effects
in the CEP of $\phi \phi$ pairs.

The odderon was introduced on theoretical grounds in \cite{Lukaszuk:1973nt, Joynson:1975az}.
For a review of the odderon, see, e.g.,~\cite{Ewerz:2003xi}.
Recent experimental results by the TOTEM Collaboration \cite{Antchev:2017yns, Antchev:2018rec}
have brought the odderon question to the forefront again.
For recent theoretical papers dealing with the odderon, see, e.g.,~\cite{Ewerz:2013kda, Bolz:2014mya},
which came out before the TOTEM results, and 
\cite{Martynov:2017zjz, Martynov:2018nyb, Khoze:2017swe, Broilo:2018qqs,
Goncalves:2018pbr, Harland-Lang:2018ytk, Csorgo:2018uyp}.

Clearly, it is of great importance in this context to study
possible odderon effects in reactions other than proton-proton elastic scattering.
We shall argue here that the CEP of a $\phi\phi$ state offers
a very nice way to look for odderon effects as suggested in \cite{Ewerz:2003xi}.

\begin{figure}[!ht]
\includegraphics[width=0.48\textwidth]{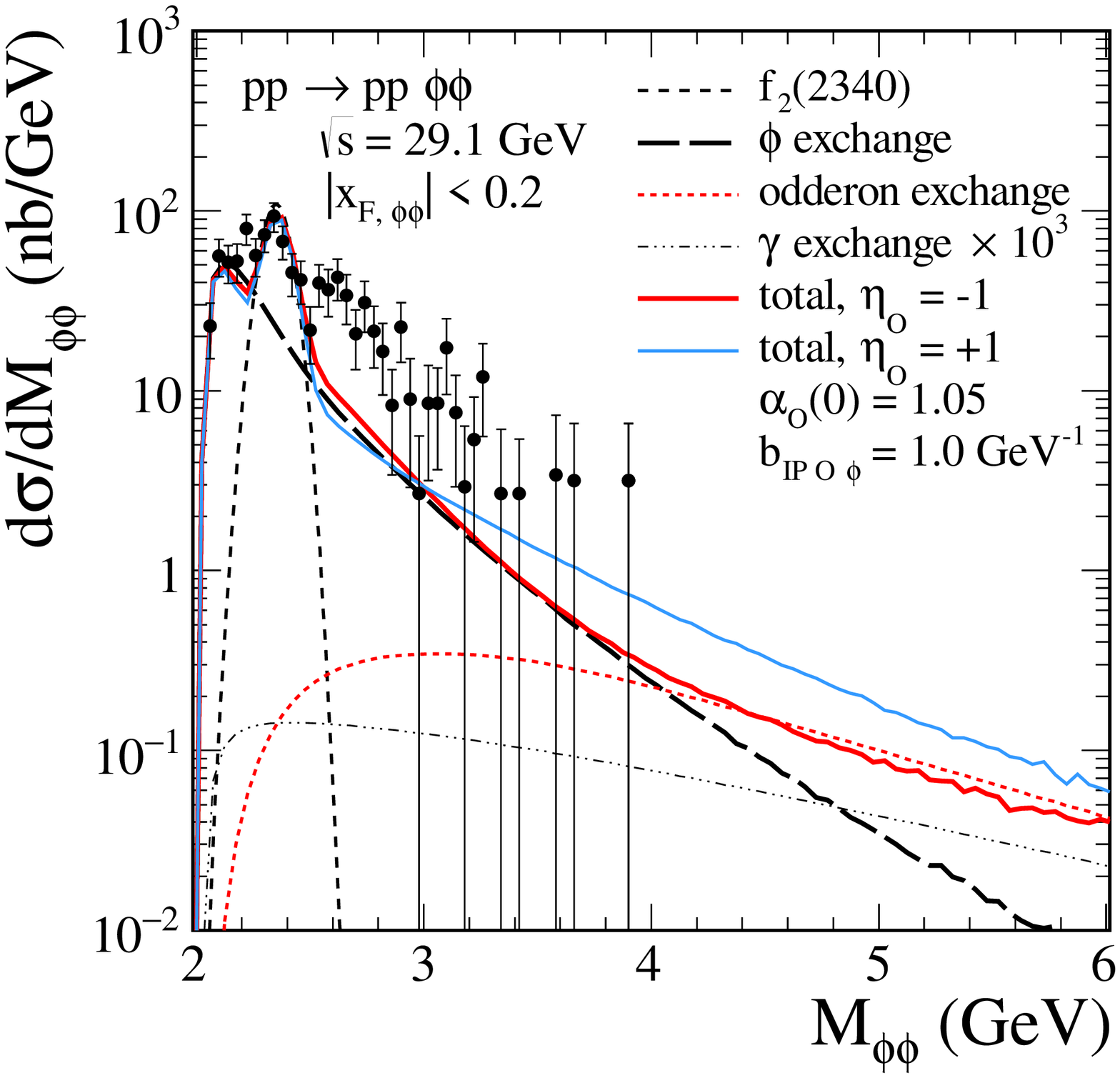}
\includegraphics[width=0.48\textwidth]{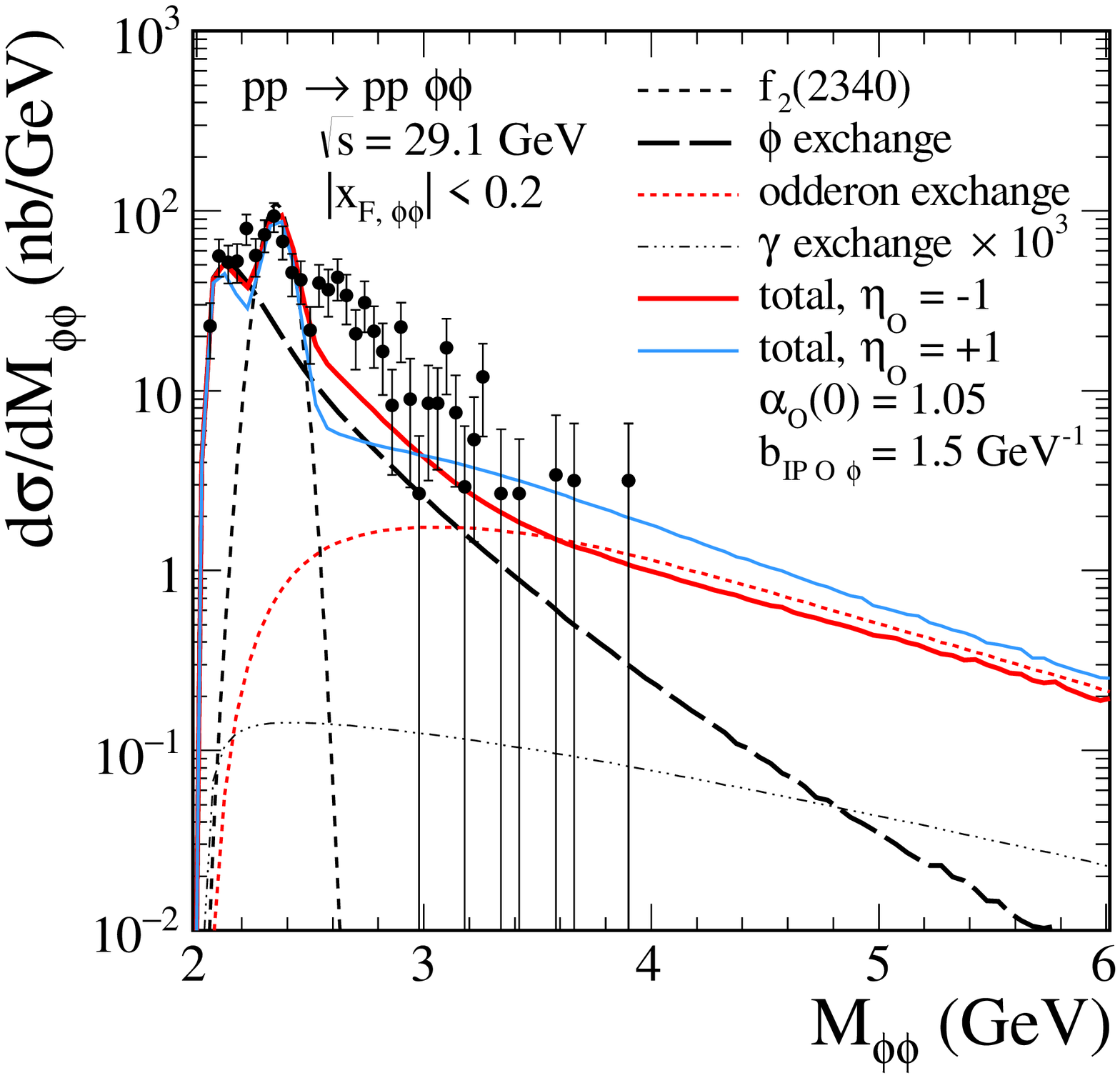}
\caption{\label{odderon_WA102}
Invariant mass distributions for the central production of $\phi \phi$ at $\sqrt{s} = 29.1$~GeV
and $|x_{F,\phi \phi}| \leqslant 0.2$ together with the WA102 data \cite{Barberis:1998bq} are shown.
The black long-dashed line corresponds to the $\phi$-exchange contribution 
and the black dashed line corresponds to the $f_{2}(2340)$ contribution.
The black dotted-dashed line corresponds to the $\gamma$-exchange contribution 
enlarged by a factor $10^{3}$.
In the calculations the parameter set~B of Table~\ref{table:parameters}
for the $\phi$-exchange and $f_{2}$ terms,
and the parameters (\ref{odd_parameters}) for the odderon term have been used.
The red dotted line represents the odderon-exchange contribution
for $a_{\Pom \Ode \phi}=0$, $b_{\Pom \Ode \phi}= 1.0$~GeV$^{-1}$ (left panel) 
and for $a_{\Pom \Ode \phi}=0$, $b_{\Pom \Ode \phi}= 1.5$~GeV$^{-1}$ (right panel).
The coherent sum of all terms is shown by the red and blue solid lines
for $\eta_{\Ode} = -1$ and $\eta_{\Ode} = +1$, respectively.
Here we take $\alpha_{\Ode}(0) = 1.05$.
The absorption effects are included in the calculations.}
\end{figure}

\begin{figure}[!ht]
\includegraphics[width=0.48\textwidth]{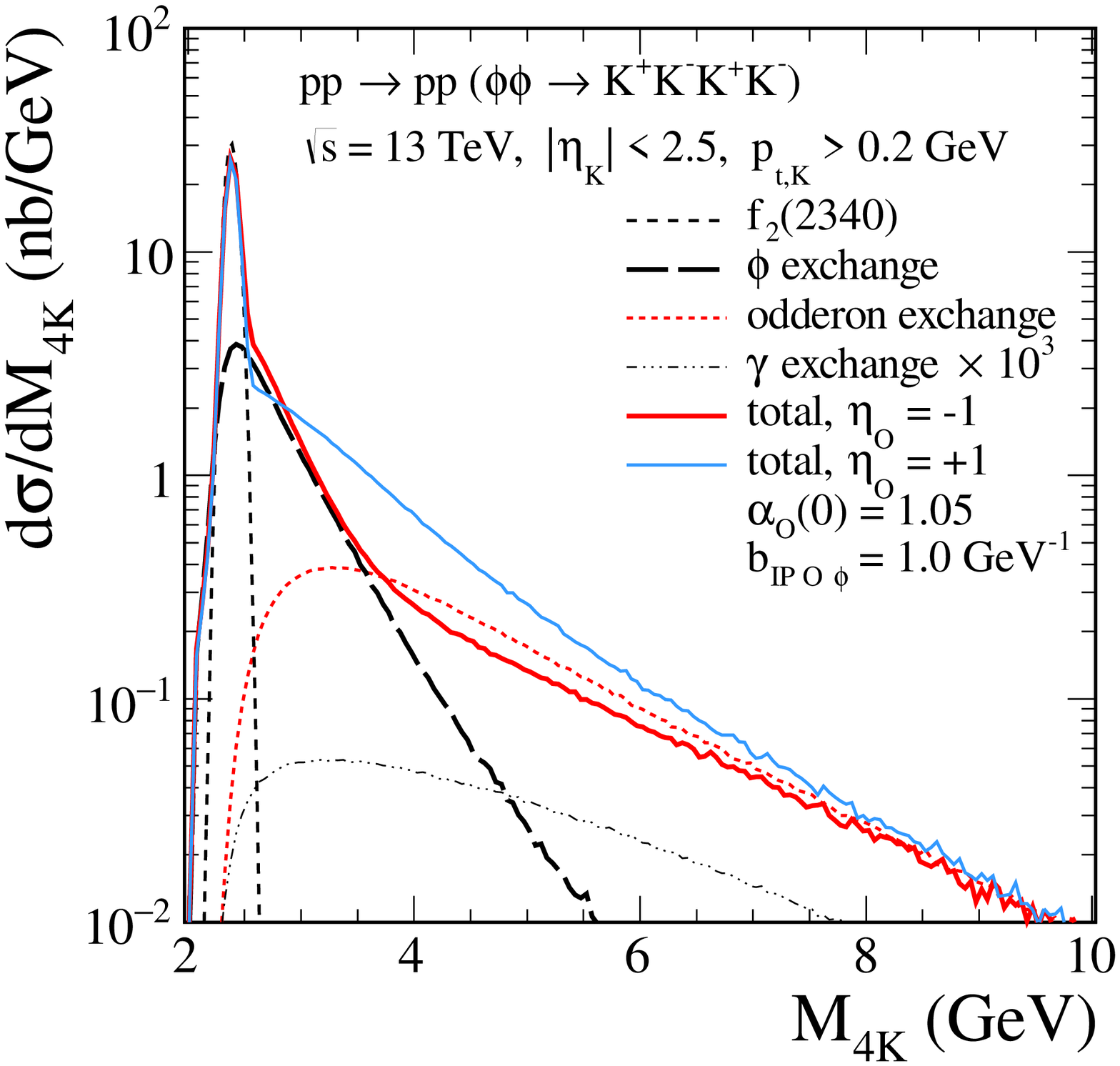}
\includegraphics[width=0.48\textwidth]{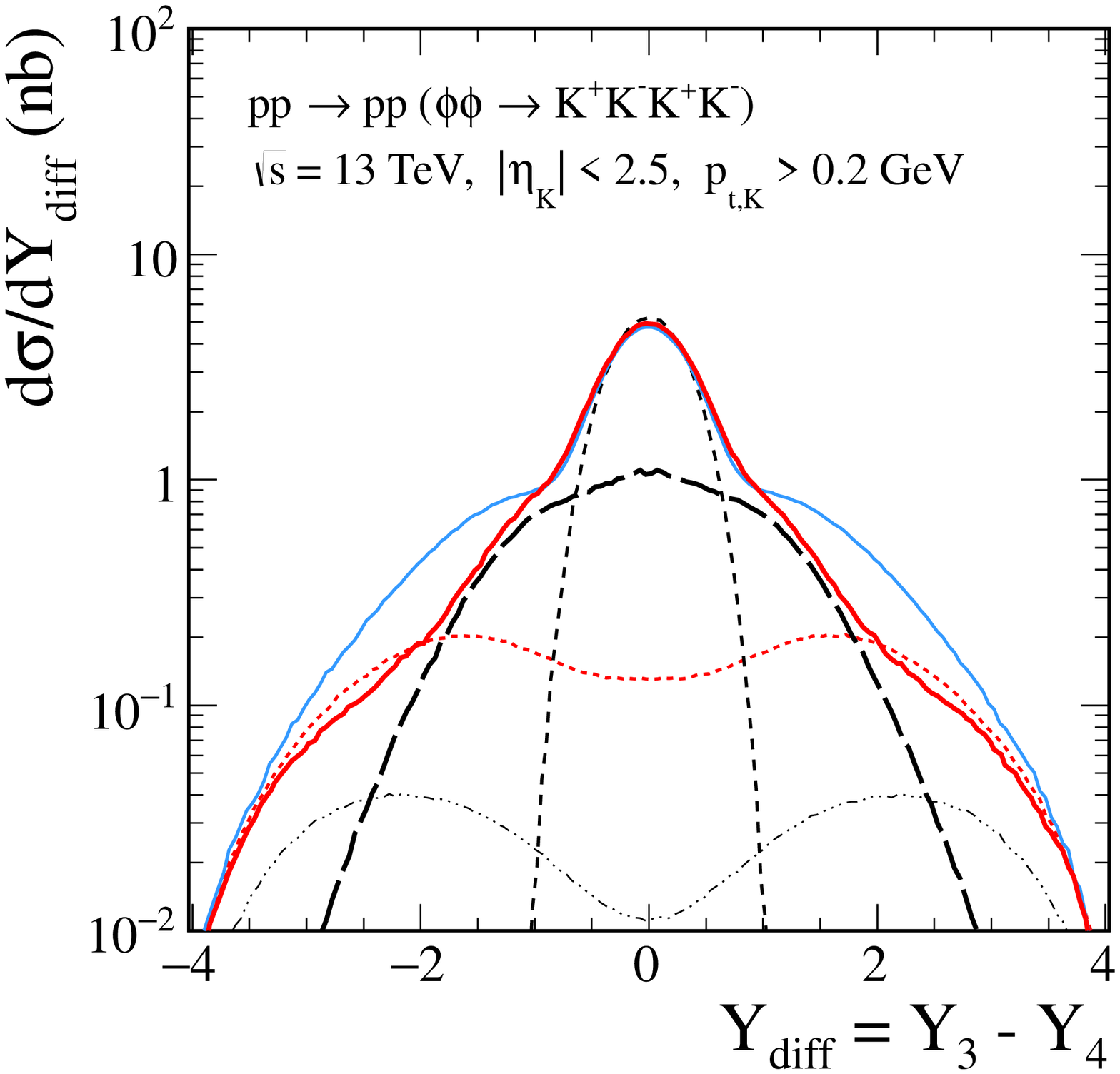}\\
\includegraphics[width=0.48\textwidth]{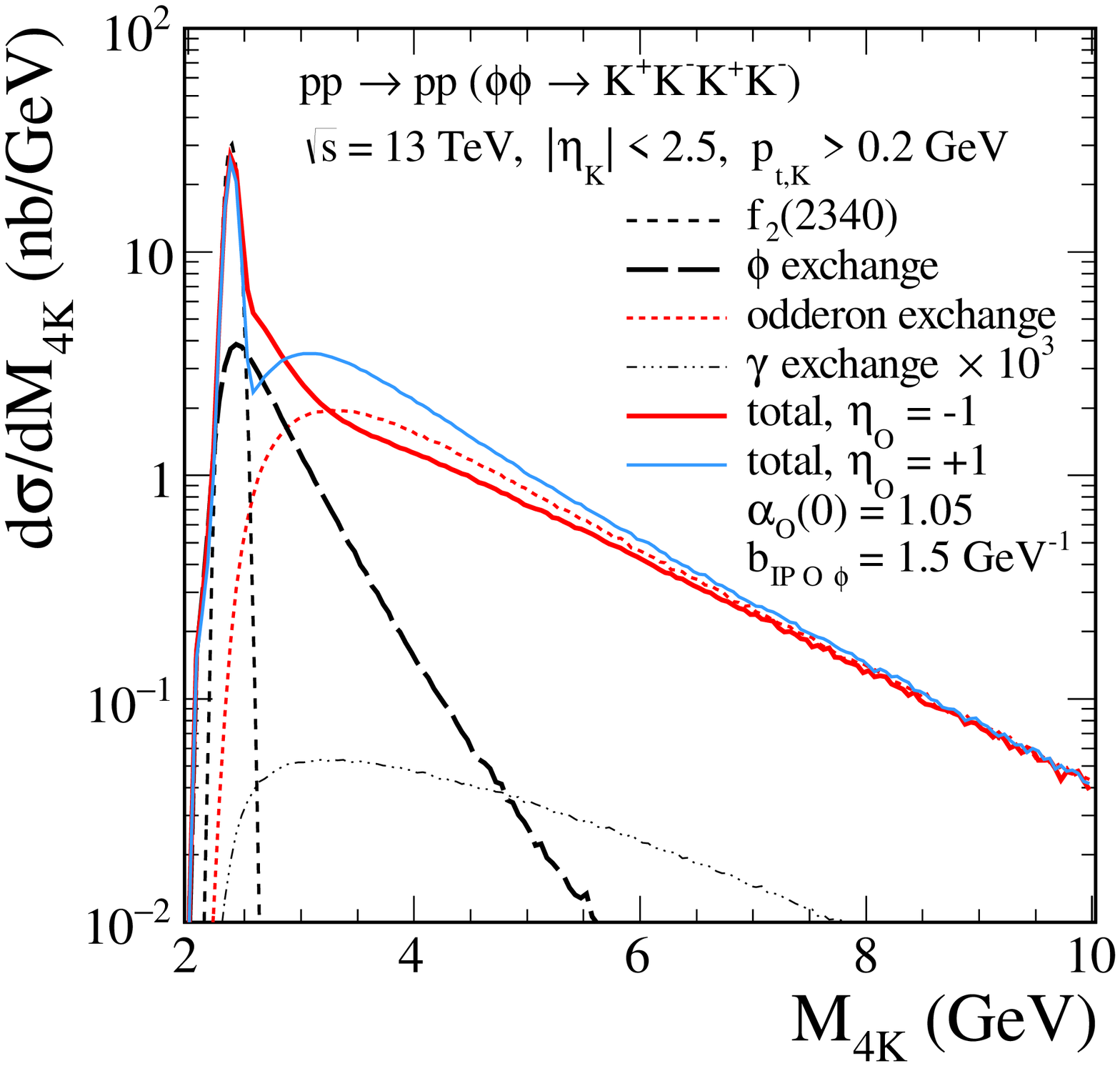}
\includegraphics[width=0.48\textwidth]{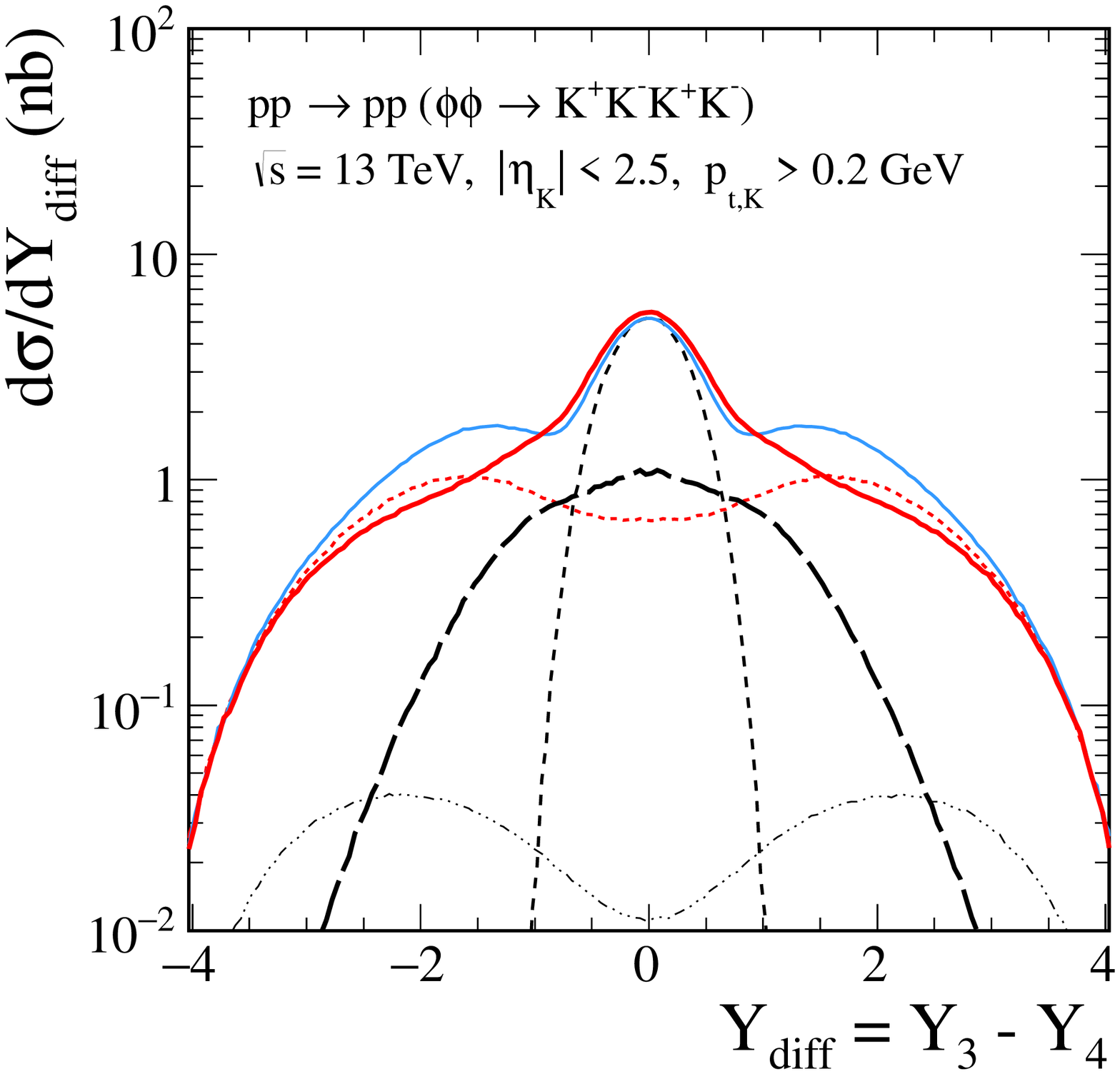}
\caption{\label{odderon_LHC}
The distributions in ${\rm M}_{4K}$ (left panels)
and in $\rm{Y_{diff}}$ (right panels)
for the $pp \to pp (\phi \phi \to K^{+}K^{-}K^{+}K^{-})$ reaction 
calculated for $\sqrt{s} = 13$~TeV and $|\eta_{K}| < 2.5$, $p_{t, K} > 0.2$~GeV.
The meaning of the lines is the same as in Fig.~\ref{odderon_WA102}.
The red and blue solid lines correspond to the complete results with
$\eta_{\Ode} = -1$ and $\eta_{\Ode} = +1$, respectively.
The results for $b_{\Pom \Ode \phi}= 1.0$~GeV$^{-1}$ (top panels)
and for $b_{\Pom \Ode \phi}= 1.5$~GeV$^{-1}$ (bottom panels) are presented.
The absorption effects are included in the calculations.}
\end{figure}

In Figs.~\ref{odderon_WA102} and \ref{odderon_LHC} 
we show results for the diffractive CEP of $\phi \phi$ pairs
including the mechanism with odderon exchange shown in Fig.~\ref{fig:odderon} (a).
Here we take the following values of the parameters for the odderon exchange:
\begin{eqnarray}
\eta_{\Ode} = \pm 1\,, \;\;
\alpha_{\Ode}(0) = 1.05\,,\;\;
a_{\Pom \Ode \phi}=0\,,\;\;
b_{\Pom \Ode \phi}=1.0,\, 1.5 \; {\rm GeV}^{-1}\,;
\label{odd_parameters}
\end{eqnarray}
see (\ref{A14}), (\ref{A15}), and $\Lambda^{2}=1.0$~GeV$^{2}$ in (\ref{Fpomodephi_ff}).
In the calculations we have used the parameter 
set~B of Table~\ref{table:parameters}
for the $\Pom \Pom f_{2}$ contribution.
For the case of $\phi$ exchange we have used the formula of reggeization (\ref{reggeization}).
In Fig.~\ref{odderon_WA102} we show the results for $\sqrt{s} = 29.1$~GeV
and compare them to the WA102 data.
Figure~\ref{odderon_LHC} shows the predictions for $\sqrt{s} = 13$~TeV using the same parameters.
We show the $\phi$-meson-exchange contribution (see the black long-dashed line),
the $f_{2}(2340)$ contribution (see the black dashed line), 
and the odderon-exchange contribution (see the red dotted line).
The black dotted-dashed line
corresponds to the photon-exchange contribution, 
represented by the diagram in Fig.~\ref{fig:gamma},
multiplied by a factor $10^{3}$ to be visible in the figure.
The coherent sum of all contributions is shown by the red and blue solid lines,
corresponding to $\eta_{\Ode} = -1$ and $\eta_{\Ode} = +1$, respectively.
Clearly, the complete result indicates a large interference effect
between the $\phi$- and odderon-exchange diagrams.
We see from the right panel of Fig.~\ref{odderon_WA102}
that for ${\rm M}_{\phi \phi} \gtrsim 2.5$~GeV the WA102 data leave room
for a possible odderon contribution which here we
normalised in such a way as not to exceed the WA102 cross section.
Such an odderon contribution
with $b_{\Pom \Ode \phi}= 1.5$~GeV$^{-1}$
can be treated then rather as an upper limit.
Of course the ``true'' odderon contribution may be much smaller.

In Fig.~\ref{odderon_LHC} we show the results
for the ATLAS experimental conditions ($|\eta_{K}| < 2.5$, $p_{t, K} > 0.2$~GeV).
For the odderon term we take here again the parameters (\ref{odd_parameters}).
With these the odderon term gives a large enhancement of the ${\rm M}_{4K}$ distribution
for \mbox{${\rm M}_{4K} \gtrsim 3$~GeV} and clearly dominates at large $|\rm{Y_{diff}}|$.
Whereas for \mbox{${\rm M}_{4K} \gtrsim 3$~GeV} and $\eta_{\Ode} = +1$ 
there is constructive interference of the $\phi$-exchange and the odderon terms, 
for $\eta_{\Ode} = -1$ the interference is destructive.
But in any case, for ${\rm M}_{4K} \gtrsim 4$~GeV and $|\rm{Y_{diff}}| \gtrsim 2$
the odderon term wins.

\begin{figure}[!ht]
\includegraphics[width=0.48\textwidth]{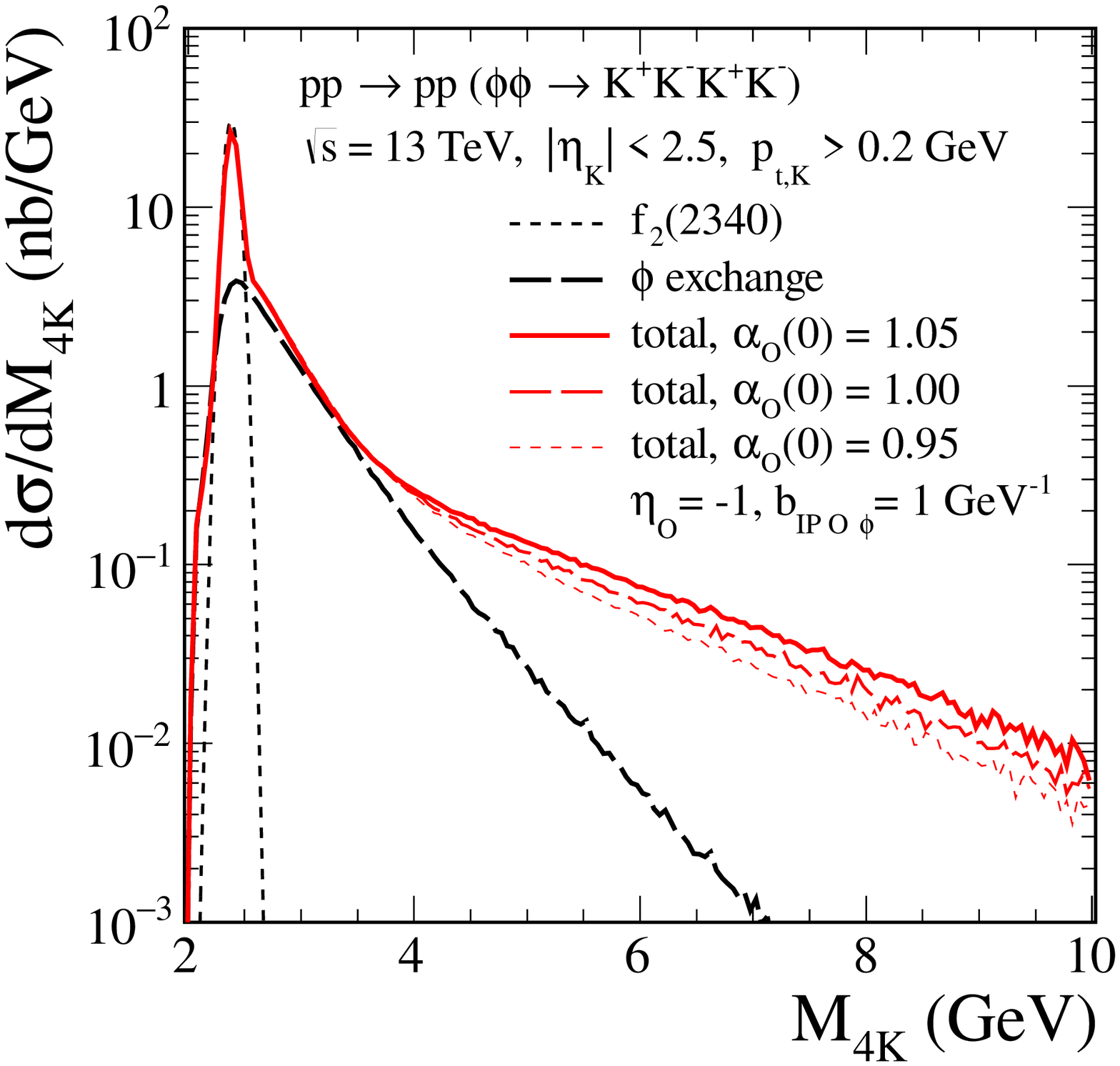}
\includegraphics[width=0.48\textwidth]{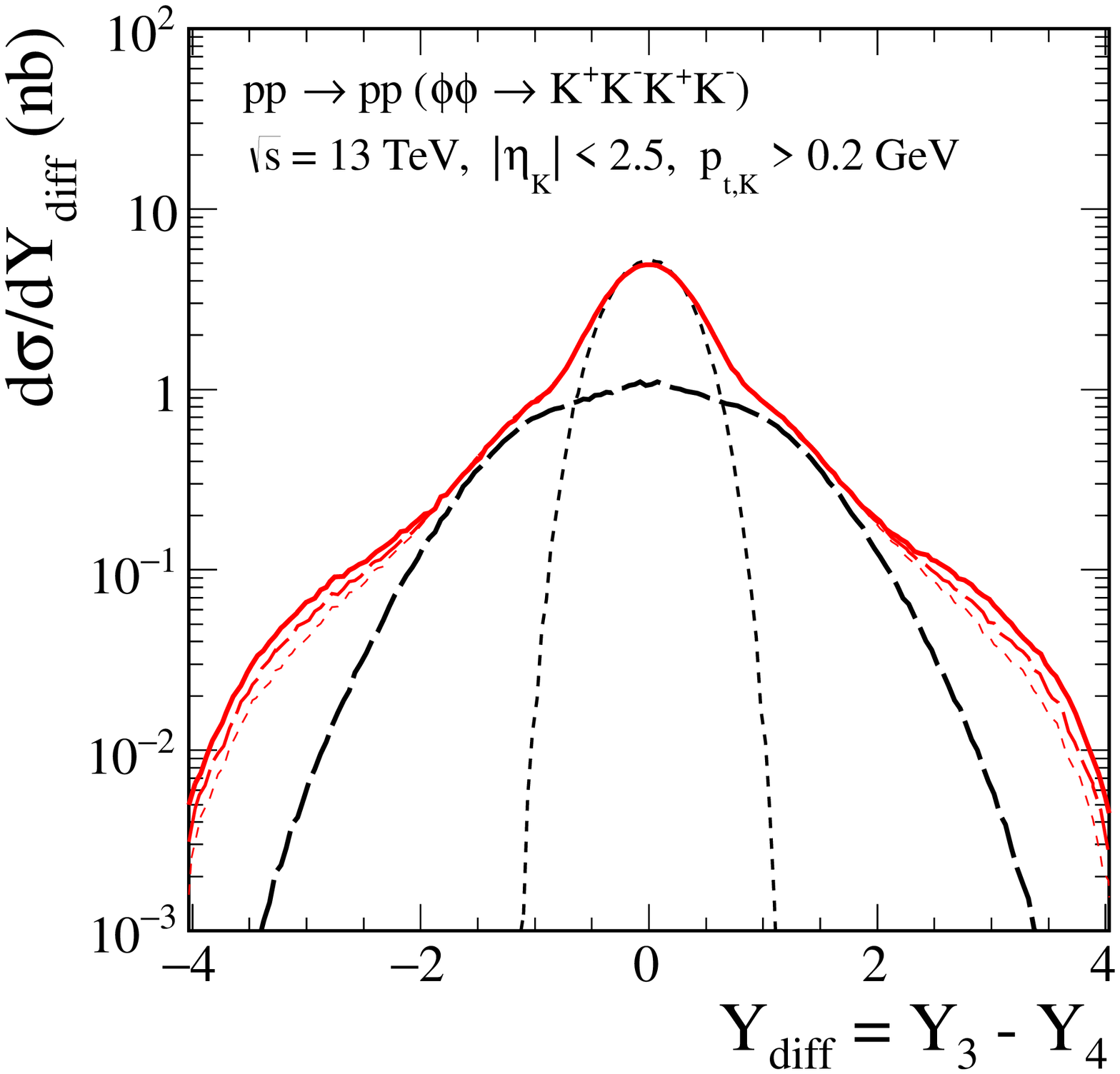}
\caption{\label{odderon_LHC_aux2}
\small
The complete results for $\sqrt{s} = 13$~TeV and $|\eta_{K}| < 2.5$, $p_{t, K} > 0.2$~GeV are shown.
Here we show results for $\eta_{\Ode} = -1$, $a_{\Pom \Ode \phi}= 0$, 
$b_{\Pom \Ode \phi}= 1$~GeV$^{-1}$,
and for various values of the odderon intercept $\alpha_{\Ode}(0)$.}
\end{figure}

In Fig.~\ref{odderon_LHC_aux2} we show the 
complete result including the odderon exchange with
$\eta_{\Ode} = -1$ and various values of the odderon intercept $\alpha_{\Ode}(0)$:
\begin{eqnarray}
\eta_{\Ode} = -1\,, \;\;
\alpha_{\Ode}(0) = 0.95,\, 1.00,\, 1.05\,.
\label{A14_aux}
\end{eqnarray}
Even a much smaller odderon contribution should be visible
for ${\rm M}_{4K} \gtrsim 5$~GeV and $|\rm{Y_{diff}}| > 3$,
provided the experimental statistics (luminosity) is sufficient.
The distributions in ${\rm M}_{4K}$ and $\rm{Y_{diff}}$ seem therefore
to offer good ways to identify the odderon exchange if it is there.

\begin{figure}
\includegraphics[width=9.cm]{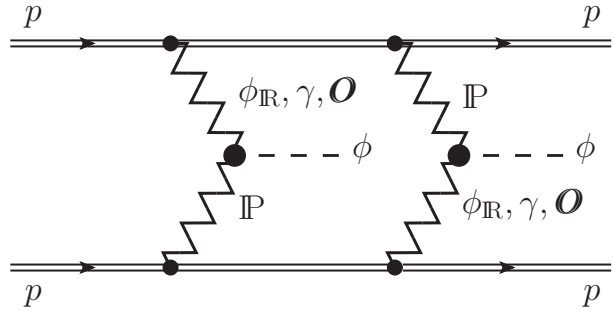} 
  \caption{\label{fig:double_phi}
Example of a diagram for the production of two $\phi$ mesons by two independent exchanges.}
\end{figure}

The small intercept of the $\phi$ reggeon exchange, 
$\alpha_{\phi}(0) = 0.1$ \cite{Collins:1977}
makes the $\phi$-exchange contribution steeply falling 
with increasing ${\rm M}_{4K}$ and $|\rm{Y_{diff}}|$.
Therefore, an odderon with an intercept $\alpha_{\Ode}(0)$ around 1.0
should be clearly visible in these distributions
if the $\Pom \Ode \phi$ coupling is of reasonable size.
This is, at least, the conclusion of our present model study.
Of course, in a real experiment many investigations of the background
will be necessary before one could claim to have seen odderon exchange.
Sources of background are $\phi_{\Reg}$ reggeon exchange as discussed in the present paper.
But one will also have to consider $\omega_{\Reg}$ reggeon exchange and
double $\phi$ production from two independent exchanges as shown in Fig.~\ref{fig:double_phi}.

\section{Conclusions}

In the present paper we have presented first estimates of 
the contributions to the reaction $p p \to p p K^+ K^- K^+ K^-$ 
via the intermediate $\phi(1020) \phi(1020)$ resonance pairs.
This reaction is being analyzed experimentally 
by the ALICE, ATLAS, CMS, and LHCb collaborations.
The analysis of the reaction $pp \to pp (\Pom\Pom \to f_{2} \to \phi \phi)$ can be used
for an identification of the tensor meson states.
We note that the states $f_{2}(2300)$ and $f_{2}(2340)$ are good candidates for tensor glueballs.

We have considered the pomeron-pomeron fusion to $\phi \phi$
through the continuum process, with the $\hat{t}$- and $\hat{u}$-channel $\phi$-meson exchange,
and through the $s$-channel resonance reaction [$\Pom \Pom \to f_{2}(2340) \to \phi \phi$].
The amplitudes for the process have been obtained 
within the tensor-pomeron approach \cite{Ewerz:2013kda}.
By comparing our theoretical results to the cross sections 
found by the WA102 Collaboration \cite{Barberis:1998bq,Barberis:2000em},
we have fixed some coupling parameters 
and the off-shell dependencies of the intermediate $\phi$ mesons.
We have discussed also the $\phi \phi$ production through the
$f_{0}(2100)$ and $\eta(2225)$ resonances,
which were observed in radiative decays of $J/\psi$ \cite{Ablikim:2016hlu}.
We have shown that the contribution of the pseudoscalar $\eta(2225)$ meson 
is disfavored by the WA102 experimental distributions.

We have made estimates of the integrated cross sections 
as well as shown several differential distributions for different experimental conditions.
The distribution in $\rm{Y_{diff}}$, 
the rapidity difference between the two $\phi$-mesons,
depends strongly on the choice of the $f_{2}(2340) \to \phi \phi$ coupling.
The general $f_{2} \phi \phi$ coupling is a sum of two
basic couplings multiplied with two coupling constants; see (\ref{vertex_f2phiphi}). 
Our default values of the coupling parameters
in the $\Pom \Pom f_{2}$ and $f_{2} \phi \phi$ vertices 
can be verified by future experimental results to be obtained at the LHC.
Future studies at the LHC could potentially determine them separately. 
Low-$p_{t,K}$ cuts are required for this purpose.
It has been shown that absorption effects change considerably the shapes
of the ``glueball-filter variable'' distributions as well as those 
for the azimuthal angle between the outgoing protons.

The study of the $p p \to p p \phi \phi$ reaction
offers the possibility to search for effects of the odderon.
Such double diffractive production of two vector mesons
with odderon exchange as a means to look for the latter was discussed in \cite{Ewerz:2003xi}.
In the present paper we have presented a concrete calculation of this process.
Odderon contributions in diffractive production of single vector mesons, e.g.,
$p p \to p p \phi$, were investigated in \cite{Schafer:1991na}.
In the diffractive production of $\phi$ meson pairs, it is possible to have
pomeron-pomeron fusion with intermediate $\hat{t}/\hat{u}$-channel $C=-1$ odderon exchange.
The presence of odderon exchange in the middle of the diagram 
should be important and distinguishable from other contributions
for relatively large rapidity separation between the $\phi$ mesons.
Hence, to study this type of mechanism one should investigate events with
rather large four-kaon invariant masses, outside of the region of resonances.
These events are then ``three-gap events'':
proton--gap--$\phi$--gap--$\phi$--gap--proton.
Experimentally, this should be a clear signature.
A study of such events should allow a determination of the pomeron-odderon-$\phi$ meson
coupling, or at least of an upper limit for it.
Of course, one will have to investigate in detail
the contribution of other exchanges like the $\phi_{\Reg}$ reggeon exchange
studied in the present work.
This could be done, for instance, by adjusting couplings and form factors
at lower ${\rm M}_{\phi \phi}$ and then studying 
the extrapolations to higher ${\rm M}_{\phi \phi}$
where $\phi_{\Reg}$ exchange is a ``background'' to odderon exchange.
Experimentally one has to make sure that one is really dealing with three-gap events.
Thus, additional meson production in the gaps, a reducible background,
must be excluded.
There is, however, also the irreducible background from the production
of two $\phi$ mesons by two independent exchanges;
see Fig.~\ref{fig:double_phi}.
This has to be estimated theoretically and, in a sense, is an absorptive correction.
If an odderon exchange is seen, then the distributions
of the four-kaon invariant mass and of the rapidity difference between
the two $\phi$ mesons will reveal the intercept of the odderon trajectory.

In conclusion we note the following. 
If the final protons in our reaction
(\ref{2to4_reaction_phiphi}) can be measured one can reconstruct
the complete kinematics of the reaction $\Pom + \Pom \to \phi + \phi$.
A detailed study of this reaction as a function of its c.m. energy
${\rm M}_{\phi \phi}$ and its momentum transfer should then be possible.
The great caveat is that one has to get the absorption corrections
under good theoretical control.
The resonances at low ${\rm M}_{\phi \phi}$ could then be investigated in detail.
The special feature of the above reaction, however,
is that the leading term at high energies must be due to a charge
conjugation $C= -1$ exchange since $C = +1$ exchanges like the pomeron 
cannot contribute. Therefore, an odderon would give the leading term
if its intercept is higher than that of the normal $C= -1$ reggeons.
Clearly, an experimental study of CEP of a $\phi$-meson pair 
should be very valuable for clarifying the status of the odderon.
Finally we note that analogous reactions which are suitable for
odderon studies, see \cite{Ewerz:2003xi}, 
are double $J/\psi$ and double $\Upsilon$ central exclusive production.

\acknowledgments
We are indebted to Carlo Ewerz for discussions and comments.
This research was partially supported by
the Polish National Science Centre Grant No. 2014/15/B/ST2/02528
and by the Center for Innovation and Transfer of Natural Sciences 
and Engineering Knowledge in Rzesz{\'o}w.

\bibliography{refs}

\end{document}